\documentclass[twocolumn,prb,showpacs,aps,superscriptaddress,preprintnumbers]{revtex4-1}

\usepackage[english]{babel}
\usepackage[ansinew]{inputenc}
\usepackage{times}
\usepackage{graphicx}
\usepackage{graphics}
\usepackage{amsmath}
\usepackage{amsfonts}
\usepackage{amssymb}
\usepackage{amsbsy}
\usepackage{makeidx}
\usepackage{ytableau}
\usepackage[export]{adjustbox}
\usepackage{algpseudocode}
\usepackage{color}
\usepackage{bm}
\usepackage{appendix}
\usepackage{caption}
\usepackage{subcaption}
\usepackage{float}

\usepackage{soul}

\makeindex

\begin{document}
\preprint{SAND2022-10403R}

\title{Permutation-adapted complete and independent basis for atomic cluster expansion descriptors}

\author{J. M. Goff}
\affiliation{Center for Computing Research, Sandia National Laboratories, Albuquerque, New Mexico 87185, USA}
\author{C. Sievers}
\affiliation{Center for Computing Research, Sandia National Laboratories, Albuquerque, New Mexico 87185, USA}
\affiliation{The Boeing Company, Seattle, Washington 98108, USA }

\author{M. A. Wood}
\affiliation{Center for Computing Research, Sandia National Laboratories, Albuquerque, New Mexico 87185, USA}
\author{A. P. Thompson}
\affiliation{Center for Computing Research, Sandia National Laboratories, Albuquerque, New Mexico 87185, USA}

\date{\today}

\begin{abstract}

 Atomic cluster expansion (ACE) methods provide a systematic way to describe particle local environments of arbitrary body order.  For practical applications it is often required that the basis of cluster functions be symmetrized with respect to rotations and permutations. Existing methodologies yield sets of symmetrized functions that are over-complete. These methodologies thus require an additional numerical procedure, such as singular value decomposition (SVD), to eliminate redundant functions. In this work, it is shown that analytical linear relationships for subsets of cluster functions may be derived using recursion and permutation properties of generalized Wigner symbols. From these relationships, subsets (blocks) of cluster functions can be selected such that, within each block, functions are guaranteed to be linearly independent. It is conjectured that this block-wise independent set of permutation-adapted rotation and permutation invariant (PA-RPI) functions forms a complete, independent basis for ACE. Along with the first analytical proofs of block-wise linear dependence of ACE cluster functions and other theoretical arguments, numerical results are offered to demonstrate this. The utility of the method is demonstrated in the development of an ACE interatomic potential for tantalum. Using the new basis functions in combination with Bayesian compressive sensing sparse regression, some high degree descriptors are observed to persist and help achieve high-accuracy models.

\end{abstract}


\maketitle

\section{\label{intro}Introduction}
%%%

In a wide range of atomistic and particle systems, the theory of quantum angular momentum, including its connection to with spherical harmonics, is useful. Systems with multiple particles or multiple coupled angular momenta are of particular interest for recently developed many body descriptors for machine-learned interatomic potentials.  The coupling of quantum angular momenta is relevant for applications such as the energy levels of electronic orbitals in atoms, fine structure in electronic spectroscopy, and wave functions of atomic nuclei.\cite{eisberg1985quantum} In these and many other cases, spherical harmonics are the natural basis. Coupling quantum angular momenta and related tasks such as the reduction of products of spherical harmonics may be accomplished through the Clebsch-Gordan (CG) coefficients or other related values such as the Wigner-3j symbols. The traditional CG coefficients and Wigner-3j symbols are used to describe the coupling of two quantum angular momenta in terms of a third, or the analogous reduction of a tensor product of two spherical harmonics. The traditional CG coefficients and Wigner-3j symbols (a.k.a. coupling coefficients) and some of their properties are highlighted in original work from Wigner.\cite{wigner2012group} Permutation symmetries and recursion properties of the Wigner-3j symbols are known, and often correspond to changes in the ordering of angular momenta to be added.\cite{yutsis_mathematical_1962} The properties  of the generalized Wigner symbols, those used to couple $N$ angular momenta, are far less studied. Understanding these symmetries would be useful in a variety of fields including quantum mechanics, acoustical analysis, and in the present use-case of interatomic potentials.\cite{edmonds_angular_1996,zotkin_regularized_2009,drautz_atomic_2019}

Spherical harmonics are commonly used in descriptors for systems of particles and particularly for systems of atoms and their local bonding environments. The spherical harmonics, or other related symmetry functions, are often used to span angular degrees of freedom in descriptors that allow for variable atomic positions. Such descriptors encode information about the atomic environment in terms of its spatial, chemical, and potentially other degrees of freedom.\cite{zuo2020performance, musil2021physics} For example, smooth overlap of atomic positions (SOAP) descriptors use spherical harmonics to describe angular character of atomic environments.\cite{bartok_representing_2013} Hyperspherical harmonics are used in the spectral neighbor analysis potential (SNAP) descriptors.\cite{thompson_spectral_2015} The preceding descriptors are restricted to specific body-orders.  For example, SNAP models are comprised of 4-body descriptors. A complete basis of $N$-body descriptors that reflects the physical and chemical interactions of atoms would allow for greater ease of use and (long sought after) interpretability of constructed ML interatomic potentials. 
One of the first examples of a $N$-body model was the moment-tensor potential (MTP).\cite{shapeev_moment_2016} The notion of angular momentum addition may be found in these more generalized $N$-body types of models as well. The atom-centered density correlations (ACDC) models produce $N$-body equivariant descriptors for arbitrary equivariant character: scalar, vectorial, tensorial, and so on with progressively higher body-orders.\cite{willatt_atom-density_2019,nigam_recursive_2020} A key feature of this method is the generation of the $N$-body equivariants through a recursive reduction of spherical harmonics. This is analogous recursive addition of quantum angular momenta up to some final angular momentum state with specified equivariant character. A set of linearly-independent ACDC descriptors is produced through a combination of recursive angular momentum addition rules, as well as principal component analysis over a set of atomic configurations. Recently, this method has been extended to message-passing networks.\cite{nigam_unified_2022}

While the reduction of spherical harmonic products in ACDC models is done by recursively reducing pairs of spherical harmonics, the atomic cluster expansion (ACE) method reduces a product of $N$ spherical harmonics at once.\cite{drautz_atomic_2019} The contraction of two spherical harmonics in the ACDC method is done using traditional CG coefficients or Wigner-3j symbols, while the contraction of $N$ spherical harmonics in ACE is treated using generalized Wigner symbols. Both preceding approaches for contracting products of spherical harmonics are analogous to coupling arbitrary numbers of quantum angular momenta but with different order of addition/coupling. The result of any coupling scheme is that a product of spherical harmonics may be reduced to a single spherical harmonic with some multiplicity. Leaving detailed justification of this multiplicity for later, it is due to the rules for addition of quantum angular momenta that allow for multiple valid couplings of one product of spherical harmonics. There are also rigorous mathematical connections between this multiplicity and the dimension of a rotationally invariant/equivariant product basis.\cite{dusson_atomic_2022,yutsis_mathematical_1962} 

In addition to invariance or equivariance with respect to rotations, invariance with respect to permutations is also often imposed in ACE and similar methods. No matter what addition/coupling scheme is used, coupling a product of spherical harmonics that is invariant with respect to permutations may result in some of the reduced functions being linearly dependent. This means that when $N$-body functions and descriptors are generated by reducing products of permutation-invariant functions (including a spherical harmonic component) the resulting set is generally over-complete. Singular value decomposition (SVD) typically follows in ACE and similar methods to address this. As evidenced by numerical results and new analytical expressions in this work, linearly dependent functions occur when the product of functions to be coupled contains duplicate quantum numbers. This can be challenging to observe and exploit in many schemes for enumerating ACE functions. In most enumeration schemes, spherical harmonics with duplicate indices are not always directly coupled. A key innovation of this approach is that analytical recursion and permutation relationships are derived for the generalized Wigner symbols. The enumeration scheme is adapted to the permutation and recursion properties of these generalized Wigner symbols. In this permutation-adapted (PA) method, cluster products are constructed such that analytical linear relationships within subsets (blocks) of ACE functions are straightforward to derive and apply. These relationships are applied within all blocks of functions to construct a set of rotation and permutation-invariant functions that is complete. Linear independence of these functions is guaranteed within each block. We refer to this as the permutation-adapted rotation and permutation invariant set (PA-RPI ACE). Below we present a theoretical description of the procedure. Since linear independence is only guaranteed within blocks of functions in the PA method, numerical evidence is also provided to support its validity. From these results, we conjecture that a set of rotation and permutation invariant functions constructed this way forms a complete and independent basis that can be used in ACE and similar methods.
 
\section{Theory Background}
\subsection{\label{sec:defs} Definitions}
\begin{itemize}

    \item $N$ : (Rank) The number of bonds in an ACE cluster (analogous to the number of vertices in a fixed-lattice cluster).

    \item $\boldsymbol{l}$ : Multiset of $N$ non-negative integer indices $l_i$, the angular momentum quantum numbers of the atomic basis functions. Angular indices may appear more than once.
    
    \item $\boldsymbol{L}$ : Multiset of $N-1$ non-negative integer indices $L_k$, the angular momentum quantum numbers of the intermediate functions that are used for pairwise reduction of $\boldsymbol{l}$. Intermediate angular indices may appear more than once.
    
    \item Two generalized Wigner symbols will be considered equivalent when, for the same multiset of intermediates $\boldsymbol{L}$, either one of the following conditions are met:
    \begin{enumerate}
        \item  $(\boldsymbol{lm})_a  =(\boldsymbol{lm})_b $ such that all tuples of angular momentum and projection quantum numbers are equal element-wise: $(l_i,m_i)_{a}=(l_i,m_i)_{b} \forall i \in N$.
        \item $(\boldsymbol{lm})_a = \sigma(\boldsymbol{lm})_b $ where a permutation of elements yields multisets of angular momentum and projection quantum numbers that are equal element-wise. This holds only for specific permutations of $S_N$.
    \end{enumerate}

    \item $G_N$: the group of equivalent permutations for a generalized Wigner symbol.

    \item $\boldsymbol{lL}$ : Combined atom-indexed angular momentum quantum numbers $l_i$ and intermediate $L_k$ indices for an angular momentum coupling, used to unambiguously index an $N$-bond angular function of rank $N$ with $N-1$ intermediates. This combined collection of $\boldsymbol{lL}$ indices may also be written unambiguously as a binary tree.

    \item $C_1, C_2$ : The two child nodes of parent node $P$ in a binary tree.
    
    \item $\triangle(l_1,l_2,l_3)$ : Triangle conditions: $|l_1-l_2| \le l_3 \le (l_1 + l_2)$ (conditions for coupling quantum angular momenta).
    
\end{itemize}
\subsection{The atomic cluster expansion}

First developed by Drautz in 2019, the ACE formalism was shown to be an extension of many interatomic potential models.\cite{drautz_atomic_2019} ACE has been used to produce accurate and efficient energy models, models of vector properties such as magnetism, and have been used in other methods such as message passing networks.\cite{bochkarev_efficient_2022,batatia_design_2022} All these methods require generation of a complete set of rotation and permutation invariant cluster functions. Much like the traditional fixed-lattice cluster expansion it was based on, a key benefit of the method is a complete description of the configurational space of an atomic system.\cite{sanchez_generalized_1984} The key distinction between ACE and the fixed lattice cluster expansion is the extension to include continuous spatial degrees of freedom. In their most simple form, ACE models are linear expansions of atomic properties in terms of ACE basis functions. However a complete orthogonal ACE basis has not been defined analytically. Previous constructions of ACE bases have employed numerical methods to eliminate dependent functions. In this work, the fundamental connection between ACE and quantum angular momentum is used to expose such linear relationships analytically.

The tensor product basis for ACE starts with a set of complete orthogonal single bond functions.\cite{drautz_atomic_2019}
\begin{equation}
    \phi_{nlm}(\boldsymbol{r}_{ij}) = R_n(r_{ij}) Y_l^m(\boldsymbol{\hat{r}_{ij}}) 
    \label{eq:single_bond}
\end{equation}
where $R_n(r_{ij})$ is a family of orthogonal radial basis functions and $Y_l^m(\boldsymbol{\hat{r}_{ij}})$ are the spherical harmonics. These span radial and angular degrees of freedom for one site/bond pair. The tensor product (cluster) basis is comprised of all possible products of the single bond basis. 
\begin{equation}
    \Phi_{\boldsymbol{{nlm}}}(1,2, \cdots N) = \prod_{j=1}^N \phi_{{(nlm)}_j}(\boldsymbol{r}_{ij})
    \label{eq:direct_prod}
\end{equation}
where the product is taken over neighbors at positions $\boldsymbol{r}_{ij}$ up to $N$ times. These cluster functions are indexed by the multiple functions comprising the product, and there may be repeated indices. Therefore multisets of non-angular and angular indices are used to label the cluster functions.
\begin{equation}
    \boldsymbol{{nlm}} = (n_1 n_2 \cdots n_N) (l_1 l_2 \cdots l_N) (m_1 m_2 \cdots m_N ).
    \label{eq:index_vector}
\end{equation}
The cluster product basis in Eq. \eqref{eq:direct_prod} spans the radial and angular degrees of freedom for a collection of site/bond pairs.
Restrictions on $\boldsymbol{n}$ and other non-angular function indices are only those imposed by the basis for that subspace, such as $\boldsymbol{n} : n_i \ge 0  \forall i \in N$ required by many radial bases. The projection quantum numbers are bound by the angular momentum quantum numbers, $-l_i \le m_i \le l_i \forall i \in N$. The allowed values for angular momentum quantum numbers are restricted by the polygon condition, which is the generalization of the triangle condition for coupling two quantum angular momenta.\cite{yutsis_mathematical_1962} It is clear that the basis in Eq. \eqref{eq:direct_prod} may be constructed such that it is complete and orthogonal. However, it is not invariant with respect to rotations and permutations. Independently enforcing permutation invariance (PI) and rotational invariance (RI) of the functions in the cluster basis is straightforward, but the combined rotation and permutation invariance (RPI) is less so. Imposing invariance with respect to operations in the joint set of rotations, elements of SO(3), and permutations, elements of $S_N$, on the cluster basis results in an over-complete set that is typically reduced numerically.\cite{drautz_atomic_2020,dusson_atomic_2022} Understanding and resolving this is of theoretical and practical interest for ACE and related methods. Such benefits have been argued in the definition of a complete, orthogonal basis for the traditional cluster expansion.\cite{sanchez_generalized_1984} As alluded to before, the challenge with doing so for ACE arises when both permutation and rotation invariance are imposed on the cluster products. 

There are a number of ways to discuss how rotation and permutation invariance are both imposed on the cluster basis.\cite{drautz_atomic_2019,dusson_atomic_2022} For now we consider imposing them one after the other. Beginning with the more straightforward condition of permutation invariance, the product functions in Eq. \eqref{eq:direct_prod} may be made symmetric with respect to exchanges of coordinates by summing over all possible permutations.\cite{dusson_atomic_2022}
\begin{equation}
    \bar{\Phi}_{\boldsymbol{nlm}}(1,2,\cdots N) = \frac{1}{\sqrt{N !}}\sum_{\sigma \in S_N} \Phi_{\boldsymbol{nlm}} \sigma \big( (1,2,\cdots N) \big) 
    \label{eq:pi_product_cluster}
\end{equation}
The sum in Eq. \eqref{eq:pi_product_cluster} runs over all permutations, $\sigma$, in the symmetric group $S_N$. Invariance of the product function with respect to permutations is now indicated with a bar. The functional form of this permutation invariant (PI) cluster basis closely resembles some parts of the traditional cluster expansion, however it is not always used in practice.\cite{sanchez_generalized_1984}

In practical applications, the cluster basis in Eq. \eqref{eq:direct_prod} and the permutation-invariant counterpart in Eq. \eqref{eq:pi_product_cluster} are avoided due to exponential scaling with rank of the cluster function. An ``atomic base" is constructed using the atomic density projection from the SOAP method, which recovers linear scaling in the number of neighbors.\cite{bartok_representing_2013,drautz_atomic_2019} The atomic density for atom $i$ is projected onto the complete single bond basis \textit{c.f.} Eq. \eqref{eq:single_bond}. 
\begin{equation}
    A_{ nlm}=\langle \rho | \phi_{nlm} \rangle
    \label{eq:projection}
\end{equation}
Note that because the atomic density is permutation-invariant by construction, the $A_{ nlm}$ density projections are also permutation-invariant. The cluster basis in Eq. \eqref{eq:direct_prod} is replaced with products of Eq. \eqref{eq:projection}, yielding,
\begin{equation}
    \boldsymbol{A}_{\boldsymbol{nlm}} = \prod_{\kappa}^N A_{ (nlm)_\kappa}.
    \label{eq:atomic_base}
\end{equation}
This product basis in Eq. \eqref{eq:atomic_base} possesses permutation invariance by construction, but not with respect to rotations. While this atomic base resolves scaling issues encountered with the traditional cluster products, it does have problematic self-interactions.\cite{dusson_atomic_2022} These self-interactions can affect the semi-numerical construction of an ACE basis.\cite{drautz_atomic_2020,dusson_atomic_2022}

Separately, rotational invariance or equivariance may be imposed on PI functions with the generalized Wigner symbols.
\begin{equation}
    B_{\boldsymbol{nlL}} = \sum_{\boldsymbol{m}}W_{\boldsymbol{l}}^{\boldsymbol{m}}(\boldsymbol{L},\boldsymbol{M})\boldsymbol{A}_{\boldsymbol{nlm}}
    \label{eq:ACE_desc}
\end{equation}
The $W_{\boldsymbol{l}}^{\boldsymbol{m}}(\boldsymbol{L},\boldsymbol{M})$ are the generalized Wigner symbols, and are indexed by a multiset of angular momentum quantum numbers $\boldsymbol{l} = {l_1 , l_2, \cdots l_N}$ a multiset of projection quantum numbers $\boldsymbol{m} = {m_1 , m_2, \cdots m_N}$, and a multiset of intermediate angular momentum quantum numbers, $\boldsymbol{L} = \{ L_1 , L_2 \cdots  L_{N-2}, L_R \}$. Summing the PI products in Eq. \eqref{eq:ACE_desc} multiplied by the corresponding generalized Wigner symbols with $L_R=0$ results in rotational invariance of the ACE descriptors. The intermediate angular momenta will be defined in more detail later, but for now it is sufficient to understand these as auxiliary quantities needed to couple four or more quantum angular momenta. It is important to note that the polygon conditions allow for more than one multiset of the intermediate angular momentum indices \textit{for the same} $\boldsymbol{l}$. To enumerate a complete set of ACE functions, one needs to obtain all valid distinct combinations of angular and non-angular indices. Many conventions can be adopted to achieve this.

One common indexing convention for ACE functions relies on ordering tuples of non-angular and angular indices, $(n_i,l_i)$, lexicographically to obtain distinct $\boldsymbol{n}\boldsymbol{l}$ labels.\cite{drautz_atomic_2019,dusson_atomic_2022} Any other ordering conventions that avoid repeated combinations of $\boldsymbol{n}$ with $\boldsymbol{l}$ could also be used. In many methods for enumerating ACE functions, it is common to also consider all valid multisets of intermediates allowed by polygon conditions of the underlying angular functions, denoted by $\{ \boldsymbol{L} : \triangle^g( \boldsymbol{l}, \boldsymbol{L}) \}$. These different collections of intermediates may correspond to linearly-independent rotation and permutation invariant functions \textit{for the same $\boldsymbol{n}$ and $\boldsymbol{l}$}. To avoid ambiguity and loss of completeness, the intermediates should also be indexed. When indexed in this way, an over-complete set of ACE functions with equivariant character corresponding to angular momentum quantum number $L_R$ can be generated with an intuitive block structure. Provided that the radial indices, $n_i$, and angular indices $l_i$ are non-negative, then the enumeration of functions may be given in terms of the blocks discussed above, as, 
\begin{equation}
\begin{split}
& \beta_{\boldsymbol{nl}} = \{  B_{\boldsymbol{n} \boldsymbol{l} \boldsymbol{L} } :  \forall \boldsymbol{L} \; if \; \triangle^{g} (\boldsymbol{l}, \boldsymbol{L}) \} \\
& \mathcal{S}^{OC}_N = \{ \beta_{\boldsymbol{nl}} : (n_i,l_i) \; ordered, ( n_i \ge 0, l_i \ge 0)  \forall \, i \in N \} 
%& \{ \{ B_{\boldsymbol{n} \boldsymbol{l} \boldsymbol{L} } \} : \triangle^g(\boldsymbol{l}) \, , (n_i,l_i) \; ordered \, \} \\
%&\{ B_{\boldsymbol{n} \boldsymbol{l} \boldsymbol{L} }  : \boldsymbol{L} \in \{ \boldsymbol{L} \}_{L_R}^{\boldsymbol{l}} \; \} 
\end{split}
\label{eq:bblocks}
\end{equation}
where a block of functions, $\beta_{\boldsymbol{nl}}$, comprises part of the (over)-complete set of all rank $N$ RPI functions indexed by lexicographically ordered non-angular and angular indices, $\beta_{\boldsymbol{nl}} \subset \mathcal{S}^{OC}_N $. The ``$OC$" in the superscript indicates that this is over-complete and its elements are indexed by all distinct combinations of angular function indices and non-angular indices obtained through lexicographical ordering. It is worth noting that the blocks are empty if the angular indices do not obey the polygon conditions, $\triangle^g$. These polygon conditions are the generalized angular momentum coupling conditions (a.k.a. triangle conditions) from Ref.~[\!\citenum{yutsis_mathematical_1962}], and are defined in detail later. The blocks of functions are defined as those sharing non-angular and angular function indices, such as those defined in Eq. \eqref{eq:bblocks}, may have different intermediates. These blocks have special significance. Once rotation invariance and permutation invariance are both imposed, functions within these blocks are not always linearly independent. For $N\le3$, this is not a concern because for such cases it is often true that, $size(\beta_{\boldsymbol{nl}})=1$.  Numerical results from Ref.~[\!\citenum{dusson_atomic_2022}] indicate that these linear dependencies cannot be neglected when considering higher rank functions and occur when there are duplicate indices in $\boldsymbol{n},\boldsymbol{l}$. Semi-numerical approaches for ACE generally rely on performing SVD over blocks of functions to eliminate redundant functions.\cite{bochkarev_efficient_2022} There is still little analytical explanation for where these linear dependencies originate from. This is, in part, due to the restricted indexing convention. Enforcing strict lexicographical ordering of indices can make it difficult to expose linear relationships between functions.

The indexing convention is ultimately arbitrary. In this work, we adopt an indexing convention that makes the derivation and application of linear relationships between RPI functions more tractable. The fundamental connection between the spherical harmonics and quantum angular momentum is exploited. By generalizing the recursion relationships for raised/lowered coupled quantum angular momenta, we are able to derive relationships for blocks of functions in Eq. \eqref{eq:bblocks} and define linearly independent RPI function sequences within these blocks. Such a task is tractable after adapting the coupled functions to the permutation symmetries of the generalized Wigner symbols. For this reason, we relax the lexicographical ordering of the non-angular and angular indices and adopt another convention. 
\begin{equation}
\begin{split}
& \beta_{\boldsymbol{nl}}^{PA} = \{  B_{\boldsymbol{n} \boldsymbol{l} \boldsymbol{L} } :  \forall \boldsymbol{L} \; if \; \triangle^{g} (\boldsymbol{l}_{fc}, \boldsymbol{L})\, ,  \\
& B_{\boldsymbol{n} \boldsymbol{l} \boldsymbol{L} } \in \mathcal{F}_{a}^{PA}(P_f(\boldsymbol{n}), P_f(\boldsymbol{l}))  \} \\
& \mathcal{S}^{OC}_N = \{ \beta_{\boldsymbol{nl}} :  \boldsymbol{n}= \varsigma(\boldsymbol{n}) \, , \boldsymbol{l} = \boldsymbol{l}_{fc}, ( n_i \ge 0, l_i \ge 0)  \forall \, i \in N \}
\end{split}
\label{eq:bblocks_pa}
\end{equation}
The blocks are still comprised of functions sharing $\boldsymbol{nl}$. In Eq. \eqref{eq:bblocks_pa}, the distinct $\boldsymbol{nl}$ labels are just defined using a different ordering convention. Apart from a different ordering of non-angular and angular indices, a new restriction is applied to the intermediates. One function with one multiset of intermediates may not be related to another by ladder relationships. These ladder relationships resemble those for raising/lowering two coupled quantum angular momentum states.\cite{rose1995elementary} They have been generalized to arbitrary $N$ and, for the application in ACE methods, account for the permutation invariance of the $N$ coupled functions. Using these, one may define a function sequence of independent RPI functions, denoted as $\mathcal{F}_{a}^{PA}$. This function sequences are obtained from the respective over-complete blocks, $\beta_{\boldsymbol{nl}}$. The sampling resulting in this linearly independent sequence is obtained from repeated application of ladder relationships. In general, the sampling depends on the numbers of duplicate non-angular indices as well as the numbers of duplicate angular indices, which are compactly described by frequency partitions of the index multisets, $P_f(\boldsymbol{n}), P_f(\boldsymbol{l})$. In practice, the angular function indices are permuted to allow for straightforward application of ladder relationships. The adapted permutation of angular indices and the ladder relationships themselves will be defined in detail after the generalized Wigner symbols are more formally introduced. For now, it is sufficient to understand that the permutation, $\sigma(\boldsymbol{l})=\boldsymbol{l}_{fc}$ comes from a frequency partition, $P_{fc}$, that has been adapted to the permutation symmetries induced by the generalized Wigner symbols. It is adapted such that it maximizes how many functions with duplicate indices are coupled, which allows for straightforward application of ladder relationships. Distinct permutations of non-angular indices $\varsigma(\boldsymbol{n})$ may be generated in a way similar to lexicographical ordering based on the multiplicity of angular indices, encoded in $P_{f}(\boldsymbol{l})$. Ordering the non-angular indices in this way produces the same number of distinct $\boldsymbol{nl}$ labels as that generated by lexicographical ordering of $(n_i,l_i)$ tuples. 

Linear relationships between RPI functions are obtained by raising/lowering the intermediates, applying permutation properties, and grouping like terms in Eq. \eqref{eq:ACE_desc}. These relationships are referred to as ladder relationships. The ladder relationships are at the core of the proposed approach, and are used to define sequences of independent RPI functions. It is important to note that ladder relationships are defined only for specific blocks of functions with fixed $\boldsymbol{n}$ and fixed $\boldsymbol{l}$. As a result, linear independence can be guaranteed within each respective block. This is done efficiently by adapting the functions within a block to the permutation symmetries of the underlying generalized Wigner symbols. Then in terms of these convenient permutation-adapted (PA) blocks, we define a complete set of functions under the condition that, within each block, no functions are related by recursions. This is referred to as the permutation-adapted rotation and permutation invariant (PA-RPI) set. In the PA-RPI set, functions are guaranteed to be independent within each block, however it is not strictly guaranteed to form a basis. Relationships are not derived and linear independence is not proven for functions in different blocks (e.g., with different $\boldsymbol{nl}$ or different ranks). Considering linear dependence only for distinct blocks of functions is supported by numerical results in this and other works, but it does not strictly guarantee an independent basis.\cite{dusson_atomic_2022}

\begin{figure}
    \centering
    \includegraphics[width=0.48\textwidth]{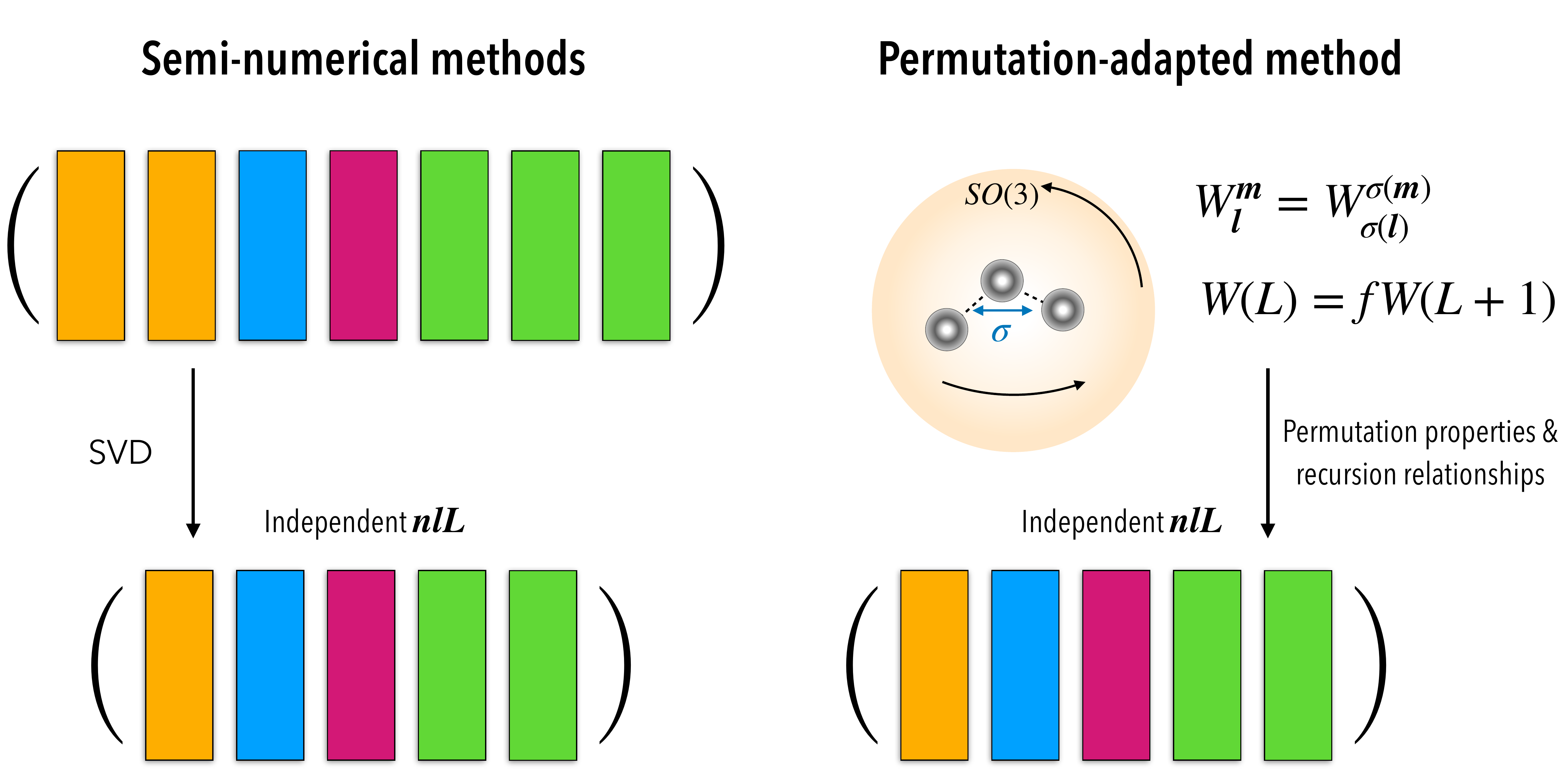}
    \caption{Previous semi-numerical methods produce linearly independent ACE descriptors by applying singular value decomposition (SVD) to a larger set of dependent lexicographically ordered descriptor labels $\boldsymbol{nl}$. The new permutation-adapted method takes advantage of permutational symmetries and recursion relationships embedded in the generalized Wigner symbols to define sets descriptors that are block-wise independent \textit{a priori}. }
    \label{fig:num_vs_rpi}
\end{figure}

The set of ACE descriptors in Eq. \eqref{eq:ACE_desc} is poorly conditioned for SVD due to self-interactions.\cite{dusson_atomic_2022} This poses a problem for some semi-numerical approaches to ACE using SVD, as it may not be numerically stable for very large $N$ and polynomial degree. However, one compelling feature of ACE models is the ability to generalize to arbitrarily large body order with varying degrees of radial and angular character.\cite{drautz_atomic_2019} It is therefore desirable to define systematic approaches for ACE with large $N$ and polynomial degree. In some cases, as we will show in this work, retaining a small number of high body-order interactions can help reduce error in models. Semi-numerical methods can produce sets of ACE descriptors that are adequate for many practical applications.\cite{lysogorskiy_performant_2021,bochkarev_efficient_2022} However, these approaches may not be stable for arbitrarily high-degree functions, and testing for such cases is difficult and possibly prohibitively expensive.\cite{dusson_atomic_2022} Symmetrized bases of atomic environment descriptors possessing \textit{either} permutational invariance (PI) or rotationally invariance/equivariance (RI), independently have already been constructed.\cite{braams2009permutationally,drautz_atomic_2019,yutsis_mathematical_1962} To our knowledge, the completeness and independence of an ACE basis and explanations of linear dependence between RPI cluster functions remains to be analytically shown. We are able to derive these analytical linear relationships within blocks of functions. While an all-analytical basis is not strictly proven in this work with a derivation of relationships between blocks of function or a formal inner product, we do conjecture that a complete and independent basis is obtained when generating the PA set. This is supported by some numerical results and our newly derived analytical relationships for functions within blocks.

\section{Methodology}
\subsection{Rotation Invariant Functions}
\subsubsection{Wigner symbols and angular momentum coupling}
To obtain a complete RPI set, it is useful to know the dimension of both the angular and the non-angular subspaces. We will begin with the angular portion and defining a set of angular functions. The angular cluster functions are comprised of a product of $N$ spherical harmonics, and is not invariant with respect to rotations.  
To achieve this, we will introduce the Wigner symbols which allow one to reduce a product of spherical harmonics to one spherical harmonic. The fact that this is analogous to coupling quantum angular momenta is a key connection that helps prove linear independence of descriptors in later sections. These generalized Wigner symbols may be used to reduce the product of $N$ spherical harmonics to a single spherical harmonic with angular momentum and projection quantum numbers both equal to zero, $Y_0^0$. This $Y_0^0$ function is rotationally invariant. The generalized Wigner symbols will be defined beginning with simple cases and theory for quantum angular momentum coupling. 

The Wigner-3j symbols are commonly used in the reduction of 3 angular momenta. Generalized Wigner symbols are comprised of multiple Wigner-3j symbols. They are closely related to the Clebsch-Gordan (CG) coefficients from quantum angular momentum theory. The traditional Clebsch-Gordan coefficients may be written in symbolic and matrix form
\begin{equation}
    C^{m_1,m_2,m_3}_{l_1,l_2,l_3}=     \begin{bmatrix}
     l_1 & l_2 & l_3\\
     m_1 & m_2 & m_3
\end{bmatrix} .
    \label{eq:cg_traditional}
\end{equation}
In symbolic and matrix form, the traditional Wigner-3j symbols are:
\begin{equation}
\begin{split}
&W^{m_1, m_2, m_3}_{l_1,l_2,l_3} =  \frac{(-1)^{l_1-l_2 -m_3}}{\sqrt{2l_3 +1}}C^{m_1,m_2,m_3}_{l_1,l_2,l_3} \\ &=  \begin{pmatrix}
     l_1 & l_2 & l_3\\
     m_1 & m_2 & m_3
\end{pmatrix} 
\end{split}.
    \label{eq:w3j_traditional}
\end{equation}
Explicit algebraic forms for Eqs. \eqref{eq:cg_traditional} and \eqref{eq:w3j_traditional} may be found elsewhere.\cite{wigner2012group,yutsis_mathematical_1962} For non-zero traditional Wigner-3j symbols, the triangle condition must be obeyed, $\triangle [l_1,l_2,l_3] = |l_1 - l_2| \le l_3 \le (l_1 + l_2)$. Additionally, there are also conditions on the $m_i$. For non-zero Wigner-3j symbols, we must have $m_1 + m_2 + m_3 = 0$. As shown in Eq. \eqref{eq:oddeven_w3j_perms}, traditional Wigner-3j symbols are equivalent under permutations of $(l_i,m_i)$ tuples (columns in matrix form). It may also be convenient to express Wigner-3j symbols as binary trees, and we will refer to these within, as coupling trees.\cite{yutsis_mathematical_1962} 

\begin{equation}
    \begin{array}{ccc}
    & l_3 & \\
    / & & \backslash \\
    l_1 & & l_2
\end{array}
\label{eq:simple_tree}
\end{equation}

The coupling tree structure in Eq. \eqref{eq:simple_tree} helps illustrate how the traditional Wigner-3j symbols are used to couple two spherical harmonics. Additionally it provides a simple graphical way to show which permutations of $l_i$ yield equivalent Wigner symbols. If one were to permute the two children, switching $l_1$ and $l_2$, the Wigner-3j symbols would remain unchanged. To couple more than two, the generalized Wigner symbols are used. The generalized Wigner symbols are contractions of multiple Wigner-3j symbols. These are constructed such that $N$ spherical harmonics coupled two at a time until only one remains. 

There are multiple ways to construct the generalized Wigner symbols and notation can be challenging. There is some ambiguity in how the $N$ spherical harmonics are coupled; the order in which they are coupled is arbitrary. One example of this arbitrary order of reduction for four spherical harmonics is analogous to adding angular momenta to form an intermediate, $l_1 + l_2 \rightarrow L_1$, and again for the next two, $l_3 + l_4 \rightarrow L_2$ then coupling the intermediates to the reduced spherical harmonic with angular momentum and projection, $L_R,M_R$. The corresponding generalized Wigner symbol in matrix form is,
\begin{equation}
\begin{split}
&\begin{pmatrix}
     l_1 & l_2 & l_3 & l_4\\
     m_1 & m_2 & m_3 & m_4
\end{pmatrix}(\boldsymbol{L},\boldsymbol{M}) =  \\ &\sum_{M_{1},M_{2}} \phi (\boldsymbol{L},\boldsymbol{M})  \cdot\begin{pmatrix}
     l_1 & l_2 & L_{1}\\
     m_1 & m_2 & -M_{1}
\end{pmatrix} \\ 
& \cdot \begin{pmatrix}
     l_3 & l_4 & L_{2} \\
     m_3 & m_4 & -M_{2}
\end{pmatrix}
\begin{pmatrix}
     L_1 & L_2 & L_R \\
     M_1 & M_2 & -M_R
\end{pmatrix} 
\end{split}
\label{eq:first_rank4}
\end{equation}
where $\phi (\boldsymbol{L},\boldsymbol{M})= \prod_k (-1)^{L_k-M_k}$ and the projection quantum numbers implicitly define the intermediate projections, $(M_1 = m_1 + m_2 , M_2= m_3+m_4)$. Another valid generalized Wigner symbol could be constructed by coupling of $l_1 + l_2 = L_{12}$, then $l_3 + L_{12} = L_{123}$, and finally $l_4 + L_{123} = L_{R}$. As one can see, there are many others. Generalized Wigner symbols constructed with different coupling schemes are not equivalent in general, but are related by some linear transformation.\cite{yutsis_mathematical_1962} A coupling scheme will often be denoted with a permutation, $\sigma_c$, of leaves and/or a binary tree.\cite{yutsis_mathematical_1962,drautz_atomic_2020,bochkarev_efficient_2022} We will refer to this permutation as the coupling permutation or coupling scheme. The permutation itself describes the order in which spherical harmonics are reduced. For any coupling scheme, the intermediates must obey the proper triangle conditions for constituent Wigner symbols. For Eq. \eqref{eq:first_rank4}, they are $\triangle[l_1,l_2,L_1]$, $\triangle[l_3,l_4,L_2]$, and $\triangle[L_1,L_2,L_R]$. These conditions restrict the intermediate angular momenta $L_1,L_2$. When represented as a binary tree, the coupling scheme is given by the structure of the tree. The $l_i$ form the leaves of the tree while the $L_k$ are the internal nodes. Some examples are given in Fig. \ref{fig:first_tree}. 

We will always reduce the products of $N$ spherical harmonics using a coupling permutation $\sigma_c$ that is constructed by coupling disjoint pairs of $l_i$. This family of pairwise coupling permutations are characterized by the partition of $S_N$ below.
\begin{equation}
    \boldsymbol{\lambda}_c = \begin{cases}
    N \; even : (2 \, , \, 2 \, , \, 2 \, \cdots  2 (N/2) ) \\
    N \; odd : (2 \, , \, 2 \, , \, 2 \, \cdots  2 (N-1/2) \, , \, 1 ) \\
    \end{cases} 
    \label{eq:coupling_partition}
\end{equation}
Any permutation element of $S_N$ that belongs to this partition could be used, so for simplicity, the coupling permutation used will always be the one associated with ordered disjoint cycles from the partition in Eq. \eqref{eq:coupling_partition}. In cyclic form, the pairwise coupling permutation is given in Eq. \eqref{eq:explicit_sigmac}.
\begin{equation}
    \sigma_c = \begin{cases} 
    N=4: \;\; (1,2)(3,4)\\
    N=5: \;\; (1,2)(3,4)(5)\\
    N=6: \;\; (1,2)(3,4)(5,6)\\
    N=7: \;\; (1,2)(3,4)(5,6)(7)\\
    \cdots
    \end{cases}
    \label{eq:explicit_sigmac}
\end{equation}
We will refer to this as the pairwise coupling scheme. The coupling permutations from Eq. \eqref{eq:explicit_sigmac} result in coupling trees that are complete binary trees where the leaves have values that are angular momentum quantum numbers, $l_i$, and the internal nodes have values that are the intermediate angular momenta, $L_k$. The coupling trees in our scheme will have a height given by the number of intermediates, $K=N-2$, and the number of leaves, $N$.
\begin{equation}
    h = {\rm{log}}_2(K + N + 2) \, - \, 1
    \label{eq:min_height}
\end{equation}

A generalized Wigner symbol may be written unambiguously with an explicitly specified coupling scheme without ambiguity.\cite{yutsis_mathematical_1962}
\begin{equation}
\begin{pmatrix}
     l_1 & l_2 & \cdots & l_N\\
     m_1 & m_2 & \cdots & m_N
\end{pmatrix}^{\sigma_c}({\boldsymbol{L}},\boldsymbol{M}),
\label{eq:generalized_wig_rep}
\end{equation}
In this work, we will omit the explicit specification of the coupling permutation in Eq. \eqref{eq:generalized_wig_rep}, because it will always be that from the pairwise coupling scheme defined in Eqs. \eqref{eq:coupling_partition} and \eqref{eq:explicit_sigmac}. Unlike the traditional Wigner-3j symbols, the notation for generalized Wigner symbols must include the multiset of intermediates $\boldsymbol{L}$. In a generalized Wigner symbol, the multiset of intermediate projections $\boldsymbol{M}$ are implicitly defined by $\boldsymbol{m}$ and may be expressed explicitly for clarity. These intermediate projections obey, $(m_1+m_2)=M_1,(m_3+m_4)=M_2, \cdots (m_{N-1}+m_{N})=M_{N/2}$ in our pairwise coupling scheme. The generalized Wigner symbols will often be given in a non-matrix form in terms of multisets of angular momentum, projection, and intermediate angular momentum quantum numbers, $\boldsymbol{l}=(l_1l_2\cdots l_N)$, $\boldsymbol{m}=(m_1m_2\cdots m_N)$, and $\boldsymbol{L}=(L_1L_2\cdots L_{(N-2)}, L_R)$, respectively.
\begin{equation}
W_{\boldsymbol{l}}^{\boldsymbol{m}}
(\boldsymbol{L},\boldsymbol{M})
\label{eq:generalized_wig_eq}.
\end{equation}
In cases where it is sufficient to implicitly define $M_k$ through $m_i$, the symbolic form in Eq. \eqref{eq:generalized_wig_eq} may be abbreviated, $W_{\boldsymbol{l}}^{\boldsymbol{m}} (\boldsymbol{L},M_R)$.

An alternative form for the generalized Wigner symbols would be in the form of a complete binary tree with height from Eq. \eqref{eq:min_height}. 
Note that for non-zero generalized Wigner symbols, the triangle conditions must be obeyed for angular function indices and all intermediates. 
%The collections of triangle conditions for generalized Wigner symbols are sometimes referred to as the polygon conditions.
The polygon conditions\cite{yutsis_mathematical_1962} for the generalized Wigner symbols in the pairwise coupling scheme may be written in terms of triangle conditions for each coupled pair of angular functions and the intermediate they are reduced to. 
\begin{equation}
\begin{split}
    &\triangle ^g(\boldsymbol{l},\boldsymbol{L}) = \{ |l_1 - l_2| \le L_1 \le (l_1+l_2) , \\
    &  |l_3 - l_4| \le L_2 \le (l_3+l_4) , \\ 
    & \cdots \\
    &  |L_{N-3} - L_{N-2}| \le L_R \le (L_{N-3}+L_{N-2}) \}
    \end{split}
    \label{eq:node_triangle_app}
\end{equation}
In Eq. \eqref{eq:node_triangle_app} triangle conditions are specified for all pairwise couplings of angular indices, including all intermediates. These are mentioned after the generalized Wigner symbols and coupling schemes are defined, because the multisets of valid intermediates may differ per coupling scheme. The total number of valid multisets of intermediates from coupling scheme to coupling scheme will be the same. At this stage, we have defined specifically which generalized Wigner symbols we will use, what form they take in symbolic, matrix, and tree forms. We have introduced permutation symmetries and recursion properties of the generalized Wigner symbols. These properties and relationships have implications and for the construction of the PA set and for the derivation of linear relationships between PI cluster functions when rotational invariance/equivariance is enforced.

\subsubsection{Angular basis}
With the generalized Wigner symbols defined, it becomes apparent that products of spherical harmonics may be coupled to form functions that are rotationally invariant. What may not be immediately clear after only defining the generalized Wigner symbols is, for arbitrary $N$, what the size of a complete RI product basis is and how it may be defined. To help show this, we may use the relationship between spherical harmonics and irreducible representations of SO(3). Without going into extraneous detail, the irreducible representations of SO(3) are the Wigner-D matrices, represented as $D_l$ within, and detailed definitions may be found elsewhere.\cite{yutsis_mathematical_1962} The size of the angular product basis should match the dimension of the corresponding representation space, which is comprised of products of representations of SO(3). The product of $N$ representations of SO(3) can be related to a sum of representations of SO(3) using Clebsch-Gordan decomposition.\cite{yutsis_mathematical_1962,dusson_atomic_2022,fulton2013representation} In matrix form, it can be written as,
    \begin{equation}
        \mathbf{M}_{\boldsymbol{l}}^{-1} D_{l_1}\times D_{l_2} \times \cdots D_{l_N}\mathbf{M}_{\boldsymbol{l}} =\sum_{R(\boldsymbol{L})} \alpha_R D_{L_R}.
        \label{eq:reduction}
    \end{equation}
Following the form of Ref.~[\!\citenum{yutsis_mathematical_1962}], $\alpha_R$ is the multiplicity of the irreducible representation with value, $L_R$. It is easy to show that $\alpha_R$ is given by the number of valid intermediate couplings (the size of $\beta_{\boldsymbol{l}}$). The elements of the matrix $\mathbf{M}_{\boldsymbol{l}}$ are typically the generalized Clebsch-Gordan coefficients. The generalized Wigner symbols may be used along with some conversion factors for the same effect.\cite{yutsis_mathematical_1962}

Given the useful relationships for representation theory of SO(3) and Eq. \eqref{eq:reduction}, it can be shown that a complete angular basis must be the same size as the dimension given by the right-hand side of Eq. \eqref{eq:reduction}. It is known that the dimension of a given irreducible representation of SO(3), $D_l$ is given by $d=2l +1$. From 7.4 of Ref.~[\!\citenum{yutsis_mathematical_1962}], we have the more general expression that gives the dimension of the product in the left hand side of Eq. \eqref{eq:reduction}.
\begin{equation}
    \prod_i^N(2l_i+1) = \sum_{L_R} \alpha_{R} (2L_R+1)
    \label{eq:basis_span}
\end{equation}
This gives the dimension of the product space in terms of irreducible subspaces, and the required dimension of the angular product basis. One may take advantage of the fact that the spherical spherical harmonic products with angular momentum indices $l_i$ are related to those coupled to a resultant, $L_R$, to construct a complete, independent angular basis. The reduction through generalized Wigner symbols to do this yields functions are invariant/equivariant with respect to rotations in SO(3), but the invariance with respect to permutations is not treated at this stage.

As a result, we may write an angular basis for products functions of arbitrary $N$. The angular basis functions can be indexed by a valid multiset of angular function indices and  a multiset of intermediate indices  that obey angular momentum coupling conditions. These may be expressed in terms of the product functions that are \textit{not} invariant with respect to permutations, the generalized Wigner symbols, and a final projection quantum number $-L_R \le M_R \le L_R$. 
\begin{equation}
\begin{split}
    &\tilde{V}_{\boldsymbol{lL}}(M_R) = \\
    &\sum_{\boldsymbol{m} = M_R }W^{m_1m_2\cdots m_N}_{l_1l_2\cdots l_N}(\boldsymbol{L},(M_1,M_2 \cdots M_R)) Y^{l_1}_{m_1}\cdots Y^{l_N}_{m_N}
\end{split}
    \label{eq:general_basis}
\end{equation}
In Eq. \eqref{eq:general_basis}, the sum is taken over all possible collections of $ -l_i \le m_i \le l_i$ such that $\sum_i {m_i}=M_R$. Similar bounds and conditions must be applied for all intermediates: $m_i+m_j = M_k$, and $-L_k \le M_k \le L_k$ for all coupled pairs of functions and/or intermediates. We will notice immediately that the functions defined in Eq. \ref{eq:general_basis} have the correct dimension, given the relationship in Eq. \eqref{eq:basis_span}. For each allowed value of $L_R$ there is a collection of intermediate multisets that obey polygon conditions. The number of intermediate multisets in this collection add up to $\alpha_R$. A complete angular basis is obtained when an independent basis function is generated for each valid multiset of intermediates. For example with $\boldsymbol{l}=(1111)$, one will find that the multisets of intermediates allowed by polygon conditions for $L_R=0$ are $\{ \boldsymbol{L}= (000) , \boldsymbol{L}=(110), \boldsymbol{L}=(220) \}$, and this correctly yields $\alpha_R=3$. This is done for all $L_R$ allowed by polygon conditions, $L_R \in \{0,1,2,3,4\}$, with a potentially different $\alpha_R$ for each. It can be shown that is achieved by blocks comprised of functions from Eq.\eqref{eq:general_basis} given the completeness and orthogonality properties of the generalized Clebsch-Gordan coefficients, and by extension, the generalized Wigner symbols.\cite{yutsis_mathematical_1962,dusson_atomic_2022} The functions defined in Eq. \eqref{eq:general_basis} include rotationally invariant functions and potentially some that are equivariant. In Eq. \eqref{eq:general_basis}, a tilde is used to distinguish these functions from functions constructed with permutation invariance. 

In many practical cases and applications of ACE, only rotationally invariant (RI) functions are considered. For this reason, we will often discuss the subset of basis functions with $L_R=0$ and define a shorthand for this RI subspace. As in Eq. \eqref{eq:bblocks}, it may be defined in terms of blocks of functions comprising the complete set. 
\begin{equation}
    \begin{split}
    & \tilde{\beta}_{\boldsymbol{l}} = \{  \tilde{b}_{\boldsymbol{l} \boldsymbol{L} } :  \forall \boldsymbol{L} \; if \; \triangle^{g} (\boldsymbol{l}, \boldsymbol{L}) \} \\
    & \tilde{\mathcal{S}}_N = \{ \tilde{\beta}_{\boldsymbol{l}} : l_i \,  ordered , \,  l_i \ge 0  \forall \, i \in N \}
    \end{split}
\label{eq:ri_product}
\end{equation} 
In Eq. \eqref{eq:ri_product}, a shorthand has been adopted for RI angular product functions constructed using Eq. \eqref{eq:general_basis}, but only for the cases with $L_R=0$. This shorthand is $\tilde{b}_{\boldsymbol{lL}}$ These RI product functions are the elements of the blocks, defined in the first line in Eq. \eqref{eq:ri_product}. In the case for the angular product basis, blocks are now comprised of the subsets of functions sharing the same $\boldsymbol{l}$. The blocks of functions are subsets of of the complete set, $\tilde{\beta}_{\boldsymbol{l}} \subset \tilde{\mathcal{S}}_N$. In the construction of these RI product functions from Eq. \eqref{eq:general_basis}, the sum is taken over all possible collections of $ -l_i \le m_i \le l_i$ such that $\sum_i {m_i}=0$. The rotation-invariance of this basis is achieved by reducing product functions to $L_R=0$ using the generalized Wigner symbols. While rotation-invariance is ensured this way, the cluster functions in practical applications of ACE often require that product functions are also symmetric with respect to permutations/exchange of coordinates as well.

To begin the construction of angular RPI functions, one typically begins with similar procedures and definitions as that in Eq. $\eqref{eq:ri_product}$, but for product functions symmetrized with respect to exchange of coordinates. These PI angular product functions are generated similarly to the functions in Eq. \eqref{eq:pi_product_cluster}. For this definition and others, an abbreviation for products of spherical harmonics (or products of more general functions later on) may be abbreviated using boldfaced function symbols, $\boldsymbol{Y}_{\boldsymbol{m}}^{\boldsymbol{l}}(1,2,\cdots N) = Y^{l_1}_{m_1}(1)\cdots Y^{l_N}_{m_N}(N)$, where coordinates of the functions are denoted with a single integer index inside parenthesis for brevity.
\begin{equation}
    \bar{\boldsymbol{Y}}_{\boldsymbol{m}}^{\boldsymbol{l}}(1,2,\cdots N) = \frac{1}{\sqrt{N !}}\sum_{\sigma \in S_N} \boldsymbol{Y}_{\boldsymbol{m}}^{\boldsymbol{l}} \sigma \big( (1,2,\cdots N) \big) 
    \label{eq:pi_product}
\end{equation}
The sum in Eq. \eqref{eq:pi_product} runs over all permutations in the symmetric group $S_N$ and invariance with respect to permutations is indicated with a bar. Applying the generalized Wigner symbols to functions in Eq. \eqref{eq:pi_product} yields functions that are both rotation and permutation invariant. The RPI functions may be expressed in terms of the permutation invariant products from Eq. \eqref{eq:pi_product}.
\begin{equation}
    b_{\boldsymbol{l}\boldsymbol{L}} = \sum_{\boldsymbol{m}} W_{\boldsymbol{l}}^{\boldsymbol{m}}(\boldsymbol{L},(M_1,M_2 \cdots 0))
    \bar{\boldsymbol{Y}}_{\boldsymbol{m}}^{\boldsymbol{l}} 
\label{eq:ri_product2}
\end{equation} 
From this construction, permutation and rotation invariance are enforced. As evidenced by numerical results in previous studies and derived for some key example in this work, the resulting functions are not always linearly independent.\cite{dusson_atomic_2022} Linear dependencies between functions occur when some $l_i$ appear more than once in a label. Enumerating all RPI functions in Eq. \eqref{eq:ri_product2} as it is done for the functions without permutation invariance results in an over-complete set.
\begin{equation}
    \begin{split}
    & \beta_{\boldsymbol{l}} = \{  b_{\boldsymbol{l} \boldsymbol{L} } :  \forall \boldsymbol{L} \; if \; \triangle^{g} (\boldsymbol{l}, \boldsymbol{L}) \} \\
    & \mathcal{S}_N^{OC} = \{ \beta_{\boldsymbol{l}} : l_i \,  ordered , \,  l_i \ge 0  \forall \, i \in N \}
    \end{split}
\label{eq:pi_ri_product}
\end{equation} 
The conventional definition of the set of angular RPI functions Eq. \eqref{eq:pi_ri_product} results in one that is over-complete. Linear dependencies arise within blocks that are subsets of this over-complete set, $\beta_{\boldsymbol{l}} \subset \mathcal{S}_N^{OC}$. Investigating these linear dependencies is one key focus of this work. An explanation of this linear dependence between angular RPI functions is needed. Many intuitive insights and relationships may be derived when using the fundamental connection between products of spherical harmonics and quantum angular momentum coupling.

\subsubsection{Properties of General Wigner Symbols and Angular Functions}
\begin{figure}[ht]
\hfill
\subfloat[$\sigma_c=(12)(34)$]{\scalebox{0.39}{\includegraphics[width=0.5\textwidth,valign=b]{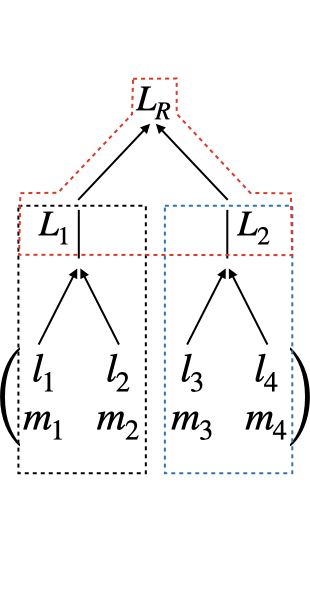}}}\quad
\hfill
\subfloat[$\sigma_c=(12)(3)(4)$]{\scalebox{0.39}{\includegraphics[width=0.5\textwidth,valign=b]{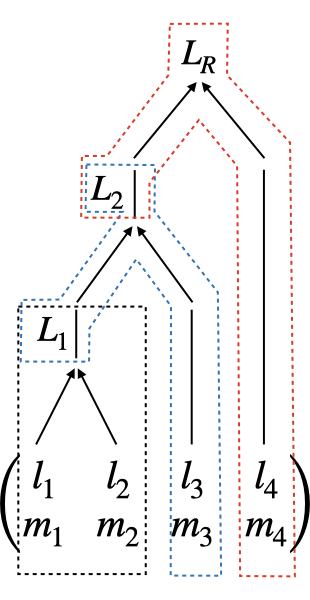}}}
\caption{ Two tree diagrams for generalized Wigner symbols of rank 4.
        Both (a) and (b) represent valid coupling schemes. The indices that are symmetric as a result of Eq. \eqref{eq:oddeven_w3j_perms} are highlighted in corresponding colors. The coupling scheme in (a) preserves permutation symmetry in more of the $(l_i,m_i)$ tuples, rather than in the intermediates $L_k$. Scheme (a) was used to couple Clebsch-Gordan coefficients in Ref.~[\!\citenum{drautz_atomic_2020}].  Scheme (b) was used to couple Clebsch-Gordan coefficients in Ref.~[\!\citenum{dusson_atomic_2022}].
        }
\label{fig:first_tree}
\end{figure}

To explore linear relationships between RPI functions, it is beneficial to begin with properties of the generalized Wigner symbols. While the choice of the coupling scheme for the generalized Wigner symbols, defined by a coupling permutation $\sigma_c$, is an arbitrary one, one may notice that certain coupling schemes allow for more equivalent permutations of product functions than others. This is easier to see graphically, as the angular product function indices comprise the leaves in a binary coupling tree. This is highlighted in the tree diagram in Fig. \ref{fig:first_tree}. The first scheme in Fig. \ref{fig:first_tree} is the pairwise coupling scheme. Permuting two children of a parent, such as permuting $l_1$ and $l_2$ while leaving $L_1$ fixed, in a generalized Wigner symbol is equal to the original generalized symbol up to a sign. This is a direct result of the permutation symmetries of the Wigner-3j symbols in Eq. \eqref{eq:oddeven_w3j_perms}. Odd permutations result in a negative sign while even permutations do not. Similarly, permutations of branches are equivalent as well; permuting $L_1$ and $L_2$ and all corresponding children gives an equivalent generalized Wigner symbol. The second scheme, though amenable to fast recursive evaluation algorithms,\cite{dusson_atomic_2022} has more complicated permutation properties. The permutation symmetries of the generalized Wigner symbols have not yet been reported in detail, so we demonstrate and prove these symmetries in Appendix \ref{appendix:theory}. To summarize the outcome of these permutation symmetries: equivalent permutations of leaves and or intermediates of a generalized Wigner symbol result from equivalent permutations of Wigner 3-j symbols, c.f. Eq. \eqref{eq:oddeven_w3j_perms} and Eq. \eqref{eq:w3j_metric}. 

While permutations of generalized Wigner symbols may be related by iterated application of Eq. \eqref{eq:oddeven_w3j_perms}, there are more elegant and efficient ways to construct the group of permutations that yield Wigner symbols equal up to a sign. Some approaches considered in this work for achieving this include methods such as tableaux fillings of Young diagrams that encode the permutation symmetries of the coupling scheme and the construction of the automorphism group for complete binary trees using wreath products of $S_2$.\cite{james2006representation,knuth1970permutations,brunner_automorphism_1997} Some of these methods are highlighted in Appendix \ref{appendix:theory}. For pedagogical purposes, we will construct the group of related permutations by applying Eq. \eqref{eq:oddeven_w3j_perms} for every coupled pair of angular momentum quantum numbers in a generalized Wigner symbol. It is straightforward to collect and combine all of these transposition permutations. In rank 4 symbol example, equivalent transposition permutations for each coupled multiset of indices are:
\begin{equation}
    \begin{cases}
        T_{(l_1,l_2)} : [(), (12)] \\
        T_{(l_3,l_4)} : [(), (34)] \\
        T_{(L_1,L_2)} : [(), (56)] 
    \end{cases}
    \label{eq:ysg_perms}
\end{equation}
where $(56)$ corresponds to a transposition of the two intermediate angular momenta, $L_1$ and $L_2$ and the blank cycles are the identity permutations for each pair. Taking all possible combinations of permutations (the direct product of the sets of transpositions) in Eq. \eqref{eq:ysg_perms} yields:
\begin{subequations}
\begin{equation}
    \begin{split}
   &T_4=\{ (), (12), (12)(34), \\
   &(12)(34)(56), (12)(56), (34), (34)(56), (56) \}
   \end{split}
   \label{eq:direct_ysg}
\end{equation}    
\begin{equation}
  \begin{split}
  &G_4 = \{ (), (12), (12)(34), \\
  &(14)(23), (1324), (34), (1423), (13)(24) \}
  \end{split}
  \label{eq:ysg_action}
\end{equation}
\end{subequations}
where $G_4$ is obtained by rewriting elements of $T_4$ that contain intermediate indices (56) in terms of their action on their children. Equivalent permutations may be generated for arbitrary rank $N$ Wigner symbols as it was done for rank 4 in Eq. \eqref{eq:ysg_perms}.
\begin{equation}
    \begin{cases}
        T_{(l_1,l_2)} : [(), (12)] \\
        T_{(l_3,l_4)} : [(), (34)] \\
        \cdots \\
        T_{(l_{N-1},l_N)} : [(), (N-1,N)] \\
        T_{(L_1,L_2)} : [(), (N+1,N+2)] \\
        \cdots  \\
        T_{(L_{N-3},L_{N-2})} : [(), (N+K-1,N+K)] \\
    \end{cases}
    \label{eq:ysg_perms_gen}
\end{equation}
From Eq. \eqref{eq:ysg_perms_gen}, the collection of all equivalent permutations, $T_N$, may be generated by taking the direct product of all sets of equivalent transpositions. In Eq. \eqref{eq:ysg_perms_gen}, there are now transpositions for all $N/2$ coupled angular momentum quantum numbers as well as for all $N-2$ intermediate angular momentum quantum numbers (excluding the fixed $L_R$). To obtain $G_N$, the transpositions of internal nodes may be written in terms of their action on leaf nodes. In this work, the automorphism group is constructed using the wreath product of $S_2$.\cite{brunner_automorphism_1997,james2006representation} 

In addition to the group of permutation automorphisms, generalized Wigner symbols may be related to others using recursion relationships. These recursion relationships follow from applying ladder operators used to raise/lower quantum mechanical angular momentum states.\cite{rose1995elementary} The recursion relationships for the generalized Wigner symbols have yet to be derived, but it is straightforward to do so by iteratively applying recursion relationships to the traditional Wigner-3j symbols. General derivations for these are provided in Appendix \ref{appendix:recursions}, and we list some results here. The first type of relationships relate generalized coupling coefficients with one multiset of intermediates $\boldsymbol{L}$ to another with intermediates that have been incremented, $\boldsymbol{L} + \boldsymbol{k}$, while $\boldsymbol{lm}$ remain fixed. This may be derived for arbitrary rank $N$ couplings with intermediates that have been incremented by integer values $k_i \cdots k_{N-2}$ times, as reported in Appendix \ref{appendix:recursions}. Note that intermediates can be incremented such that $L_R\in \boldsymbol{L}$ at the index $k_{N-1}$ remains constant. For brevity in this section, the results are listed for rank 4 only and for the case that $L_R=0$.
\begin{equation}
\begin{split}
&\begin{pmatrix}
     l_1 & l_2 & l_3 & l_4\\
     m_1 & m_2 & m_3 & m_4
\end{pmatrix}\big( (L_1 , L_2, 0) (\boldsymbol{M}) \big) = \\
& A_{-,-}(\boldsymbol{l},\boldsymbol{m})\begin{pmatrix}
     l_1 & l_2 & l_3 & l_4\\
     m_1 & m_2 & m_3 & m_4
\end{pmatrix}\big((L_1 -1, L_2 -1, 0)(\boldsymbol{M}) \big)\\
&+ \\
&A_{+,+}(\boldsymbol{l},\boldsymbol{m})\begin{pmatrix}
     l_1 & l_2 & l_3 & l_4\\
     m_1 & m_2 & m_3 & m_4
\end{pmatrix}\big((L_1 +1, L_2 +1, 0)(\boldsymbol{M}) \big) 
\end{split}
\label{eq:generalized_rank4_recur_lr0}
\end{equation}
In Eq. \eqref{eq:generalized_rank4_recur_lr0}, we provide exact expressions for generalized Wigner symbols with incremented intermediates. The factors, $A_{\pm,\pm}$ follow from Eq. \eqref{eq:wigner_recursion_full}, and may be written entirely in terms of $l_i$ and $m_i$. These factors are defined when $m_i$ are appropriately bound by $-l_i \le m_i \le l_i$ and polygon conditions are met by both $\boldsymbol{L}$ and $\boldsymbol{L}+\boldsymbol{k}$. The result is that a generalized Wigner symbol with intermediates $\boldsymbol{L}$ may be related to a generalized Wigner symbol with different intermediates $\boldsymbol{L}'$, provided that intermediates obey polygon conditions with $\boldsymbol{l}$ and the $\boldsymbol{lm}$ are the same for both Wigner symbols.

A second type of relationship shows how generalized Wigner symbol with incremented intermediate projection quantum numbers, $\boldsymbol{M}_{\pm}$, relates to those with opposing increments of projection quantum numbers, $\boldsymbol{M}_{\mp}$. These relationships may be defined for fixed angular indices (e.g., fixed $\boldsymbol{lL}$). Similar to the relationships for angular momentum quantum numbers, these are derived by iteratively applying the analogous relationship for the traditional Wigner-3j symbols. Detailed derivations and the generalized expressions are given in Appendix \ref{appendix:recursions}, and for brevity here it is shown for rank 4.
\begin{equation}
\begin{split}
&W_{l_1 l_2 l_3 l_4}^{m_1 m_2 m_3 m_4}( \boldsymbol{L} ,( M_1  , M_2 , M_R))\\ &=\sum_{\boldsymbol{m}_{\mp}} c_{\boldsymbol{m}_{\mp}}(\boldsymbol{l}) W_{l_1 l_2 l_3 l_4}^{\boldsymbol{m}_{\mp}}( \boldsymbol{L} ,\boldsymbol{M}_{\mp})
\end{split}
\label{eq:generalized_raise_lower3_main}
\end{equation}
In Eq. \eqref{eq:generalized_raise_lower3_main}, the multiset of incremented intermediate projections $\boldsymbol{M}_{\pm} = ( M_1=(m_1 +m_2) , M_2=(m_3+m_4) , M_R)$ are implicitly defined by the multiset of projection quantum numbers of the state to be coupled, $\boldsymbol{m}_{\pm} = (m_1 m_2 m_3 m_4)$ . For our purposes, projection quantum numbers are incremented such that $M_R$ is conserved. A Wigner symbol with projections, $\boldsymbol{M}_{\pm}$ and $\boldsymbol{m}_{\pm}$, are described in terms of Wigner symbols with lowered/raised projections. These lowered/raised projections, denoted by $\boldsymbol{M}_{\mp}$ and $\boldsymbol{m}_{\mp}$, have increments that are the opposite sign of those in $\boldsymbol{M}_{\pm}$. The coefficients, $c_{\boldsymbol{m}_{\mp}}(\boldsymbol{l})$ follow from generalizing Eqs. \eqref{eq:full_m_wig_recur} - \eqref{eq:m3_cg_recur}. Specific relationships for this example are given in Eqs. \eqref{eq:generalized_raise_lower} and \eqref{eq:generalized_raise_lower2}. It is worthwhile to note that for important practical cases when there are duplicate $l_i$, the $c_{\boldsymbol{m}_{\mp}}(\boldsymbol{l})$ are often the same. Used in conjunction with the permutation invariance of product functions in ACE, this allows one to combine many terms in the construction of RPI functions/descriptors, Eq. \eqref{eq:ri_product2} and Eq. \eqref{eq:ACE_desc}. Combining terms using relationships between Wigner symbols with raised/lowered projections, Eq. \eqref{eq:generalized_raise_lower3_main} and its generalized counterpart in Eq. \eqref{eq:generalized_raise_lower3}, allows one to tractably derive relationships between RPI functions with different intermediates. Particularly, once the terms are combined in terms of common PI function indices, it is often just a matter of applying relationships for raised/lowered intermediates to derive ladder relationships.

A detailed derivation of ladder relationships between RPI functions of rank 4 is given in Appendix \ref{appendix:recursions}. We highlight the resulting analytical relationship between RPI functions with different intermediates.
\begin{equation}
\begin{split}
&b_{(1111)(22)} = \frac{2}{5\sqrt{5}} b_{(1111)(00)}
\end{split}
    \label{eq:final_rank4_explicit_proof_main}
\end{equation}
The result in Eq. \eqref{eq:final_rank4_explicit_proof_main} can be verified numerically. These may be derived for all possible values of intermediates and extended to other values of angular momentum quantum numbers. To show that it is not restricted to the first non-trivial case of $N=4$, other exapmles are provided for $N=4$ in the appendix where we also include an $N=5$ case. Analogous relationships for arbitrary rank and arbitrary $\boldsymbol{l}$ are straightforward following the generalized relationships for raising/lowering the quantum numbers in the Wigner symbols that we derived. With such relationships, we may reconsider each block of angular RPI functions and derive relationships for all functions within the block. By repeatedly incrementing the $l_i$, we may construct function sequences of independent RPI produts. 
\begin{equation}
    \begin{split}
    & \beta_{\boldsymbol{l}}^{PA} = \{  b_{\boldsymbol{l} \boldsymbol{L} } :  \forall \boldsymbol{L} \; if \; \triangle^{g} (\boldsymbol{l}, \boldsymbol{L}), b_{ \boldsymbol{l} \boldsymbol{L} } \in \mathcal{F}_{a}^{PC}(P_{f}(\boldsymbol{l}) \} \\
    & \mathcal{S}_N^{OC} = \{ \beta_{\boldsymbol{l}}^{PA} : l_i \; \sigma_{fc}-ordered   , \,  l_i \ge 0  \forall \, i \in N \}
    \end{split}
\label{eq:PA_ri_set}
\end{equation} 
In the PA approach, we apply the ladder relationships, to obtain a function sequence of independent RPI functions within each block, $\mathcal{F}_{a}^{PC}(P_{f}(\boldsymbol{l}))$. This ensures that angular RPI functions within each block are independent. One distinction from other approaches is that the angular indices are ordered according to a permutation $\sigma_{fc}(\boldsymbol{l}) \in P_{fc}(l)$ that aids in the application of the ladder relationships. This permutation will be defined in the next section; the next section describes the construction of angular functions in practice. We conjecture that when the set of angular RPI functions is constructed in this way, it forms a complete independent RPI basis of rank $N$. It is conjectured because, while proofs and sequences are provided that ensure linear independence \textit{within} each block, proofs of independence \textit{between} different blocks are not provided. For example, we do not prove independence between functions in with the form: $b_{(1111)(LL)}$ and $b_{(2222)(LL)}$. There are some strong theoretical arguments such as the orthogonality properties of the generalized Wigner symbols and numerical results that support only considering blocks of functions, however we do not define an inner product that guarantees a basis is obtained.\cite{yutsis_mathematical_1962} The linearly independent function sequences, ladder relationships, and other properties derived within may help achieve this when combined with rigorous definitions from other works.\cite{dusson_atomic_2022} For now, the PA approach provides a procedure for the construction of a complete set of functions that is block-wise independent. This still presents a practical improvement in ACE methods and theoretical advances that allow for the derivation of analytical relationships between RPI functions or ACE descriptors. It will be outlined how the construction of the PA set is done in practice, and how it compares to other methods to enumerate angular functions.

\subsubsection{Constructing the angular PA-RPI set in practice}

Two types of rotationally invariant functions will be considered before including non-angular degrees of freedom. One will be referred to as the canonical basis (C-RI). The canonical RI basis starts with a fixed, ordered multiset of $l_i$, and is \textit{not} invariant with respect to permutations. The number of valid multisets of intermediates give the size of the canonical RI basis according to Eq. \eqref{eq:reduction}. The canonical basis functions may be indexed on the fixed $l_i$ and corresponding intermediate angular momenta that obey iterated triangle conditions, $\{ \boldsymbol{l}\boldsymbol{L} : \triangle^g(\boldsymbol{l}\boldsymbol{L}) \, , \, l_i \ge 0 \forall i \in N \} $. The function labels for an exhaustive list of $\boldsymbol{l}$ up to $l_{max}=3$ is given in Table \ref{tab:canonical_RI}. 

The precursor to the PA-RPI set, is the over-complete set of RPI functions before constraints from ladder relationships are applied within blocks. This over-complete set has the same elements as those in Eq. \eqref{eq:PA_ri_set}, but without the condition $b_{\boldsymbol{lL}}  \in \mathcal{F}_{a}^{PC}(P_{f}(\boldsymbol{l}))$. This is referred to as PA-RPI$^*$. From this over-complete set, the set of angular PA-RPI functions that is block-wise independent may be obtained. This PA-RPI set leverages both the permutation symmetries and recursion relationships of the generalized Wigner symbols to eliminate linearly dependent angular RPI functions within every block.

The canonical RI procedure is straightforward. Obtaining an RI basis that is not permutation invariant has been done.\cite{dusson_atomic_2022} For all distinct multisets of angular function indices, the multisets of intermediate angular momenta are generated based on polygon conditions. The elements of this basis are the functions from Eq. \eqref{eq:ri_product}. Note again that these are not invariant with respect to permutations.

\begin{figure*}[htbp]
\hfill
\subfloat[]{\scalebox{1.}{\includegraphics[width=0.45\textwidth,valign=b]{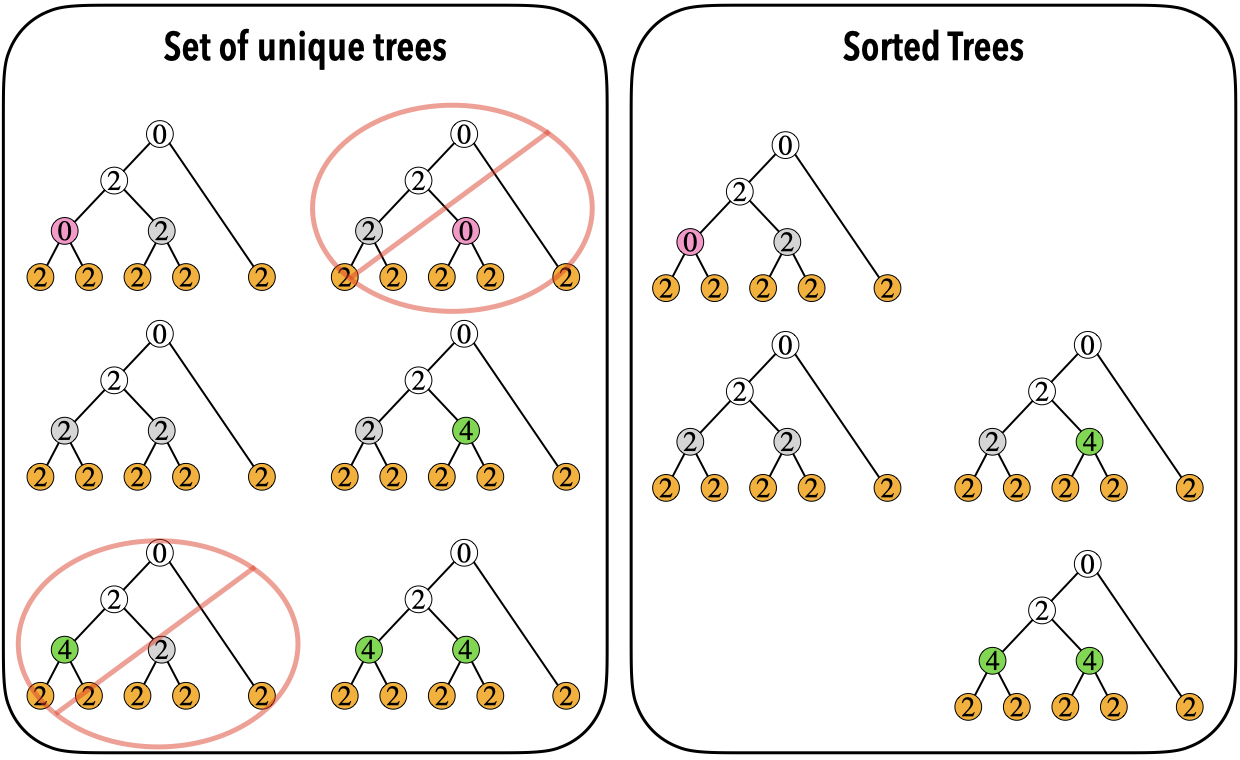}}}\quad
\hfill
\subfloat[]{\scalebox{1.}{\includegraphics[width=0.49\textwidth,valign=b]{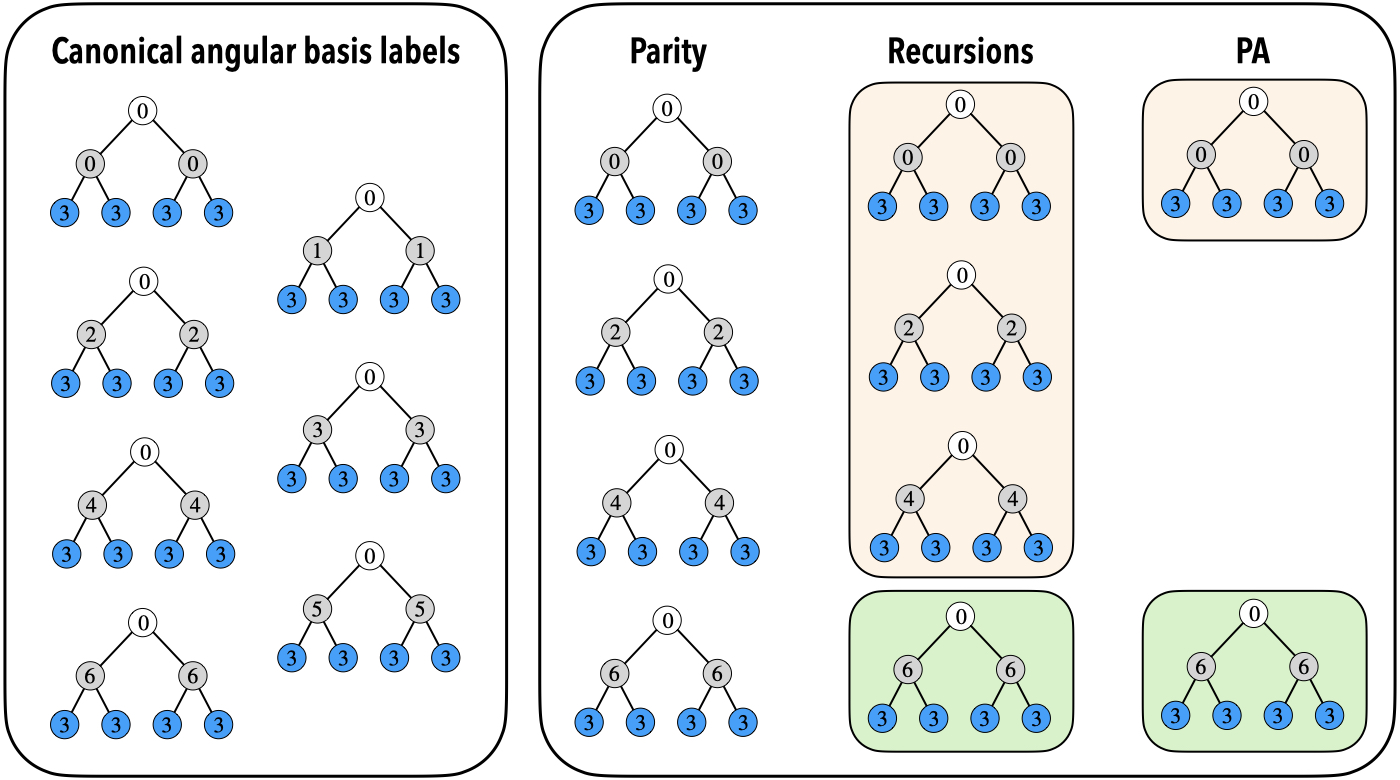}}}
\caption{ The diagrams highlight how linear dependence manifests within blocks of rotation and permutation invariant functions. (a) highlights functions are related by permutations of intermediates are linearly dependent. In (b), it is highlighted that functions with different intermediates (where sets of intermediates are not a symmetric permutation of the others) can also result in linearly dependent functions. Some of these may be addressed with intuitive heuristics, but they may all be remedied through the derivation and application of ladder relationships.}
\label{fig:degen_reduce}
\end{figure*}

To construct the angular PA-RPI set, we begin the same way as with the canonical RI basis. A key difference is that the angular function indices may not be ordered strictly by increasing value. When there are duplicate $l_i$, there are often permutations of angular indices that are more amenable to the application of ladder relationships. One of these convenient permutations of angular indices can be adopted, because a permutation of angular momentum quantum numbers yields a multiset of intermediates with the same size, provided the intermediates obey polygon conditions. That is $size ( \beta_{\boldsymbol{l}} ) = size( \beta_{\sigma(\boldsymbol{l})})$. This maintains the dimension requirements from Eq. \eqref{eq:reduction} and Eq. \eqref{eq:basis_span}. Reasoning for using a different permutation of indices is related to the way linear dependence is proven between certain functions. 

Linear dependence between angular RPI functions typically manifests in two seemingly different ways. The first occurs when there are many duplicate indices, $\# l_i \ge 4$. In such cases permutations of intermediates yield an equivalent function label, for example permuting $L_1$ and $L_2$ and the attached children of $\boldsymbol{lL}=(22222)(022)$ yields $\boldsymbol{lL}'=(22222)(202)$. In this example, both multisets of intermediates are valid according to polygon conditions, but it is clear that these yield the exactly the same function given the permutation symmetries of the underlying Wigner symbols and the permutation invariance of the angular product functions. This can remedied by sorting binary coupling trees as in Fig. \ref{fig:degen_reduce} (a). The second is linear dependence between functions are not permutations of one another, for example the functions indexed by $\boldsymbol{lL}=(1111)(00)$ and $\boldsymbol{lL}=(1111)(22)$. Such relationships are typically observed numerically. Through our newly derived ladder relationships, these cases may be considered analytically. Linear dependencies of this kind may be resolved by applying ladder relationships as highlighted in Fig. \ref{fig:degen_reduce} (b). Ladder relationships expose linear dependencies between functions with duplicate $l_i$. These relationships are derived by grouping like terms in the construction of RPI functions. This grouping of terms depends on the permutation symmetries induced by multiplicity of $l_i\in \boldsymbol{l}$ and the permutation symmetries of the generalized Wigner symbols. It is more straightforward when duplicate $l_i$ are coupled. For this reason, we adapt the ordering of the angular indices to the structure of the coupling tree. This is at the core of the PA-RPI method; as we allow for a specific permutation of indices that is adapted to properties of the generalized coupling coefficients. The permutation-adapted method may be summarized in two steps. In the first step, blocks of functions are defined such that they adapted to the permutation symmetries of the generalized Wigner symbols and how many duplicate $l_i$ are present. This first step is optional, but it allows for straightforward derivation and application of ladder relationships. The second step ensures linear independence of functions is achieved within those blocks using ladder relationships.

The first step in the permutation-adapted approach is to obtain/define the blocks. Blocks of functions for the canonical basis are defined by functions with the same multiset of angular indices, that are ordered by increasing value. This choice of ordering is arbitrary, and is just one method for enumerating the distinct multisets of angular function indices. In the PA method, we choose an ordering of $l_i$ to maximize the pairwise coupling of duplicate $l_i$. This ordering is the most compatible with the permutation and recursion properties of generalized Wigner symbols in our pairwise coupling scheme from Eq. \eqref{eq:explicit_sigmac}. Such an ordering of indices is defined based on the properties of the generalized Wigner symbols and the inherent permutation symmetries within a multiset containing duplicate indices. An intuitive place to begin defining this permutation would be with the angular function indices sorted within the frequency partition of $\boldsymbol{l}$. Note that any permutation of duplicate indices results in the same function. The frequency partition, $P_f$, is the partition that reflects the permutation symmetries induced by duplicate indices in $\boldsymbol{l}$. This is easy to obtain in practice by sorting first by the frequency of duplicate elements and secondly on the values of the elements. Some examples of this are: $\boldsymbol{l}=(1223)\rightarrow \boldsymbol{l}_{f} = (2213)$ generated by $P_f=(2,1,1)$, $\boldsymbol{l}=(11222)\rightarrow \boldsymbol{l}_{f}=(22211)$ generated by $P_f=(3,2)$, and $\boldsymbol{l}=(2222233333)\rightarrow \boldsymbol{l}_{f}=(2222233333)$ generated by $P_f=(5,5)$. While this groups duplicate $l_i$ near one another, it does not always maximize the pairwise coupling of duplicate indices as seen in the final example. This shortcoming can be addressed by considering a permutation from a different partition. To address this shortcoming, we find the permutation of $\boldsymbol{l}$ that maximizes the automorphisms of the coupling tree and use the corresponding partition, $P_{fc}$. The group of automorphisms for the coupling tree is initially generated by transpositions of pairs. More permutations that yield equivalent couplings are generated when duplicate nodes are coupled. We therefore find the permutation of $\boldsymbol{l}$ that sorts $l_i$ within parts that are the largest powers of two and use that partition.  
\begin{equation}
\begin{split} 
& P_{fc} = \max_{P} \sum_{\mathcal{O} \in P} \mathcal{O} :\\
&\forall \mathcal{O} \in P, \exists l_i \in \boldsymbol{l}: \# l_i \geq  2^{k} \\
& \sum \mathcal{O} = N
\end{split}
\label{eq:sym_couple_parts}
\end{equation}
In Eq. \eqref{eq:sym_couple_parts}, it can be seen that the part sizes are those that can be composed into perfect binary trees (parts that are powers of 2). Given a multiset of angular indices, we may obtain $P_{fc}$ based on the frequency of duplicate indices. Once obtained, the previous examples where $l_i$ were ordered by frequency will instead be: $\boldsymbol{l}=(1223)\rightarrow \boldsymbol{l}_{fc} = (2213)$ generated by $P_{fc}=(2,2)$, $\boldsymbol{l}=(11222)\rightarrow \boldsymbol{l}_{fc}=(11222)$ generated by $P_{fc}=(2,2,1)$, and $\boldsymbol{l}=(2222233333)\rightarrow \boldsymbol{l}_{fc}=(2222333323)$ generated by $P_{fc}=(4,4,2)$. In practice, $\boldsymbol{l}_{fc}$, may be obtained by taking advantage of the correspondence between semi-standard Young Diagrams and partitions of $S_N$.\cite{fomin_generalized_1988,schensted_longest_1961} Doing so can avoid a full search over all permutations over $S_N$ to find an ordering of indices that has our desired properties and efficient approaches for doing so scale like $\mathcal{O}(N^2)$.\cite{knuth1970permutations} Alternatively, one could rely algorithms for binary tree sorting.

In the second step of the PA method, a sequence of independent RPI functions is obtained by sampling the over-complete sequence of RPI functions $\mathcal{F}_{i}^{OC}(\boldsymbol{l})$ for each block. The over-complete sequence, $\mathcal{F}_{i}^{OC}(\boldsymbol{l})$ as discussed in Appendix \ref{appendix:recursions}, is comprised of RPI functions with all possible multisets of intermediates, $b_{\boldsymbol{lL}}: \forall \boldsymbol{L} \, if \, \triangle^g(\boldsymbol{l},\boldsymbol{L})$. The sampling of the over-complete sequence that yields a linearly independent sequence, $\mathcal{F}_{a}^{PA}(P_f(\boldsymbol{l}))$, is obtained by deriving relationships from multiple ladder relationships. Ladder relationships for functions with angular indices incremented such that the multiplicity of $l_i \in \boldsymbol{l}_{fc}$ is conserved (e.g., such that the frequency partition $P_f(\boldsymbol{l})$ is the the same for each increment) are used to do this. The second step in the angular PA-RPI procedure ensures that the RPI functions within a block form a linear independent function sequence. While this sampling is derived from analytical relationships, it is not the same for different types of blocks. The "type of block" in this context refers to sequences with the same numbers of duplicate angular indices $\# l_i \in \boldsymbol{l}$ or more succinctly with the same frequency partition $P_f(\boldsymbol{l})$. The sampling to obtain independent functions must be derived for each possible $P_f(\boldsymbol{l})$. Our newly derived properties of the generalized Wigner symbols allow one to do this in principle for arbitrary $N$, but defining $\mathcal{F}_{a}^{PA}(P_f(\boldsymbol{l}))$ for large $N$ becomes intensive. For this reason, independent RPI function sequences have only been obtained up to $N=8$. This is sufficient for many practical applications of ACE, but applications with much larger $N$ may require more general approaches to obtain $\mathcal{F}_{a}^{PA}(P_f(\boldsymbol{l}))$.

The resulting set of angular PA-RPI functions is complete and, within blocks defined by functions with shared $\boldsymbol{l}$, independent. Similarly angular RPI descriptors may be defined that are block-wise independent. A comparison between the canonical angular RI basis, the over-complete RPI set generated by all intermediates obeying polygon conditions for $\boldsymbol{l}_{fc}$ , and the block-wise independent  PA-RPI labels are given in Table \ref{tab:canonical_RI}.

\begin{table}[h]
    \centering
    \begin{tabular}{|p{1.5cm}|p{1.7cm}|p{1.7cm}|p{1.7cm}|}
        \hline
         & \,\,\,\,C-RI & \,\,\,\,RPI$^*$ & \,\,\,\, PA-RPI\\
         \,\,\,\,$\boldsymbol{l}$ & \,\,\,\,$\tilde{\beta}_{\boldsymbol{l}}$ &  \,\,\,\, $\beta_{\boldsymbol{l}}^{OC}$ &  \,\,\,\,$\beta_{\boldsymbol{l}}^{PA}$\\
\hline
(1111) &  $(1111) (00)$
$(1111) (11)$
$(1111) (22)$ &
$(1111) (00)$
$(1111) (11)$
$(1111) (22)$ &
$(1111) (00)$ \\
\hline
(1113) & $(1113) (22)$ &  $(1113) (22)$ &  $(1113) (22)$\\
\hline
(1122) & $(1122)(00)$
$(1122)(11)$
$(1122)(22)$&
$(1122)(00)$
$(1122)(11)$
$(1122)(22)$&
$(1122)(00)$
$(1122)(22)$ \\
\hline
(1133) &  $(1133) (00)$
$(1133) (11)$
$(1133) (22)$&
$(1133) (00)$
$(1133) (11)$
$(1133) (22)$&
$(1133) (00)$
$(1133) (22)$\\
\hline
(1223) & $(1223) (11)$
$(1223) (22)$
$(1223) (33)$ &
$(2213) (22)$
$(2213) (33)$
$(2213) (44)$&
$(2213) (22)$
$(2213) (44)$\\
\hline
(1333) & $(1333) (22)$
$(1333) (33)$
$(1333) (44)$&
$(3331) (22)$
$(3331) (33)$
$(3331) (44)$&
$(3331) (22)$\\
\hline
(2222) & 
$(2222) (00)$
$(2222) (11)$
$(2222) (22)$
$(2222) (33)$
$(2222) (44)$ &
$(2222) (00)$
$(2222) (11)$
$(2222) (22)$
$(2222) (33)$
$(2222) (44)$ &
$(2222) (00)$\\
\hline
(2233) & 
$(2233) (00)$
$(2233) (11)$
$(2233) (22)$
$(2233) (33)$
$(2233) (44)$ & 
$(2233) (00)$
$(2233) (11)$
$(2233) (22)$
$(2233) (33)$
$(2233) (44)$& 
$(2233) (00)$
$(2233) (22)$
$(2233) (44)$\\
\hline
(3333) & $(3333) (00)$
$(3333) (11)$
$(3333) (22)$
$(3333) (33)$
$(3333) (44)$
$(3333) (55)$
$(3333) (66)$ & 
$(3333) (00)$
$(3333) (11)$
$(3333) (22)$
$(3333) (33)$
$(3333) (44)$
$(3333) (55)$
$(3333) (66)$ & 
$(3333) (00)$
$(3333) (66)$\\
\hline
    \end{tabular}
    \caption{An exhaustive listing of angular function blocks for all valid $\boldsymbol{l}$ up to $l_{max}=3$ with $L_R=0$ are given in the first column. The canonical RI labels (C-RI) for angular functions without permutation invariance is given in the second column. The over-complete of permutation invariant RI labels is given in the third column, (RPI$^*$), and the block-wise independent PA-RPI angular function labels are given in the fourth.}
    \label{tab:canonical_RI}
\end{table}
In Table \ref{tab:canonical_RI}, it can be seen that the C-RI and PA-RPI bases are equivalent in some cases. In cases where they are not the same, labels corresponding to dependent functions have been eliminated, resulting in the fourth column. The functions eliminated are those missing from the over-complete set in third column. 

\subsection{RPI functions with non-angular degrees of freedom}

This section will focus on exposing linear dependencies for blocks of RPI functions with angular and non-angular degrees of freedom. Similar to the permutation invariant angular product functions, imposing rotational invariance on permutation invariant radial + angular product functions may also result in linearly dependent RPI functions.
\begin{equation}
    B_{\boldsymbol{nlL}} = \sum_{\boldsymbol{m}} W_{\boldsymbol{l}}^{\boldsymbol{m}}(\boldsymbol{L},\boldsymbol{M})
    \bar{\boldsymbol{\Phi}}_{\boldsymbol{nlm}}(\boldsymbol{r}) 
\label{eq:rpi_produc12}
\end{equation} 
In Eq. \eqref{eq:rpi_produc12}, the 'cluster basis' from Eq. \eqref{eq:direct_prod} has been made symmetric with respect to exchange of coordinates as in Eq. \eqref{eq:pi_product_cluster}. Detailed procedures on this symmetrization may also be found in Ref.~[\!\citenum{dusson_atomic_2022}]. It is made rotationally invariant by contracting the permutation invariant cluster basis with the generalized Wigner symbols. Similar to the purely angular basis, the functions defined by Eq. \eqref{eq:rpi_produc12} are also over-complete. As with the procedure to generate a set of permutation-adapted angular functions, a set of permutation-adapted radial and angular RPI functions may be determined that is block-wise independent. For the angular PA-RPI procedure, it was demonstrated through results and derivations of ladder relationships that the permutation symmetries induced by the multiplicity of $l_i$ result in linearly dependent angular functions. For some blocks of RPI functions with both angular and non-angular indices, the non-angular indices may break the permutation symmetries induced by the multiplicity of $l_i$ and/or those induced by the generalized Wigner symbols. For these reasons, blocks of RPI functions with non-angular degrees of freedom must be defined by both the angular and non-angular indices. The ladder relationships must be derived and applied within these blocks as well. For example, when including radial degrees of freedom, the blocks need to be defined by $\boldsymbol{nl}$.

To define the blocks for RPI functions with radial and angular degrees of freedom, the enumeration and indexing scheme for $\boldsymbol{nl}$ labels needs to be specified. Previous approaches using lexicographical ordering of $(n_i,l_i)$ tuples to define $\boldsymbol{nl}$ labels is a natural first choice. This approach ensures that distinct labels are chosen and is straightforward to implement in algorithms. Although lexicographical ordering does not always align the angular functions with the permutation symmetries imposed by the generalized Wigner symbols, as observed for the case with purely angular functions in Table \ref{tab:canonical_RI}. Alternatively the indexing be done within the binary tree structures of the generalized Wigner symbols. One such way to do this would be composing $(n_i,l_i)$ from distinct $\boldsymbol{nl}$ labels as leaves in binary trees. This way, many ordering schemes may be use that adapt the distinct $\boldsymbol{nl}$ labels to the symmetries of the underlying generalized Wigner symbols. Instead, we adopt the more compact definition in Eq. \eqref{eq:bblocks_pa}. The multiset of angular indices is adapted to permutation symmetries induced by generalized Wigner symbols and the multiplicities of $l_i$ by applying a specific permutation/ordering to the angular indices. This is the same as that for the purely angular functions in Eq. \eqref{eq:PA_ri_set}. This adapted ordering, denoted previously as $\boldsymbol{l}_{fc}$ is obtained from the permutation in Eq. \eqref{eq:sym_couple_parts} that maximizes equivalent permutations of angular indices given the underlying Wigner symbols. The permutations of non-angular indices, $\varsigma$, used in Eq. \eqref{eq:bblocks_pa} are those that generate distinct $\boldsymbol{nl}$ labels up to the multiplicity of $\boldsymbol{l}$ and the permutation automorphisms of the generalized Wigner symbols. In short, a valid $\varsigma(\boldsymbol{n})$ is obtained when not related by another permutation in $P_{fc}$ or $P_f$. It is also possible to construct labels with these properties using binary trees. In general cases, these permutations are not element-wise equivalent to those generated by lexicographical ordering of $(n_i,l_i)$ tuples. However, the total number of $\boldsymbol{nl}$ labels does not exceed that obtained from lexicographical ordering. 

In order to construct these labels, the following procedure may be used. First, all $\boldsymbol{nl}$ permutations are constructed, and the corresponding intermediates are added to form the blocks. In pseudocode, this first step is:
\begin{algorithmic}
\State OC\_blocks
\State $\boldsymbol{l}_{fc} \leftarrow P_{fc} \leftarrow \# l_j  \forall j \in set(\boldsymbol{l})$
\State $distinct\_\boldsymbol{n} \leftarrow \{ \varsigma \} \leftarrow P_{fc}, P_f$
\State $\boldsymbol{nl}$\_blocks = \{ $\varsigma (\boldsymbol{n})\boldsymbol{l}_{fc} : \forall \varsigma(\boldsymbol{n}) \in distinct\_\boldsymbol{n}$
\For{$\boldsymbol{nl}$\_block in $\boldsymbol{nl}$\_blocks}
    \For{$\boldsymbol{L}$ \textbf{in} Generate\_intermediates($\boldsymbol{l},L_R$) }
        \State $\boldsymbol{nl}$\_block $\leftarrow$ $(\boldsymbol{n}$ $\boldsymbol{l}$ $\boldsymbol{L}$)
    \EndFor
    \State OC\_blocks $\leftarrow$ $\boldsymbol{nl}$\_block
\EndFor
\end{algorithmic}
First, the multiset of angular indices is adapted to the permutation symmetries of the generalized Wigner symbols. It results from an element of the partition, $P_{fc}$ that maximizes the permutation automorphisms of the coupled angular functions. The distinct permutations of non-angular indices are then generated based on the frequency partition of angular indices. This gives the unique $\boldsymbol{nl}$ that will used to make blocks of RPI functions. For each of these $\boldsymbol{nl}$ labels, the valid intermediates are appended to the label and collected into the $\boldsymbol{nl}$\_block. These intermediates are obtained from Generate\_intermediates($\boldsymbol{l},L_R$), which is a function that obtains all multisets of intermediates obeying $\triangle^g(\boldsymbol{l},\boldsymbol{L})$. Each over-complete block is collected into, OC\_blocks, and linear dependence needs to be treated. In practice, step one is repeated for all combinations of radial and angular indices up to some specified $n_{max}$ and $l_{max}$, respectively. The second step is to sample the over-complete blocks according to the function sequence of independent RPI functions, $\mathcal{F}_{a}^{PA}$ that is derived from ladder relationships. This may be done by with the following procedure.
\begin{algorithmic}
\State $PA\_set$
\For { OC\_block in OC\_blocks}
    \State $PA\_block$
    \For {$B_i \in$ OC\_block}
        \State $PA\_block \gets B_i$ if $B_i \in \mathcal{F}_{a}^{PA}$
    \EndFor
    \State $PA\_set \gets PA\_block$
\EndFor
\end{algorithmic}
In this step each (potentially over-complete) block of functions is sampled according to $\mathcal{F}_{a}^{PA}$, which defines a sequence of linearly independent RPI functions derived from analytical ladder relationships. Some archetypal cases of this are given in Appendix \ref{appendix:recursions} and the relationships necessary to generalize beyond that case are also provided. 

There is an important note on cost of step 1 in the procedure, in which one may expect the use of brute-force searches over $S_{N}$. Brute-force searches over this group can cumbersome very quickly, and a brute-force search over all of $S_N$ is not what is done in practice. In practice, other approaches are used such as the Robinson-Schensted-Knuth correspondence theorem.\cite{schensted_longest_1961,knuth1970permutations} Such approaches and the corresponding algorithms allow one to take advantage of the relationship between Young diagrams and partitions of $S_N$. Young Diagrams are graphical tools commonly used in the representation theory for the symmetric group, and methods for manipulating them, using their relationship to irreducible representations of $S_N$, and the standard fillings (a.k.a. Young Tableaux) for these diagrams can be found elsewhere.\cite{fulton2013representation,knuth1970permutations} The result of using these in the PA method is that, rather than considering all permutations in $S_N$ to find $\boldsymbol{l}_{fc}$, we may consider a subset of permutations related to the Young Diagrams for the partition, $P_{fc}$. The allowed permutations of non-angular indices are found in a similar way. We also emphasize that many of the steps only needs to be done once for a multiset of indices with a certain number of duplicates. For example, the construction of blocks for $\boldsymbol{nl}=(1122)(1122)$ may be applied to $\boldsymbol{nl}=(4455)(1122)$ and $\boldsymbol{nl}=(4455)(4466)$, etc.

\subsubsection{Highlighted results and examples}

Sorting the angular momentum quantum numbers is not sufficient to produce a complete and independent basis when certain constraints are enforced. An illustrative example within the $\sigma_c=(12)(34)$ coupling scheme is $\boldsymbol{n}=(1122)$ and $\boldsymbol{l}=(1122)$. We show where using lexicographical ordering of $(n_i,l_i)$ for this fixed permutation of $\boldsymbol{l}$ while also imposing parity constraints does not work, and where the PA-RPI method remedies it. Forcing $\boldsymbol{nl}$ to be ordered first on $l_i$ in $\boldsymbol{nl}$ tuples gives certain blocks of functions.
\begin{equation}
    \{\boldsymbol{nl}\} =  \{ (1122)(1122) \, , \, (1212)(1122) \, , \, (2211)(1122) \}.
    \label{eq:lfirst}
\end{equation}
There are two multisets of intermediates allowed after imposing parity constraints, $(l_i + l_j + L_k) :\; even$, for all coupled pairs of angular momenta. Excluding the $L_R=0$, these multisets are $\{\boldsymbol{L}\}=\{(00),(22)\}$. The multiset $(11)$ is often eliminated because it does not obey even parity constraints. All possible descriptor labels generated by this example are given by: 
\begin{equation}
\begin{split}
    &\{  (1122)(1122)(00) \, , \, (1212)(1122)(00) \, , \, \\
    & (2211)(1122)(00) \, , \, (1122)(1122)(22) \, , \, \\
    &(1212)(1122)(22) \, , \, (2211)(1122)(22) \}
    \end{split}
    \label{eq:only_6}
\end{equation}
The problematic result of Eq. \eqref{eq:only_6} is that, at a maximum there are 6 descriptor labels. However, for the same multisets of $\boldsymbol{nl}$ blocks we find, and it is also reported in Table 3 of Ref.~[\!\citenum{dusson_atomic_2022}], that there are 7 independent functions. The result observed in the construction of some ACE descriptor sets (and the reason parity constraints are sometimes used in ACE methods) that couplings with odd parity, $(l_i + l_j + L_k) :\; odd$, yield zero-valued functions or do not transform correctly with respect to rotations is not true for all cases. This is actually the result of ladder relationships, and the transformation properties are determined by the value of $L_R$. In this case of $L_R=0$, they will still be rotation-invariant.

Returning to the PA-RPI method, we will consider this example within the context of our newly derived properties for generalized Wigner symbols. Following the first part of the PA procedure, we find the blocks of functions are the same as those in Eq. \eqref{eq:lfirst}. The result that couplings with odd parity yield zero-valued functions is only the case when permutation symmetries induced by duplicate angular indices are not broken by non-angular indices. In general, applying parity constraints separately from ladder relationships may not be necessary. Keeping this in mind, we will generate the blocks of functions used in the PA method regardless of coupling parity. Before removing redundancies resulting from raised/lowered intermediates of PI functions, we obtain 3 blocks containing 3 functions. The labels for these blocks of functions are given as
\begin{equation}
\begin{split}
    & \{  (1122)(1122)(00) \, , \, (1122)(1122)(11) \, , \, (1122)(1122)(22)  \} \\
    & \{  (1212)(1122)(00) \, , \, (1212)(1122)(11) \, , \, (1212)(1212)(22)  \} \\
    & \{  (2211)(1122)(00) \, , \, (2211)(1122)(11) \, , \, (2211)(1122)(22)  \} \\
    \end{split}.
    \label{eq:all_8}
\end{equation}
The second step reduces the set of functions given in Eq. \eqref{eq:all_8} to a set of functions that is, within each block, linearly independent. Doing so gives the function labels
\begin{equation}
\begin{split}
    & \{  (1122)(1122)(00) \, , \, (1122)(1122)(22)  \} \; : \;   \\
    & \{  (1212)(1122)(00) \, , \, (1212)(1122)(11) \, , \, (1212)(1212)(22)  \}  \\
    & \{  (2211)(1122)(00) \, , \, (2211)(1122)(22)  \} \\
    \end{split}
    \label{eq:final_7}
\end{equation}
These are the labels with all distinct $\boldsymbol{nl}=(1122)(1122)$ blocks in the PA-RPI set, and these match the semi-numerical basis size reported in Table 3 of Ref.~[\!\citenum{dusson_atomic_2022}]. 
\begin{equation}
    \begin{split}
    &\mathcal{F}_a^{PA}((2,2)(2,2)) = \{ B_i :  i  \; even, B_i \in  \mathcal{F}_i^{OC} \}\\
    &\mathcal{F}_a^{PA}((2,2)(2,2)) = \{ B_i :  B_i \in  \mathcal{F}_i^{OC} \}\\
    &\mathcal{F}_a^{PA}((2,2)(2,2)) = \{ B_i :  i  \; even, B_i \in  \mathcal{F}_i^{OC} \}
    \end{split}
    \label{eq:fcn_seq_PA}
\end{equation}
The corresponding linearly independent function sequences applied to the respective over-complete blocks in Eq. \eqref{eq:all_8}, to obtain the labels in Eq. \eqref{eq:final_7} are given in Eq. \eqref{eq:fcn_seq_PA}. 

There are some other notable examples of the PA method that can be instructive and directly compared with previously reported results. In this example, we highlight the use of an analytically derived ladder relationship in a non-trivial case. For this, we will consider $\boldsymbol{nl}=(1111)(1111)$. In this simple example, there is only one block of functions given by
\begin{equation}
    \begin{split}
    \{ (1111)(1111)(00) \, , \, (1111)(1111)(11) \, , \, (1111)(1111)(22) \}
    \end{split}
    \label{eq:utrees_1111}
\end{equation}
and the PA-RPI labels in the linearly-dependent function sequence along with the sampling that yields it are,
\begin{equation}
    \begin{split}
    \{ (1111)(1111)(00) \},= \{ B_i :  i \mod{7} = 0, B_i \in  \mathcal{F}_i^{OC} \}
    \end{split}
    \label{eq:parpi_1111}
\end{equation}
which is reduced from the block in Eq. \eqref{eq:utrees_1111}. It is noted that this is also the same size as that obtained with the semi-numerical construction of the RPI basis in Table 3 of Ref.~[\!\citenum{dusson_atomic_2022}]. We provide a proof to show that the descriptors with labels from Eq. \eqref{eq:utrees_1111} are linearly dependent in Appendix \ref{appendix:recursions}. Specific linear relationships, such as that for the angular RPI functions shown in Eq. \eqref{eq:final_rank4_explicit_proof_main}, may also be derived for functions with non-angular indices. In the case here where all $n_i$ are equivalent, the ladder relationships are the same as those for purely angular functions. As a result, the relationship between the functions in Eq. \eqref{eq:utrees_1111} is given below.
\begin{equation}
    B_{(nnnn)(1111)(22)} = \frac{2}{5\sqrt{5}} B_{(nnnn)(1111)(00)}
    \label{eq:final_rank4_explicit_proof_main_n}
\end{equation}
In general, ladder relationships may be derived for cases where radial indices are present and for any block of functions containing duplicate angular indices. While we do not prove linear independence between different blocks of functions within some rank $N$, we do conjecture that this construction of the PA-RPI set forms a complete and independent basis for rank $N$. If one wishes to add additional indices for additional degrees of freedom such as chemical labels $\boldsymbol{\mu}$, the PA-RPI set and the corresponding procedures may still be used. In these cases, unique multisets of non-angular indicex tuples may be used in place of the radial indices used in our examples.

\section{Results}
\subsection{Wigner symbol permutations}
To further demonstrate the permutational symmetries of the generalized Wigner symbols, Table ~\ref{tab:explicit_perm} provides an example for an $N=4$ generalized symbol with no degeneracy in $(l_i,m_i)$ tuples. The numerically calculated generalized symbol is given for elements of $G_N$ operating on $(l_i,m_i)$ that preserve the coupling tree structure. The numerical results are provided to help show examples of the permutations used to eliminate redundant functions in the PA method.
\begin{table}[]
    \centering
    \begin{tabular}{|p{1.8cm}|p{3.0cm}|p{1.7cm}|}
\hline
 Permutation  & $\sigma(\boldsymbol{Llm})$ & $W_{\boldsymbol{m}}^{\boldsymbol{l}}(\boldsymbol{L})$ \\
\hline
\hline
(1)(2)(3)(4) & $(22)(1234)(1-2-34)$ & $1/\sqrt{1125}$ \\
\hline
(12)(3)(4)   & $(22)(2134)(-21-34)$ & -$1/\sqrt{1125}$ \\
\hline
(1)(2)(34)   & $(22)(1243)(1-24-3)$ & -$1/\sqrt{1125}$ \\
\hline
(12)(34)   & $(22)(2143)(-214-3)$ & $1/\sqrt{1125}$ \\
\hline
(13)(24)   & $(22)(3412)(-341-2)$ & $1/\sqrt{1125}$ \\
\hline
(14)(23)   & $(22)(4321)(4-3-21)$ & $1/\sqrt{1125}$ \\
\hline
(1423)   & $(22)(3421)(-34-21)$ & -$1/\sqrt{1125}$ \\
\hline
(1324)   & $(22)(4312)(4-31-2)$ & -$1/\sqrt{1125}$ \\
\hline
    \end{tabular}
    \caption{Example of permutational symmetry for a generalized Wigner symbol with $\boldsymbol{lL} = (1234)(22)$. The first column gives the permutation in cyclic notation for equivalent permutations $\sigma \in G_N$. In the second column, the permutations are applied to $\boldsymbol{l}$ as well as the corresponding $\boldsymbol{m}$. The final column provides the exact value of the generalized Wigner symbols for each permutation (cf. Eq.~\eqref{eq:single_rank4c}.) }
    \label{tab:explicit_perm}
\end{table}

The second column in Table \ref{tab:explicit_perm} shows the first column of permutations applied to the original arrangement of $(l_i,m_i)$ indices. The result is always a permutation of the indices that is: 1) a permutation between two $(l_i,m_i)$ that are children of the same parent, 2) a permutation between two branches that are children of the same parent 3) some combination thereof. The exact numerical value of each permuted symbol is given in the final column (cf. Eq.~\eqref{eq:single_rank4c}.) 

\subsection{Permutation-adapted RPI function counts}
\begin{table}[]
    \centering
    \begin{tabular}{|p{1.5cm}|p{1.3cm}|p{1.3cm}|p{1.2cm}|p{1.2cm}|}
\hline
$N$  & $deg./N$ & \# All $S_N$ & lexico. & \# PA-RPI  \\
\hline
4 & 2 & 3 & 3 & 1 \\
$\,$ & 4 & 4668 & 976 & 745 \\
$\,$ & 6 & 113795 & 27228 & 23739 \\
$\,$ & 8 & 491004 & 121054 & 106667 \\
$\,$ & 10 & 689129 & 166311 & 143938 \\
$\,$ & 12 & 699840 & 168537 & 145287 \\ 
\hline
5 & 2 & 6 & 6 & 1 \\
$\,$ & 3 & 1150 & 244 & 84 \\
$\,$ & 4 & 28080 & 2773 & 1375 \\
$\,$ & 5 & 140370 & 9714 & 5573 \\
$\,$ & 6 & 268260 & 16479 & 9543 \\
$\,$ & 8 & 311040 & 19152 & 10674 \\
\hline
    \end{tabular}
    \caption{Cumulative descriptor/basis function counts as a function of polynomial degree and rank, $N$. The second column gives the degree of the functions, $\sum_i (n_i+l_i)$, divided by the rank. The third column gives the descriptor counts for all unique $\boldsymbol{nl}$ permutations in $S_N$. The fourth column gives the number of functions, ($\boldsymbol{nl}$ with lexicographically ordered $(n_i,l_i)$ and all intermediates) needed to begin SVD for some semi-numerical approaches. The final column gives the PA-RPI counts. }
    \label{tab:cum_desc_counts}
\end{table}

The function counts for the PA-RPIset and other relevant sets of functions are given for different polynomial degree, where degree is defined as ${\rm{deg.}} = \sum_i (n_i+l_i) $, in Table \ref{tab:cum_desc_counts}. The number of PA functions, (column 5) in Table \ref{tab:cum_desc_counts}, is smaller than the full set of functions with lexicographically ordered labels one would use to obtain the independent functions numerically (column 4). The third column in Table \ref{tab:explicit_perm} gives all possible $\boldsymbol{nlL}$ permutations for reference. Practical implementations of ACE use atom-centered basis functions that are poorly conditioned for numerical reduction due to self-interactions. This limits the accuracy/independence of numerically derived ACE bases for large polynomial degree for some semi-numerical approaches. Additionally, a numerical reduction of basis elements by definition requires excess computation of dependent descriptors or evaluation of a Gramian.\cite{dusson_atomic_2022} Extensive numerical validation of the PA set is provided in Tables \ref{tab:table1_a} - \ref{tab:table1_d}. These semi-numerical function counts are provided for comparison, and to evaluate support the conjecture that the block-wise independent PA set forms a complete, independent basis.

The results in Table \ref{tab:cum_desc_counts} demonstrate that the number of functions needed for numerical construction of the ACE basis using SVD grows rapidly with polynomial degree and rank. This includes function counts for large $n_{max}$ and $l_{max}$; they are $n_{max}=6$ and $l_{max}=6$ for rank 4 and $n_{max}=6$ and $l_{max}=2$ for rank 5. The minimum value of $l_i$ is one in all entries. Different values of $l_{max}$ and $n_{max}$ are used to help demonstrate limiting behavior. For rank 4, the reduction seems to plateau with increasing polynomial degree. This is the case because $n_{max}$ and $l_{max}$ are relatively high. There are fewer multisets of $\boldsymbol{n}$ and $\boldsymbol{l}$ containing duplicate indices compared to cases where $n_{max}$ and $l_{max}$ are small. The cases where there are duplicate radial and angular momentum indices are give rise to the linear dependencies, so the reduction is less significant. This limiting behavior is expected based on previous results. For cases where $n_{max}$ and/or $l_{max}$ are limited to smaller values, we have more sets of $\boldsymbol{n}$ and $\boldsymbol{l}$ with duplicate entries. As a result, the reduction in the size of the set of functions provided with the PA method is more significant (as seen for rank 5 in Table \ref{tab:cum_desc_counts}). Reduction becomes more significant with increasing rank in general. Another clear demonstration of the saturated reduction of the over-complete RPI set could also be observed in tabluated functions, given in the next section.

The PA-RPI function counts in Tables \ref{tab:table1_a} - \ref{tab:table1_d} are given exhaustively for all valid multisets of angular momentum quantum numbers from 1 to 3. Many additional sets of angular momentum quantum numbers are provided, including some $l_i=0$ cases for practical purposes. The function counts are usually trivial when some $l_i=0$, because the block sizes are one in such cases. These extra combinations help show patterns in the PA-RPI blocks arising from the linearly independent function sequences comprising them. For example, the cases where all $l_i$ are equivalent (e.g., with $P_f(\boldsymbol{l})=(4)$) in Table \ref{tab:table1_a}, all multisets for $P_f(\boldsymbol{l})=(4)$ are given up to $l_i$=7. This is to better demonstrate patterns arising from $\mathcal{F}_a^{PA}$. Descriptors with large angular momentum quantum numbers are rarely reported on, but our procedures highlight which ones may be used in the construction of a linearly independent set.

To help demonstrate the need for the PA approach, additional values have been provide in the Tables with the PA-RPI function counts, Tables \ref{tab:table1_a} - \ref{tab:table1_d}. The columns titled 'lex' in these tables provide the maximum number of function labels one may obtain when using a single fixed ordering of angular momentum quantum numbers while imposing parity constraints. It can be seen that in some cases, it is smaller than the number of PA-RPI labels. There are cases where there are up to 40\% of the functions missing for a given combination of radial and angular momentum indices. Using a single fixed ordering of angular function indices may yield an incomplete basis in such cases if parity constraints are imposed as they are in some implementations for ACE. 

The columns titled 'OCB' in Tables \ref{tab:table1_a} - \ref{tab:table1_d}, short for over-complete blocks, show the numbers of functions in the over-complete blocks generated in step 1 of the permutation-adapted method (before the block-wise independence is enforced). While these counts are high in some cases, these are still much smaller than counts one would obtain when naively using all coupling trees generated by all permutations of $n_i$ and $l_i$ with all intermediates. One may also note that there are, in many cases, very simple relationships between the number of PA-RPI functions and the size of the ovecomplete blocks obtained after the first step of the PA approach. These patterns arise as a result of the ladder relationships.

The final two columns in PA-RPI function count tables, Tables \ref{tab:table1_a} - \ref{tab:table1_d}, are labeled 'N' and 'PA'. These correspond to the basis counts from the semi-numerical approach and the size of the corresponding PA set, respectively. The counts for the PA set are obtained by applying ladder relationships and/or equivalent permutations of labels within blocks of RPI functions that share the same non-angular and angular function indices generated by step 1. Numerical results are provided for comparison to previously reported work, and to validate the PA method. Without the definition of an inner product to define a formal basis of RPI functions, numerical validation is an important step for assessing the validity of the PA method. In all reported cases, the PA-RPI set is the same size as rigorously defined semi-numerical basis sets. This includes key cases above $N=4$, and it is expected that the PA method may be used for arbitrary rank provided that the ladder relationships are defined. 

A table for PA-RPI function counts is not given for the case where no $l_i$ are equivalent. It was suggested in proposition 7 of Ref.~[\!\citenum{dusson_atomic_2022}], for the case when no $n_i$ are equivalent and no $l_i$ are equivalent, the number of complete and independent RPI functions may be obtained directly. All unique labels give an independent function in such cases. This is also the case with the PA-method. The new results presented within provide some justification for why this is the case though. For cases where no $l_i$ are equivalent, then ladder relationships show that all valid multisets of intermediate angular momenta yield a new independent function for $\boldsymbol{nl}$ blocks with no duplicate angular indices.

\subsection{Computationally Efficient Descriptor Generation}
For general usage, the permutation-adapted rotational basis and permutation-adapted rotational/permutational basis are implemented in the sym\_ACE library. This python library may be used to generate the set of PA-RPI and PA-RI descriptor labels, as well as evaluate generalized Wigner symbols for other software packages. It takes advantage of tabulated automorphism groups from the Groups, Algorithms, Programming (GAP) code for computational group theory.\cite{noauthor_gap_2021} The ladder relationships needed to construct PA-RPI functions are included as well. 

Our derivations of linear relationships, numerical validation, and theoretical arguments such as the orthogonality of the generalized Wigner symbols, support the conjecture that the PA-RPI set forms a complete, independent basis.\cite{yutsis_mathematical_1962} As such, it is proposed that the PA method be used to obtain sets of ACE descriptors, as these sets can be defined without SVD. The utility and properties of PA descriptor sets that are guaranteed to be block-wise independent is of practical and theoretical importance for ACE methods, including interatomic potential applications. The use of the PA method facilitates the efficient generation of ACE potentials with high degree descriptors. The standard usage of sym\_ACE to generate PA sets relies on tabulated automorphism groups for ranks $N \le 8$. There are tools in sym\_ACE to generate these on-the-fly for larger ranks, but the tabulated automorphisms are used by default for efficiency. Additionally, there are recursive algorithms that can be used to obtain generalized Wigner symbols and generalized Clebsch-Gordan coefficients of arbitrary rank but are not used in sym\_ACE by default. In a similar vein, the sequences of independent RPI functions obtained by sampling over-complete blocks are defined for limited ranks. This is because, as mentioned before, they must be derived for each type of block. For many practical cases, especially in the field of interatomic potentials, $N\le 8$ is sufficient; $N=8$ terms correspond to 9-body interactions.  

The PA approach could theoretically be applied for broader applications of ACE methods that use RPI functions/descriptors with larger ranks and for arbitrary $L_R$, provided block-wise independence is enforced as prescribed in this work. Extension of the PA method to these high body-order regimes may require adopting recursive algorithms to construct the generalized Wigner symbols, using on-the-fly generation of the permutation automorphisms, or relying more exclusively on Young Diagrams. The upper bound of $N$ for which a PA set can be obtained has not been extensively tested, however efficient algorithms for defining the specific permutations needed in the PA method have $\mathcal{O}(N^2)$ scaling.\cite{knuth1970permutations} 

\begin{figure}
    \centering
    \includegraphics[width=0.45\textwidth]{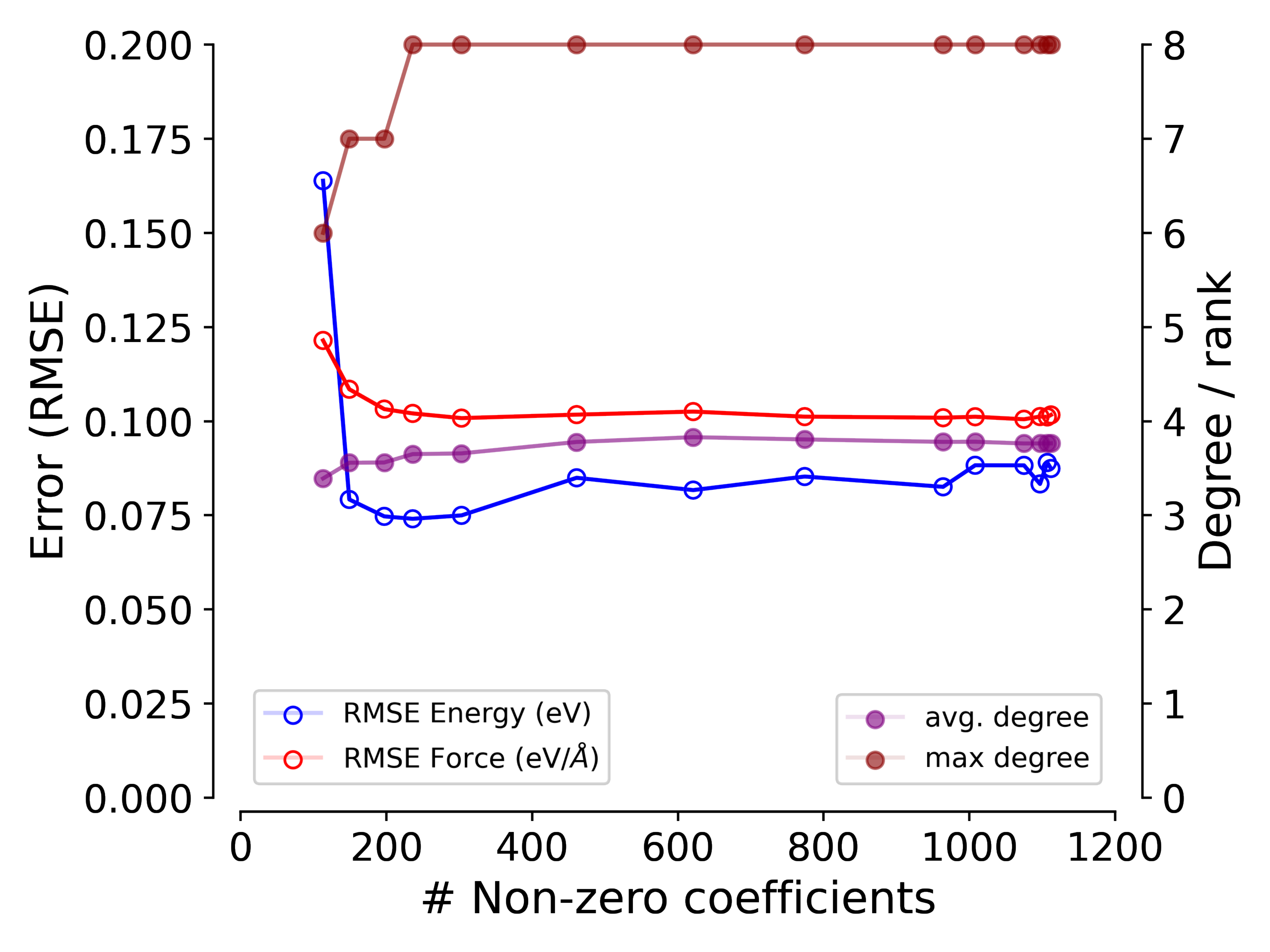}
    \caption{Effect of regularization on all $N=4$ descriptors in the PA set up to polynomial degree 24. For decreasing sparsification parameter, the root mean square training error in the energy and forces (open blue and red circles, left axis) is plotted against number of active descriptors. On the right axis, the maximum (dark red, filled circle) and average (purple, filled circle) degree is plotted.  The tantalum training dataset from Ref.~[\!\citenum{thompson_spectral_2015}] was used.}
    \label{fig:en_reg}
\end{figure}

For some ACE models, the descriptors with high rank and degree are important. An example is provided for a metallic tantalum system; linear energy models are trained using energies and forces from the data set in Ref.~[\!\citenum{thompson_spectral_2015}] using FitSNAP with a sparse regression method, Bayesian compressive sensing. This was done using 6 single-bond descriptors along with a set of rank 2-4 ACE descriptors with $n_{max}=3$ and $l_{max}=3$. Without sparse regression, the maximum degree of any $N=4$ descriptor included in the fit is 24. As shown in Fig. \ref{fig:en_reg}, many high-degree descriptors remain after heavy feature pruning. The minimum error is achieved with a sparse model containing some high-degree descriptors. The pruning of descriptors in Fig. \ref{fig:en_reg} is done using Bayesian compressive sensing, and coefficients with highest uncertainty are eliminated first. This suggests that for this small tantalum data set, some of the highest degree descriptors correspond to an important signal in the potential that is predicted with low uncertainty compared to other descriptor coefficients. Depending on the system and the training data, high degree functions may be important. The PA-RPI procedures facilitate the use of these and other high-degree descriptors, allowing users to explore these trade-offs.

\begin{table*}
\begin{subtable}[t]{0.45\textwidth}
\centering
\begin{tabular}[t]{p{1.5cm}p{1.0cm}p{1.0cm}p{1.0cm}p{1.0cm}p{1.0cm}}
\hline \hline
$\boldsymbol{n}$ & $\boldsymbol{l}$ &  lex. & OCB & Num. & PA\\
\hline                                                 
(aaaa) & (0000) & 1 & 1 & 1 & 1 \\
&        (1111) & 2 & 3 & 1 & 1 \\
&        (2222) & 3 & 5 & 1 & 1 \\
&        (3333) & 4 & 7 & 2 & 2 \\
&        (4444) & 5 & 9 & 2 & 2 \\
&        (5555) & 6 & 11 & 2 & 2 \\
&        (6666) & 7 & 13 & 3 & 3 \\
&        (7777) & 8 & 15 & 3 & 3 \\
\hline
(aaab) & (0000) & 1 & 1 & 1 & 1 \\
&        (1111) & 2 & 3 & 1 & 1 \\
&        (2222) & 3 & 5 & 1 & 1 \\
&        (3333) & 4 & 7 & 2 & 2 \\
&        (4444) & 5 & 9 & 2 & 2 \\
&        (5555) & 6 & 11 & 2 & 2 \\
&        (6666) & 7 & 13 & 3 & 3 \\
&        (7777) & 8 & 15 & 3 & 3 \\
\hline
(aabb) & (0000) & 1 & 1 & 1 & 1 \\
&        (1111) & 2 & 3 & 2 & 2 \\
&        (2222) & 3 & 5 & 3 & 3 \\
&        (3333) & 4 & 7 & 4 & 4 \\
&        (4444) & 5 & 9 & 5 & 5 \\
&        (5555) & 6 & 11 & 6 & 6 \\
&        (6666) & 7 & 13 & 7 & 7 \\
&        (7777) & 8 & 15 & 8 & 8 \\
\hline
(aabc) & (0000) & 1 & 1 & 1 & 1 \\
&        (1111) & 2 & 3 & 2 & 2 \\
&        (2222) & 3 & 5 & 3 & 3 \\
&        (3333) & 4 & 7 & 4 & 4 \\
&        (4444) & 5 & 9 & 5 & 5 \\
&        (5555) & 6 & 11 & 6 & 6 \\
&        (6666) & 7 & 13 & 7 & 7 \\
&        (7777) & 8 & 15 & 8 & 8 \\
\hline
(abcd) & (0000) & 1 & 1 & 1 & 1 \\
&        (1111) & 2 & 3 & 3 & 3 \\
&        (2222) & 3 & 5 & 5 & 5 \\
&        (3333) & 4 & 7 & 7 & 7 \\
&        (4444) & 5 & 9 & 9 & 9 \\
&        (5555) & 6 & 11 & 11 & 11 \\
&        (6666) & 7 & 13 & 13 & 13 \\
&        (7777) & 8 & 15 & 15 & 15 \\
\hline
\end{tabular}
\caption{\footnotesize $P_f(\boldsymbol{l})=(4)$}
\label{tab:table1_a}
\end{subtable}%
\begin{subtable}[t]{0.45\textwidth}
\centering
\begin{tabular}[t]{p{1.5cm}p{1.0cm}p{1.0cm}p{1.0cm}p{1.0cm}p{1.0cm}}
\hline \hline
$\boldsymbol{n}$ & $\boldsymbol{l}$ &  lex. & OCB & N & PA\\
\hline
(aaaa) & (0222) & 1 & 1 & 1 & 1 \\
&        (1113) & 1 & 1 & 1 & 1 \\
&        (1333) & 2 & 3 & 1 & 1 \\
&        (2224) & 2 & 3 & 1 & 1 \\
&        (3555) & 4 & 7 & 2 & 2 \\
\hline
(aaab) & (0222) & 2 & 2 & 2 & 2 \\
&        (1113) & 2 & 2 & 2 & 2 \\
&        (1333) & 4 & 6 & 3 & 3 \\
&        (2224) & 4 & 6 & 3 & 3 \\
&        (3555) & 8 & 14 & 6 & 6 \\
\hline
(aabb) & (0222) & 2 & 2 & 2 & 2 \\
&        (1113) & 2 & 2 & 2 & 2 \\
&        (1333) & 4 & 6 & 4 & 4 \\
&        (2224) & 4 & 6 & 4 & 4 \\
&        (3555) & 8 & 14 & 8 & 8 \\
\hline
(aabc) & (0222) & 3 & 3 & 3 & 3 \\
&        (1113) & 3 & 3 & 3 & 3 \\
&        (1333) & 6 & 9 & 7 & 7 \\
&        (2224) & 6 & 9 & 7 & 7 \\
&        (3555) & 12 & 21 & 14 & 14 \\
\hline
(abcd) & (0222) & 4 & 4 & 4 & 4 \\
&        (1113) & 4 & 4 & 4 & 4 \\
&        (1333) & 8 & 12 & 12 & 12 \\
&        (2224) & 8 & 12 & 12 & 12 \\
&        (3555) & 16 & 28 & 28 & 28 \\
\hline
\end{tabular}
\caption{\footnotesize $P_f(\boldsymbol{l})=(3,1)$}
\label{tab:table1_b}
\end{subtable}%
\label{tab:exhaust1}
\caption{\footnotesize For all possibilities of equivalent $n_i \in \boldsymbol{n}$, we provide the $\boldsymbol{n}$ (1$^{st}$ column), $\boldsymbol{l}$ (2$^{nd}$ column), the number of $\boldsymbol{nlL}$ labels one obtains for a \textit{single} lexicographically ordered $\boldsymbol{l}$ with couplings forced to have even parity (3$^{rd}$ column), the number of functions in blocks generated by step 1 of the PA-RPI procedure (4$^{th}$ column), the number of independent functions as determined by semi-numerical methods (5$^{th}$ column), and the number PA-RPI functions after applying step 2 of the procedure, 6$^{th}$ column. This is repeated for various possible $\boldsymbol{l}$ where any 2 or more $l_i$ are equivalent. Variables are given in place of the actual values of $l_i$ for the sub-tables \ref{tab:table1_a}, \ref{tab:table1_b}.}
\end{table*}

\begin{table*}
\begin{subtable}[t]{0.45\textwidth}
\centering
\begin{tabular}[t]{p{1.5cm}p{1.0cm}p{1.0cm}p{1.0cm}p{1.0cm}p{1.0cm}}
\hline \hline
$\boldsymbol{n}$ & $\boldsymbol{l}$ &  lex. & OCB & N & PA\\
\hline
(aaaa) & (0022) & 1 & 1 & 1 & 1 \\
&        (1122) & 2 & 3 & 2 & 2 \\
&        (1133) & 2 & 3 & 2 & 2 \\
&        (2233) & 3 & 5 & 3 & 3 \\
&        (4455) & 5 & 9 & 5 & 5 \\
\hline
(aaab) & (0022) & 2 & 2 & 2 & 2 \\
&        (1122) & 4 & 6 & 4 & 4 \\
&        (1133) & 4 & 6 & 4 & 4 \\
&        (2233) & 6 & 10 & 6 & 6 \\
&        (4455) & 10 & 18 & 10 & 10 \\
\hline
(aabb) & (0022) & 3 & 3 & 3 & 3 \\
&        (1122) & 6 & 9 & 7 & 7 \\
&        (1133) & 6 & 9 & 7 & 7 \\
&        (2233) & 9 & 15 & 11 & 11 \\
&        (4455) & 15 & 27 & 19 & 19 \\
\hline
(aabc) & (0022) & 4 & 4 & 4 & 4 \\
&        (1122) & 8 & 12 & 10 & 10 \\
&        (1133) & 8 & 12 & 10 & 10 \\
&        (2233) & 12 & 20 & 16 & 16 \\
&        (4455) & 20 & 36 & 28 & 28 \\
\hline
(abcd) & (0022) & 6 & 6 & 6 & 6 \\
&        (1122) & 12 & 18 & 18 & 18 \\
&        (1133) & 12 & 18 & 18 & 18 \\
&        (2233) & 18 & 30 & 30 & 30 \\
&        (4455) & 30 & 54 & 54 & 54 \\
\hline
\end{tabular}
\caption{\footnotesize $P_f(\boldsymbol{l})=(2,2)$}
\label{tab:table1_c}
\end{subtable}%
\begin{subtable}[t]{0.45\textwidth}
\centering
\begin{tabular}[t]{p{1.5cm}p{1.0cm}p{1.0cm}p{1.0cm}p{1.0cm}p{1.0cm}}
\hline \hline
$\boldsymbol{n}$ & $\boldsymbol{l}$ &  lex. & OCB & N & PA\\
\hline
(aaaa) & (0112) & 1 & 1 & 1 & 1 \\
&        (1223) & 2 & 3 & 2 & 2 \\
&        (1344) & 2 & 3 & 2 & 2 \\
&        (2455) & 3 & 5 & 3 & 3 \\
&        (3445) & 4 & 7 & 4 & 4 \\
\hline
(aaab) & (0112) & 3 & 3 & 3 & 3 \\
&        (1223) & 6 & 9 & 7 & 7 \\
&        (1344) & 6 & 9 & 7 & 7 \\
&        (2455) & 9 & 15 & 11 & 11 \\
&        (3445) & 12 & 21 & 15 & 15 \\
\hline
(aabb) & (0112) & 4 & 4 & 4 & 4 \\
&        (1223) & 8 & 12 & 10 & 10 \\
&        (1344) & 8 & 12 & 10 & 10 \\
&        (2455) & 12 & 20 & 16 & 16 \\
&        (3445) & 16 & 28 & 22 & 22 \\
\hline
(aabc) & (0112) & 7 & 7 & 7 & 7 \\
&        (1223) & 14 & 21 & 19 & 19 \\
&        (1344) & 14 & 21 & 19 & 19 \\
&        (2455) & 21 & 35 & 31 & 31 \\
&        (3445) & 28 & 49 & 43 & 43 \\
\hline
(abcd) & (0112) & 12 & 12 & 12 & 12 \\
&        (1223) & 24 & 36 & 36 & 36 \\
&        (1344) & 24 & 36 & 36 & 36 \\
&        (2455) & 36 & 60 & 60 & 60 \\
&        (3445) & 48 & 84 & 84 & 84 \\
\hline
\end{tabular}
\caption{\footnotesize $P_f(\boldsymbol{l})=(2,1,1)$}
\label{tab:table1_d}
\end{subtable}%
\label{tab:exhaust}
\caption{\footnotesize For all possibilities of equivalent $n_i \in \boldsymbol{n}$, we provide the $\boldsymbol{n}$ (1$^{st}$ column), $\boldsymbol{l}$ (2$^{nd}$ column), the number of $\boldsymbol{nlL}$ labels one obtains for a \textit{single} lexicographically ordered $\boldsymbol{l}$ with couplings forced to have even parity (3$^{rd}$ column), the number of functions in blocks generated by step 1 of the PA-RPI procedure (4$^{th}$ column), the number of independent functions as determined by semi-numerical methods (5$^{th}$ column), and the number PA-RPI functions after applying step 2 of the procedure, 6$^{th}$ column. This is repeated for various possible $\boldsymbol{l}$ where any 2 or more $l_i$ are equivalent. Variables are given in place of the actual values of $l_i$ for the sub-tables \ref{tab:table1_c}, \ref{tab:table1_d}.}
\end{table*}

\section{\label{conclusions}Conclusions}
The typical construction of RPI functions in ACE above $N=4$ is over-complete. Subsets (blocks) of functions that share the same non-angular and angular indices (regardless of ordering) may be linearly dependent. Previous numerical results suggest that linear dependencies exist within these blocks of functions, and we show that linear relationships within blocks of functions may be derived analytically and used in the definition of linearly independent RPI function sequences. This is done effectively by relaxing the common convention constraint that all basis labels, $\boldsymbol{nl}$, must be lexicographically ordered and adapt the indexing to the permutation and recursion properties of the generalized Wigner symbols. The PA procedure to construct RPI functions eliminates linear dependence within the blocks of functions that share the same non-angular and angular indices. While orthogonality or linear dependence for the complete set of RPI functions has not been proven, this PA procedure addresses linear dependencies within all blocks of functions. We provide proofs of this for some simple, non-trivial cases as well as the relationships to do this for more general cases. In other implementations of ACE, this reduction is done numerically. The size of the PA-RPI set is the same size as semi-numerical RPI basis constructed in other works. The PA-RPI methodology may help avoid numerical instabilities in some numerical methods for high-degree, high-rank descriptors. 

This PA method results from analytical relationships and properties of Wigner symbols. Permutation and recursion properties have been explored in depth for Wigner-3j symbols, but the corresponding properties of the generalized Wigner symbols were needed for the PA method. Therefore, we have presented generalized recursion relationships for $N$ coupled angular momentum states and permutation symmetries of the generalized Wigner symbols. Together, these properties may be used to derive ladder relationships between functions with incremented intermediate angular momenta. These relationships show that when two functions duplicate indices are coupled, linearly dependent RPI functions/ACE descriptors can be the result. Deriving all ladder relationships for multisets of intermediates produced by raised/lowered angular indices allows one to define a sampling of over-complete blocks that yields a sequence of linearly independent RPI functions. In general, this sampling is defined for all types of blocks,  where block types are defined as functions with the same number of duplicate indices, compactly described by the frequency partitions of both non-angular and angular indices. Adapting the function indices to the properties of the Wigner symbols allows one to apply this sampling in a straightforward way. The set of PA functions, is therefore constructed out of blocks of functions where linear independence is guaranteed within each block. Linear independence is not guaranteed/proven between different blocks, and the PA set has not been proven form a complete, independent RPI basis. We do conjecture that the PA method can be used to construct a complete independent set of RPI functions of rank $N$, and we provide some theoretical and numerical evidence to support this. 

We emphasized rank 4 functions for examples throughout; they are the most simple, non-trivial cases. This method, even though it was a smaller point of discussion, was also applied to rank 5. The properties and relationships needed to apply this method to arbitrary rank have been provided. Additional testing and validation of the PA method for cases arbitrary/large $N$ would be beneficial. This may require the derivation of generalized recursion relationships for RPI functions and/or the sampling giving the linearly independent function sequence $\mathcal{F}_{a}^{PA}$ for arbitrary rank and multiplicity of indices. Obtaining such expressions may now be possible with the properties derived for generalized Wigner symbols in this work.

One common concern for ACE models is the large increase in the size of the ACE basis for multi-element systems and/or systems with additional degrees of freedom. Though it is not discussed in depth in this work, the same principles of the PA-RPI procedure apply for atomic systems with multiple element types, or with other degrees of freedom. Significant reductions in the ACE basis indexed on additional indices, such as chemical indices $\mu_i$, may be achieved using the properties of the generalized Wigner symbols as well. For descriptor sets containing chemical indices, initial tests indicate that there are significant reductions in the PA-RPI set size compared to set one needs to start with to construct semi-numerical bases.

In the most general applications, the ACE functions only need to be \textit{equivariant} with respect to rotations. In this work, we considered primarily the case of invariance with respect to rotations. This was done to help address the immediate application of ACE in machine-learned interatomic potentials, but the permutation-adapted method could be used to define sets of equivariant functions as well. Ladder relationships may also be derived for sets of permutation-invariant equivariant functions and the independent function sequences defined.

%%%

\begin{acknowledgments}
All authors gratefully acknowledge funding support from the U.S. Department of Energy, Office of Fusion Energy Sciences (OFES) under Field Work Proposal Number 20-023149.
Sandia National Laboratories is a multi-mission laboratory managed and operated by National Technology and Engineering Solutions of Sandia, LLC, a wholly owned subsidiary of Honeywell International, Inc., for the U.S. Department of Energy's National Nuclear Security Administration under contract DE-NA0003525. 
This paper describes objective technical results and analysis. Any subjective views or opinions that might be expressed in the paper do not necessarily represent the views of the U.S. Department of Energy or the United States Government.
\end{acknowledgments}

\bibliography{main.bib}

\onecolumngrid
\appendix
\section{\label{appendix:theory} Equivalence of generalized Wigner symbols}

\subsection{Wigner-3j symbols }
The Wigner-3j symbols obey simple relationships with one another under permutations of $(l_i,m_i)$ tuples (columns in matrix form). The traditional Wigner symbols obey the following relationships for odd and even permutations.
\begin{equation}
    \begin{pmatrix}
     l_1 & l_2 & l_3\\
     m_1 & m_2 & m_3
\end{pmatrix}=
(-1)^{l_1+l_2+l_3}
    \begin{pmatrix}
     l_1 & l_3 & l_2\\
     m_1 & m_3 & m_2
\end{pmatrix}=
\begin{pmatrix}
     l_3 & l_1 & l_2\\
     m_3 & m_1 & m_2
\end{pmatrix}
\label{eq:oddeven_w3j_perms}
\end{equation}
Similar permutations of $(l_i,m_i)$ tuples in traditional Clebsch-Gordan coefficients are also equivalent, potentially with the introduction of a phase and/or scaling factor. An additional property of the traditional Wigner-3j symbols that is important when considering symmetric permutations is 
\begin{equation}
\begin{pmatrix}
     l_1 & l_2 & 0\\
     m_1 & m_2 & 0
\end{pmatrix} = \delta(l_1,l_2)\delta(m_1,-m_2)\frac{(-1)^{l_1-m_1}}{\sqrt{2l_1+1}}
    \label{eq:w3j_metric}
\end{equation}
where $\delta(1,2)$ is the Kronecker delta.\cite{edmonds_angular_1996} By Eq. \eqref{eq:w3j_metric}, the value of non-zero Wigner-3j symbols is a phase with a scale factor when any one of the $l_i$ is zero.

\subsection{Generalized Wigner symbols}
\label{subsec:coupling}

Due to the permutation symmetries of the Wigner-3j symbols, it may be shown that the pairwise coupling scheme preserves more equivalent permutations of angular function indices rather than intermediates in some other schemes. This is highlighted for the case of $N = 4$ and $L_R=0$. As it may be done for arbitrary, $N$, the collection of all equivalent permutations may be generated using the repeated application of permutation relationships for traditional Wigner-3j symbols. In practice, we make use of this to obtain permutation symmetries of generalized Wigner symbols for arbitrary rank. While these permutation automorphisms have been defined in the main text, they are also provided here with specific examples and proofs.

In matrix form, the permutation symmetries of generalized Wigner symbols of rank 4 are:
\begin{equation}
\begin{split}
   \begin{pmatrix}
     l_2 & l_1 & l_3 & l_4\\
     m_2 & m_1 & m_3 & m_4
    \end{pmatrix}
    \leftrightarrow
    \begin{pmatrix}
     l_1 & l_2 & l_4 & l_3\\
     m_1 & m_2 & m_4 & m_3
     \end{pmatrix}\\
        \leftrightarrow
\begin{pmatrix}
     l_1 & l_2 & l_3 & l_4\\
     m_1 & m_2 & m_3 & m_4
\end{pmatrix} 
    \leftrightarrow
    \begin{pmatrix}
     l_2 & l_1 & l_4 & l_3\\
     m_2 & m_1 & m_4 & m_3
     \end{pmatrix}\\
    \leftrightarrow
\begin{pmatrix}
     l_3 & l_4 & l_1 & l_2\\
     m_3 & m_4 & m_1 & m_2
\end{pmatrix}
    \leftrightarrow
    \begin{pmatrix}
     l_4 & l_3 & l_2 & l_1\\
     m_4 & m_3 & m_2 & m_1
     \end{pmatrix}\\
    \leftrightarrow
\begin{pmatrix}
     l_4 & l_3 & l_1 & l_2\\
     m_4 & m_3 & m_1 & m_2
\end{pmatrix} 
    \leftrightarrow
    \begin{pmatrix}
     l_3 & l_4 & l_2 & l_1\\
     m_3 & m_4 & m_2 & m_1
     \end{pmatrix}.
\end{split}
\label{eq:rank4syms}
\end{equation}
Note that some of the permutations in Eq. \eqref{eq:rank4syms} include permutations of intermediates and the corresponding action on the leaves is given. In practice, these permutation symmetries hold true when any intermediate permutations involved in the automorphism are performed and/or when intermediates to be permuted are duplicates. Explicit values of such permutations were provided in Table \ref{tab:explicit_perm}. Though the permutation symmetries could be constructed for any arbitrary coupling scheme, it is better to choose one that belongs to the partition in Eq. \eqref{eq:coupling_partition}. Otherwise the group of permutation automorphisms may include permutations between $l_i$ and $L_k$.

Apart from the outline provided in the main text for obtaining equivalent permutations of generalized Wigner symbols, other methods can be used to obtain them. Combining permutation symmetries for the leaves and intermediates yields a group of equivalent binary coupling trees related by a permutation operation acting on $N$ leaves and $K=N-2$ intermediates. These permutation symmetries are known; they are elements of the automorphism group of the complete binary tree. This automorphism group is given by an iterated wreath product of the symmetric group $S_2$; the wreath product is performed $h$ times where $h$ is the height of the binary tree, given from Eq. \eqref{eq:min_height}.\cite{brunner_automorphism_1997}
\begin{subequations}
\begin{equation}
   H_{1} = S_2 
   \label{eq:wreath1}
\end{equation}    
\begin{equation}
  H_{2} = S_2 \wr S_2
  \label{eq:wreath2}
\end{equation}
\begin{equation}
  H_{h} = H_{h-1} \wr S_2
  \label{eq:wreath}
\end{equation}
\end{subequations}
The group of automorphisms generated in Eq. \eqref{eq:wreath1} may be defined for any arbitrary binary tree of height $h$, by iterating the wreath product $h$ times, Eq. \eqref{eq:wreath}. However, only a subset of $H_h$ is needed when considering angular momentum coupling trees. Some elements of $H_h$ include permutations of the root node, $L_R$ in our case, with internal nodes $L_k$. This could result in couplings that are not rotationally invariant. To address this, aone could select a subset of $H_h$ is selected that preserves the level of the node values (e.g. $l_i$ cannot be permuted with $L_k$). Additionally, this collection of permutation automorphisms may change with the number of duplicate leaf indices, $(l_i,m_i)$, and/or the number of duplicate intermediates $L_k$. As a result, a subset of $H_h$ that is grown to account for duplicate indices may be collected into $G_N$. It is constructed in general by starting with $H_h$ and removing elements that permute between levels of the coupling tree. This collection of automorphisms can be generated for arbitrary coupling rank, $N$ with height given by Eq. \eqref{eq:min_height}. This is one of the convenient features of the pairwise coupling scheme used for generalized Wigner symbols in this work; the coupling tree for the generalized Wigner symbols has the same structure as the complete binary tree.

\subsection{Explicit proof for rank 4 equivalences}
\label{subsec:proof}

The rank 4 generalized Wigner symbols with a $\sigma_c = (12)(34)$ coupling scheme may be explicitly written, from Eq. 10.3 of Ref.~[\!\citenum{yutsis_mathematical_1962}] as,
\begin{equation}
\begin{split}
    \begin{pmatrix}
     l_1 & l_2 & l_3 & l_4\\
     m_1 & m_2 & m_3 & m_4
\end{pmatrix}(\boldsymbol{L},\boldsymbol{M}) = \sum_{M_{1},M_{2}}
(-1) ^ {L_1 - M_1 } (-1) ^ {L_2 - M_2}
\begin{pmatrix}
     l_1 & l_2 & L_{1}\\
     m_1 & m_2 & -M_{1}
\end{pmatrix} \\
\begin{pmatrix}
     l_3 & l_4 & L_{2} \\
     m_3 & m_4 & -M_{2}
\end{pmatrix}
\begin{pmatrix}
     L_1 & L_2 & L_R \\
     M_1 & M_2 & M_R
\end{pmatrix}
\end{split}
\label{eq:single_rank4}
\end{equation}
We often consider the case of $L_R=0$, for which the generalized symbols are non-zero only when $M_R = 0$, $L_1 = L_2 = L$,  $M_1 = -M_2 = M$, $m_1+m_2 = M$, and $m_3+m_4 = -M$. As a result of these restrictions, only a single non-zero term remains

\begin{equation}
\begin{split}
    \begin{pmatrix}
     l_1 & l_2 & l_3 & l_4\\
     m_1 & m_2 & m_3 & m_4
\end{pmatrix}((L_1 L_2 0),(M_1 M_2 0)) =
\frac{1}{\sqrt{2L + 1}}(-1)^{L - M}
\begin{pmatrix}
     l_1 & l_2 & L\\
     m_1 & m_2 & -M
\end{pmatrix}
\begin{pmatrix}
     l_3 & l_4 & L \\
     m_3 & m_4 & M
\end{pmatrix}
\end{split}
\label{eq:single_rank4b}
\end{equation}

where we have used Eq.~\eqref{eq:w3j_metric} to eliminate the third Wigner-3j symbol. By Eq.~\eqref{eq:oddeven_w3j_perms}, the values of the two traditional Wigner coefficients are unaffected by permutation of the columns, up to a phase change. Hence the full expression is invariant up to a phase change under any permutation in the collection of automorphisms, $ \forall \sigma \in G_N$.  The phase is given simply by the parity of the permutation: odd permutations yield a negative sign while even permutations do not. These permutation symmetries of the generalized Wigner symbols may be categorized a few ways. The first: permutation between pairs of $l_i$ that are coupled (e.g. $l_1$ and $l_2$ may be permuted under the parent, $L_1$). The second: a permutation of coupled branches (e.g. permuting children of $L_R$ for rank 4 trees: $L_1$ and $L_2$ may be permuted, then all corresponding children of $L_1$ and $L_2$ such that the order in the sub-trees $[l_1,l_2]$ and $[l_3,l_4]$ are preserved). The third: some combination of the first and second kinds of equivalent permutations. For the case $\boldsymbol{lL} = (1234)(22)$ and $\boldsymbol{m} = (1,-2,-3,4)$ shown in Table  ~\ref{tab:explicit_perm}, we can use the above expression to obtain

\begin{equation}
\begin{split}
\begin{pmatrix}
     1 & 2 & 3 & 4\\
     1 & -2 & -3 & 4
\end{pmatrix}((2 2 0),\boldsymbol{M}) =
\frac{-1}{\sqrt{5}}
\begin{pmatrix}
     1 & 2 & 2\\
     1 & -2 & 1
\end{pmatrix}
\begin{pmatrix}
     3 & 4 & 2\\
     -3 & 4 & -1
\end{pmatrix}
= \frac{1}{\sqrt{1125}}
\end{split}
\label{eq:single_rank4c}
\end{equation}

where $M_k$ are given implicitly by $m_i$. Conversely, for permutations across the intermediate coupling such as $(13) \notin G_N$, we get:

\begin{equation}
\begin{split}
    \begin{pmatrix}
     l_3 & l_2 & l_1 & l_4\\
     m_3 & m_2 & m_1 & m_4
\end{pmatrix}((L L 0),\boldsymbol{M}) =
\frac{1}{\sqrt{2L + 1}}(-1)^{L - M}
\begin{pmatrix}
     l_3 & l_2 & L\\
     m_3 & m_2 & -M
\end{pmatrix}
\begin{pmatrix}
     l_1 & l_4 & L \\
     m_1 & m_4 & M
\end{pmatrix}
\end{split}
\label{eq:single_rank4d}
\end{equation}
which is not equivalent to a Wigner symbol with ordered angular momentum and projection indices, $W_{l_1 l_2 l_3 l_4}^{m_1 m_2 m_3 m_4} ((L L 0)(-M,M,0)$. Note that symmetry breaking occurs with certain permutations that span indices in distinct symmetric pairs. Similar proofs of permutational symmetries are straightforward to construct for generalized Wigner symbols of arbitrary rank. This is why it is often just mentioned in other work.\cite{yutsis_mathematical_1962} In the main text, the group of all permutation automorphisms for the coupling trees are obtained from Eq. \eqref{eq:wreath}. 

\section{\label{appendix:recursions} Ladder relationships}
\subsection{Recursion relationships for Wigner-3j symbols}
In this section we provide a detailed and thorough background for the properties of the Wigner symbols when quantum numbers are raised or lowered. From p. 224 of Rose's "Elementary Theory of Angular Momentum" (1957), there are raising/lowering (a.k.a ladder) operations that yield relationships between Clebsch-Gordan coefficients of increasing/decreasing principal quantum numbers.
\begin{equation}
    \begin{split}
    \bigg[ m_1 - m_3 \bigg( \frac{l_1(l_1 +1) - l_2(l_2 +1) + l_3(l_3 +1)}{2 l_3(l_3 +1)} \bigg) \bigg]  
    \begin{bmatrix}
     l_1 & l_2 & l_3\\
     m_1 & m_2 & m_3
     \end{bmatrix} = \\ 
     \bigg[  \frac{\big( (l_3)^2 - (m_3)^2 \big) ( l_2 + l_3 -l_1)(l_1+l_3-l_2)(l_1 +l_2 -l_3 +1)(l_1 +l_2 +l_3 + 1)   }{4(l_3)^2(2l_3-1)(2l_3+1)} \bigg]^{1/2}  \begin{bmatrix}
     l_1 & l_2 & l_3-1\\
     m_1 & m_2 & m_3
     \end{bmatrix} + \\
     \bigg[ \frac{ ((l_3+1)^2 - (m_3)^2) (1 + l_2+l_3 -l_1)(1+ l_1 +l_3 - l_2)(l_1+l_2-l_3)(l_1+l_2+l_3 +2)  }{4(l_3+1)^2(2l_3+1)(2l_3+3)} \bigg]^{1/2}
     \begin{bmatrix}
     l_1 & l_2 & l_3+1\\
     m_1 & m_2 & m_3
     \end{bmatrix}
     \end{split} 
\label{eq:cg_recursion_full}
\end{equation}
and in the form of Wigner-3j symbols:
\begin{equation}
    \begin{split}
    &\bigg[ m_1 - m_3 \bigg( \frac{l_1(l_1 +1) - l_2(l_2 +1) + l_3(l_3 +1)}{2 l_3(l_3 +1)} \bigg) \bigg]  
\begin{pmatrix}
     l_1 & l_2 & l_3\\
     m_1 & m_2 & -m_3
\end{pmatrix}
(\sqrt{2l_3 +1})  = \\ 
     &\bigg[  \frac{\big( (l_3)^2 - (m_3)^2 \big) ( l_2 + l_3 -l_1)(l_1+l_3-l_2)(l_1 +l_2 -l_3 +1)(l_1 +l_2 +l_3 + 1)   }{4(l_3)^2(2l_3-1)(2l_3+1)} \bigg]^{1/2}  \begin{pmatrix}
     l_1 & l_2 & l_3 -1\\
     m_1 & m_2 & -m_3
\end{pmatrix}
(\sqrt{2l_3})  + \\
     &\bigg[ \frac{ ((l_3+1)^2 - (m_3)^2) (1 + l_2+l_3 -l_1)(1+ l_1 +l_3 - l_2)(l_1+l_2-l_3)(l_1+l_2+l_3 +2)  }{4(l_3+1)^2(2l_3+1)(2l_3+3)} \bigg]^{1/2}
\begin{pmatrix}
     l_1 & l_2 & l_3+1\\
     m_1 & m_2 & -m_3
\end{pmatrix}
(\sqrt{2(l_3 +1)}) 
     \end{split}
     \label{eq:wigner_recursion_full}
\end{equation}
For the purposes of comparing different intermediate angular momenta, it will often be sufficient to express Eq. \eqref{eq:wigner_recursion_full} as 
\begin{equation}
    \begin{pmatrix}
     l_1 & l_2 & l_3\\
     m_1 & m_2 & -m_3
\end{pmatrix} = A_-(l_1,l_2,m_1,m_2) \begin{pmatrix}
     l_1 & l_2 & l_3 -1\\
     m_1 & m_2 & -m_3
\end{pmatrix} + A_+(l_1,l_2,m_1,m_2)\begin{pmatrix}
     l_1 & l_2 & l_3+1\\
     m_1 & m_2 & -m_3
\end{pmatrix},
\label{eq:wigner_recursion}
\end{equation}
because only the principle quantum number in the third column changes. The most general notation suggests that coupling coefficients incremented by an integer value are related to others by factors $A_+$ and $A_-$. These factors in Eq. \eqref{eq:wigner_recursion} depend on all quantum numbers in the coupling coefficient (all $l_i$ and all $m_i$). However, this can be simplified when one takes triangle conditions into account as well as the fact that $m_3 = m_1+m_2$.

Note that triangle conditions state: $|l_1 - l_2| \le l_3 \le l_1 + l_3$, therefore one may express the properties in Eq. \eqref{eq:wigner_recursion} beginning with the minimum value of principle angular momentum $\min (l_3) = |l_1 - l_2| \equiv \mathcal{L}$. From here on, this minimum value of $l_3$ will be denoted as $\mathcal{L}$. It depends only on $l_1$ and $l_2$. Incrementing the third principle quantum number from this minimum value gives,
\begin{equation}
\begin{split}
&\begin{pmatrix}
     l_1 & l_2 & \mathcal{L}\\
     m_1 & m_2 & -m_3
\end{pmatrix}=A_+(l_1,l_2, m_1,m_2,1)\cdot{} \begin{pmatrix}
     l_1 & l_2 & \mathcal{L}+1\\
     m_1 & m_2 & -m_3
\end{pmatrix}
\end{split}
\label{eq:min_iter}
\end{equation}
where the first term is zero due to triangle conditions. For the dependence of the $A_+$ and $A_-$ coefficients, they now only depend on the first two angular momentum quantum numbers and an increment $k=0$, that $\mathcal{L}$ is incremented by. The result of Eq. \eqref{eq:min_iter} shows that the Wigner symbol with $l_3=\mathcal{L}$ is linearly related to that for $\mathcal{L}+1$. While it is obviously true that relationships such as that in Eq. \eqref{eq:min_iter} would be obeyed given that Wigner-3j symbols are constants used to couple states, such recursions are needed for the derivation of relationships for RPI functions with raised/lowered intermediates.

To consider other increments, the relationship in Eq. \eqref{eq:wigner_recursion} will be applied multiple times from the minimum, $\mathcal{L}$. For the next principal quantum number:
\begin{equation}
\begin{split}
&\begin{pmatrix}
     l_1 & l_2 & \mathcal{L}+1\\
     m_1 & m_2 & -m_3
\end{pmatrix}  = A_-'(l_1,l_2,m_1,m_2,1)A_+(l_1,l_2,m_1,m_2,1) \cdot \begin{pmatrix}
     l_1 & l_2 & \mathcal{L} \\
     m_1 & m_2 & -m_3
\end{pmatrix}\\ &+ A_+'(l_1,l_2,m_1,m_2,1)\cdot \begin{pmatrix}
     l_1 & l_2 & \mathcal{L}+2\\
     m_1 & m_2 & -m_3
\end{pmatrix}.
\end{split}
\label{eq:first_itera}
\end{equation}
Here, the increment is one, $k=1$, and is given as the last term in the parenthesis of the $A_\pm'$.
Substituting the relationship from Eq. \eqref{eq:min_iter} yields:

\begin{equation}
\begin{split}
&\begin{pmatrix}
     l_1 & l_2 & \mathcal{L}+1\\
     m_1 & m_2 & -m_3
\end{pmatrix}  = A_-'(l_1,l_2,m_1,m_2,1) \bigg( A_+(l_1,l_2,m_1,m_2,1) \cdot \begin{pmatrix}
     l_1 & l_2 & \mathcal{L}+1\\
     m_1 & m_2 & -m_3
\end{pmatrix} \bigg) \\
&+A_+'(l_1,l_2,m_1,m_2,1)\cdot \begin{pmatrix}
     l_1 & l_2 & \mathcal{L}+2\\
     m_1 & m_2 & -m_3
\end{pmatrix}
\end{split}
\label{eq:first_iterb}
\end{equation}

As a result, the Wigner symbol for $\mathcal{L} +1$ is linearly related to that of $\mathcal{L} + 2$. The expression in Eq. \eqref{eq:first_iterb} simplifies further upon substitution of Eq. \eqref{eq:min_iter} to
\begin{equation}
\begin{split}
&\begin{pmatrix}
     l_1 & l_2 & \mathcal{L}\\
     m_1 & m_2 & -m_3
\end{pmatrix}  =
 \frac{A_+(l_1,l_2,m_1,m_2,1) A_+'(l_1,l_2,m_1,m_2,1)}{(1- A_-'(l_1,l_2,m_1,m_2,1)A_+(l_1,l_2,m_1,m_2,1))}\cdot \begin{pmatrix}
     l_1 & l_2 & \mathcal{L}+2\\
     m_1 & m_2 & -m_3
\end{pmatrix}\\
& =
f_1(l_1,l_2,m_1,m_2,k=1) \begin{pmatrix}
     l_1 & l_2 & \mathcal{L}+2\\
     m_1 & m_2 & -m_3
\end{pmatrix}
\end{split}
\label{eq:first_iterc}
\end{equation}
The factor on the right hand side of Eq. \eqref{eq:first_iterc} is defined, and the denominator non-zero, when the triangle conditions are obeyed and projection quantum numbers are bounded by respective angular momentum quantum numbers, $-l_i \le m_i \le l_i $.

This may be iterated for all remaining principal quantum numbers up to the maximum value of $l_1+l_2$. It should also be noted that all terms in the linear factor may be written explicitly in terms of $\mathcal{L},l_1,l_2$, and $m_1,m_2$. As a result, with $l_1$, $l_2$, $m_1$, and $m_2$, held constant, all possible values of the principal quantum number yield linearly dependent Wigner-3j symbols. All Wigner-3j symbols related by this operation may be expressed in terms of the Wigner-3j symbol for $\mathcal{L}$ multiplied by some factor. By simple relationships between CG coefficients and Wigner-3j symbols, these linear dependencies extend to Clebsch-Gordan coefficients. 

A simplified expression will be used for an arbitrary increment $k_1$ from the lowest allowed value of the respective intermediate, $\mathcal{L}_1$.
\begin{equation}
 \begin{pmatrix}
     l_1 & l_2 & \mathcal{L}_1\\
     m_1 & m_2 & M_1
\end{pmatrix} =  f_1(l_1,l_2,m_1,m_2,k_1)\begin{pmatrix}
     l_1 & l_2 & \mathcal{L}_1 + k_1\\
     m_1 & m_2 & M_1
\end{pmatrix}
\label{eq:general_linear}
\end{equation}
which follows from the recursion relationships described above. An index on the third principal quantum number will be necessary when considering generalized coupling coefficients. In addition, it is important to note that the projection, $M_1$, is constrained by the values $m_1$ and $m_2$ as well as the final projection quantum number $M_R$.

Additional relationships exist for incremented projection quantum numbers in the Wigner-3j Wigner symbols. The first type of relationship is:
\begin{equation}
    \begin{split}
    & \begin{pmatrix}
     l_1 & l_2 & l_3\\
     m_1 & m_2 & m_3 \pm 1
     \end{pmatrix} =\\
     & \mathcal{J}\bigg[\frac{l_1(l_1+1) - m_1(m_1 \mp 1)}{l_3(l_3+1) - m_3(m_3 \pm 1)}\bigg]^{1/2}    \begin{pmatrix}
     l_1 & l_2 & l_3\\
     m_1 \mp 1 & m_2 & m_3
     \end{pmatrix} +
     \mathcal{J}\bigg[\frac{l_2(l_2+1) - m_2(m_2 \mp 1)}{l_3(l_3+1) - m_3(m_3 \pm 1)}\bigg]^{1/2}     \begin{pmatrix}
     l_1 & l_2 & l_3\\
     m_1 & m_2 \mp 1 & m_3
     \end{pmatrix}
    \end{split}.
    \label{eq:full_m_wig_recur}
\end{equation}
To compare different intermediate angular momenta in generalized Wigner symbols, it will be sufficient to express the prefactors as constants with respect to $l_3$. This takes the form:
\begin{equation}
    \begin{split}
    & \begin{pmatrix}
     l_1 & l_2 & l_3\\
     m_1 & m_2 & m_3 \pm 1
     \end{pmatrix} = G(l_1,l_3,m_1,m_3)
     \begin{pmatrix}
     l_1 & l_2 & l_3\\
     m_1 \mp 1 & m_2 & m_3
     \end{pmatrix} +
     H(l_2,l_3,m_2,m_3)\begin{pmatrix}
     l_1 & l_2 & l_3\\
     m_1 & m_2 \mp 1 & m_3
     \end{pmatrix}
    \end{split},
    \label{eq:m_wig_recur}
\end{equation}
where $G$ and $H$ are factors that depend on both angular momentum quantum numbers and projection quantum numbers, and $\mathcal{J}=\sqrt{2l_3 + 1}$. Similar to the recursion relationships derived for the angular momentum quantum numbers, there are ways to reduce the number of variables that $G$ and $H$ depend on.

The final type of recursion relationships for Wigner-3j symbols are for raised/lowered projection quantum numbers with fixed $m_3$. One may derive, as in Rose 1957, that:
\begin{equation}
    \begin{split}
    & \begin{pmatrix}
     l_1 & l_2 & l_3\\
     m_1 & m_2 & m_3
     \end{pmatrix} = P_{+}
     \begin{pmatrix}
     l_1 & l_2 & l_3\\
     m_1 - 1 & m_2 + 1 & m_3
     \end{pmatrix} +
     P_{-}\begin{pmatrix}
     l_1 & l_2 & l_3\\
     m_1 +1 & m_2 -1 & m_3
     \end{pmatrix}
    \end{split},
    \label{eq:m3_cg_recur}
\end{equation}
where $P_{\pm}$ is given by,
\begin{equation}
    \mathcal{J}\;\frac{\sqrt{(l_1 \pm m_1)(l_2 \mp m_2)(l_1 \mp m_1 + 1)( l_2 \pm m_2 + 1)}} { l_3(l_3 + 1) - l_1(l_1 + 1) - l_2(l_2 + 1)  - 2 m_1 m_2 }.
\end{equation}
These are the key recursion relationships for the traditional Wigner-3j symbols. The generalized Wigner symbols used to construct ACE invariants are comprised of contractions of multiple traditional Wigner-3j symbols. The recursion relationships for generalized Wigner symbols have not yet been derived, but one may obtain them by applying the recursion relationships for each traditional Wigner-3j symbol in the generalized Wigner symbol.

\subsection{Recursion relationships between generalized Wigner symbols}

These recursion relationships will now be derived for the rank 4 generalized Wigner symbols. Note that the rank 4 generalized Wigner symbol with the lowest possible intermediates is given as,
\begin{equation}
\begin{split}
&\begin{pmatrix}
     l_1 & l_2 & l_3 & l_4\\
     m_1 & m_2 & m_3 & m_4
\end{pmatrix}\big(\boldsymbol{\mathcal{L}}, \boldsymbol{M} \big)=\\ 
&\sum_{M_{1},M_{2}} \prod_p (-1)^{(\mathcal{L}_p - M_p )}
\begin{pmatrix}
     l_1 & l_2 & \mathcal{L}_1 \\
     m_1 & m_2 & -(m_1 + m_2)
\end{pmatrix}
\begin{pmatrix}
     l_3 & l_4 & \mathcal{L}_2 \\
     m_3 & m_4 & -(m_3 + m_4)
\end{pmatrix}\\
&\begin{pmatrix}
     \mathcal{L}_1 & \mathcal{L}_2 & L_R \\
     (m_1 + m_2) & (m_3 + m_4) & -M_R
\end{pmatrix} 
\end{split}
\label{eq:first_rank4b}
\end{equation}
The phase, $\phi(\boldsymbol{l},\boldsymbol{m}) = \prod_p (-1)^{(\mathcal{L}_p - M_p )}$ may change based on increments. If the intermediate angular momenta are incremented by $\boldsymbol{k}=(k_1,k_2,\cdots k_N)$, it may be given by $\phi(\boldsymbol{l},\boldsymbol{m},\boldsymbol{k}) =\prod_p (-1)^{(\mathcal{L}_p + k_p - M_p )}$. If written as, $\phi(\boldsymbol{l},\boldsymbol{m})$ the phase does not explicitly depend on the increments. The form of this phase is derived for arbitrary rank in Yutsis.\cite{yutsis_mathematical_1962} It is sufficient here to know that in general, it depends on all intermediate angular momentum quantum numbers and all intermediate projection quantum numbers.
For all increments allowed by triangle conditions, one obtains:
\begin{equation}
\begin{split}
&\begin{pmatrix}
     l_1 & l_2 & l_3 & l_4\\
     m_1 & m_2 & m_3 & m_4
\end{pmatrix}\big( (\mathcal{L}_1 +k_1 , \mathcal{L}_2+k_2, L_R), \boldsymbol{M} \big)=\\ 
& \phi(\boldsymbol{l},\boldsymbol{m},\boldsymbol{k}) f_{12}(l_1,l_2,m_1,m_2,k_1)
\begin{pmatrix}
     l_1 & l_2 & \mathcal{L}_1\\
     m_1 & m_2 & -(m_1 + m_2)
\end{pmatrix}
f_{34}(l_3,l_4,m_3,m_4,k_2) \begin{pmatrix}
     l_3 & l_4 & \mathcal{L}_2 \\
     m_3 & m_4 & -(m_3 + m_4)
\end{pmatrix}\\
&\begin{pmatrix}
     \mathcal{L}_1 + k_1 & \mathcal{L}_2 + k_2 & L_R \\
     (m_1 + m_2) & (m_3 + m_4) & -M_R
\end{pmatrix}
\end{split}
\label{eq:first_rank4c}
\end{equation}
Where the recursion relationship in Eq. \eqref{eq:general_linear} has been substituted for the first two constituent Wigner-3j symbols, and the sum over $M_1,M_2$ is simplified by replacing $M_1$ and $M_2$ with $(m_1 + m_2)$ and $(m_3+m_4)$, respectively. Other projection quantum numbers would result in zero-valued summands. Here, each traditional Wigner-3j symbol that couples $l_i$ has a factor $f_{12}$ and $f_{34}$ for increments of $L_1$ and $L_2$.
The final factor in Eq. \eqref{eq:first_rank4c} will also obey ladder relationships with its principal quantum numbers. Applying the recursion relationship for the index of $\mathcal{L}_1$ and again for the index of $L_2$ gives:
\begin{equation}
\begin{split}
&\begin{pmatrix}
     \mathcal{L}_1 + k_1 & \mathcal{L}_2 + k_2 & L_R \\
     (m_1 + m_2) & (m_3 + m_4) & -M_R
\end{pmatrix}=\\
&f_{12}'(l_1,l_2,m_1,m_2,k_1,k_2)f_{34}'(l_3,l_4,m_3,m_4,k_2)\begin{pmatrix}
     \mathcal{L}_1 & \mathcal{L}_2 & L_R \\
     (m_1 + m_2) & (m_3 + m_4) & -M_R
\end{pmatrix}
\end{split},
\label{eq:first_rank4e}
\end{equation}
where the factors $f_{12}'$ and $f_{34}'$ depend on the order in which the relationships are applied. For simplicity, these will be applied in order, $L_1$ first then $L_2$ second in Eq. \eqref{eq:first_rank4e}, but this choice is arbitrary. Finally, a linear relationship is substituted applied with the third Wigner symbol. Combining the recursion relationships from Eq. \eqref{eq:general_linear} for all three traditional Wigner-3j symbols that comprise the generalized symbol gives the following.
\begin{equation}
\begin{split}
&F(\boldsymbol{l},\boldsymbol{m},k_1,k_2)F'(\boldsymbol{l},\boldsymbol{m},k_1,k_2) \begin{pmatrix}
     l_1 & l_2 & l_3 & l_4\\
     m_1 & m_2 & m_3 & m_4
\end{pmatrix}\big( (\mathcal{L}_1 , \mathcal{L}_2, L_R), \boldsymbol{M} \big) = \begin{pmatrix}
     l_1 & l_2 & l_3 & l_4\\
     m_1 & m_2 & m_3 & m_4
\end{pmatrix}\big((\mathcal{L}_1 +k_1 , \mathcal{L}_2+k_2 , L_R), \boldsymbol{M}  \big) \\
&\begin{pmatrix}
     l_1 & l_2 & l_3 & l_4\\
     m_1 & m_2 & m_3 & m_4
\end{pmatrix}\big((\mathcal{L}_1 , \mathcal{L}_2, L_R), \boldsymbol{M} \big) =\mathcal{A}(\boldsymbol{l},\boldsymbol{m},k_1,k_2)\begin{pmatrix}
     l_1 & l_2 & l_3 & l_4\\
     m_1 & m_2 & m_3 & m_4
\end{pmatrix}\big((\mathcal{L}_1 +k_1 , \mathcal{L}_2+k_2 , L_R), \boldsymbol{M} \big) 
\end{split}
\label{eq:first_rank4_final}
\end{equation}
where the substitutions of $F(\boldsymbol{l},\boldsymbol{m},k_1,k_2)=f_{12}(l_1,l_2,m_1,m_2,k_1)f_{34}(l_3,l_4,m_3,m_4,k_2)$ and $F'(\boldsymbol{l},\boldsymbol{m},k_1,k_2)=f_{12}'(l_1,l_2,m_1,m_2,k_1,k_2)f_{34}'(l_3,l_4,m_3,m_4,k_2)$ have been made.
Despite the complicated form of the prefactor, $\mathcal{A}(\boldsymbol{l},\boldsymbol{m},k_1,k_2)$, Eq. \eqref{eq:first_rank4_final} gives the linear relationship between rank 4 Wigner symbols with incremented intermediates. All terms in $\mathcal{A}_{+,+}$ may be written in terms of $l_1,l_2,l_3,l_4$ and $m_1,m_2,m_3,m_4$. This is derived here for the pairwise coupling scheme, but it is easily extended to others.

The recursion relationships for generalized Wigner symbols with arbitrary intermediates may be derived by applying the analogous relationships for each constituent traditional Wigner-3j symbol, as was done to obtain Eq. \eqref{eq:first_rank4_final}. In general one may write:
\begin{equation}
\begin{split}
&\begin{pmatrix}
     l_1 & l_2 & l_3 & l_4\\
     m_1 & m_2 & m_3 & m_4
\end{pmatrix}\big(\boldsymbol{L}, \boldsymbol{M} \big) = \\
&A_{-,-}(\boldsymbol{l},\boldsymbol{m}) \begin{pmatrix}
     l_1 & l_2 & l_3 & l_4\\
     m_1 & m_2 & m_3 & m_4
\end{pmatrix}\big( (L_1 -1, L_2 -1, L_R) ,\boldsymbol{M} \big) + \\
&A_{-,+}(\boldsymbol{l},\boldsymbol{m})\begin{pmatrix}
     l_1 & l_2 & l_3 & l_4\\
     m_1 & m_2 & m_3 & m_4
\end{pmatrix}\big((L_1 -1 , L_2 +1 , L_R) , \boldsymbol{M} \big) +\\ 
& A_{+,-}(\boldsymbol{l},\boldsymbol{m})\begin{pmatrix}
     l_1 & l_2 & l_3 & l_4\\
     m_1 & m_2 & m_3 & m_4
\end{pmatrix}\big((L_1 +1, L_2 -1, L_R), \boldsymbol{M} \big) +\\ 
& A_{+,+}(\boldsymbol{l},\boldsymbol{m})\begin{pmatrix}
     l_1 & l_2 & l_3 & l_4\\
     m_1 & m_2 & m_3 & m_4
\end{pmatrix}\big((L_1 +1, L_2 +1, L_R), \boldsymbol{M} \big) 
\end{split}
\end{equation}
which, for $L_R=0$ reduces to:
\begin{equation}
\begin{split}
&\begin{pmatrix}
     l_1 & l_2 & l_3 & l_4\\
     m_1 & m_2 & m_3 & m_4
\end{pmatrix}\big( \boldsymbol{L}, \boldsymbol{M} \big) = \\
& A_{-,-}(\boldsymbol{l},\boldsymbol{m})\begin{pmatrix}
     l_1 & l_2 & l_3 & l_4\\
     m_1 & m_2 & m_3 & m_4
\end{pmatrix}\big((L_1 -1, L_2 -1, 0), \boldsymbol{M} \big)
+ A_{+,+}(\boldsymbol{l},\boldsymbol{m})\begin{pmatrix}
     l_1 & l_2 & l_3 & l_4\\
     m_1 & m_2 & m_3 & m_4
\end{pmatrix}\big((L_1 +1, L_2 +1,0) , \boldsymbol{M} \big) 
\end{split}
\label{eq:generalized_rank4_recur}
\end{equation}
In Eq. \eqref{eq:generalized_rank4_recur} the $A_{\pm,\pm}(\boldsymbol{l},\boldsymbol{m})$ factors, similar to those defined in terms of $\mathcal{L}_1,\mathcal{L}_2$ are products of factors from recursion relationships of the Wigner-3j symbols.

In addition to recursion relationships for generalized Wigner symbols with incremented intermediate angular momentum quantum numbers, one can also derive raising and lowering relationships for generalized Wigner symbols with incremented intermediate projection quantum numbers. The relationships between different projection quantum numbers will also be demonstrated for rank 4, and are easily generalized. In these recursion relationships all angular momentum quantum numbers will be fixed, including the intermediates. For some intermediate projections $M_1$ and $M_2$, one may raise and/or lower the projection quantum numbers according to,
\begin{equation}
\begin{split}
&\begin{pmatrix}
     l_1 & l_2 & l_3 & l_4\\
     m_1 & m_2 & m_3 & m_4
\end{pmatrix}\big(\boldsymbol{L}, (M_1 , M_2 , M_R) \big) =\\ 
& \phi(\boldsymbol{l},\boldsymbol{m}) \bigg[ G_1(l_1,L_1,m_1,M_1)
     \begin{pmatrix}
     l_1 & l_2 & L_1\\
     m_1 \mp 1 & m_2 & M_1 \mp 1
     \end{pmatrix} +
     H_1(l_2,L_1,m_2,M_1)\begin{pmatrix}
     l_1 & l_2 & L_1\\
     m_1 & m_2 \mp 1 & M_1 \mp 1
     \end{pmatrix} \bigg ] \\
&\bigg[ G_2(l_3,L_2,m_3,M_2)
     \begin{pmatrix}
     l_3 & l_4 & L_2\\
     m_3 \pm 1 & m_4 & M_2 \pm 1
     \end{pmatrix} +
     H_2(l_4,L_2,m_4,M_2)\begin{pmatrix}
     l_3 & l_4 & L_1\\
     m_3 & m_4 \pm 1 & M_2 \pm 1
     \end{pmatrix} \bigg ] \\
&\bigg[ P_{\pm}
     \begin{pmatrix}
     L_1 & L_2 & L_R\\
     M_1 \mp 1 & M_2 \pm 1 & M_R
     \end{pmatrix}  \bigg ],
\end{split}
\label{eq:generalized_raise_lower}
\end{equation}
where the phase $\phi(\boldsymbol{l},\boldsymbol{m}_{\pm})$ depends on projections \textit{have} been incremented, defined by $\boldsymbol{m}_\pm$.
Notice that in Eq. \eqref{eq:generalized_raise_lower}, the $M_k$ are incremented such that the final projection quantum number $M_R$ is conserved. The third bracketed term in this equation only has one value rather than the two from Eq. \eqref{eq:m3_cg_recur}; one of the terms will always be zero due to the valid domain of $m_i$ and $M_k$. One particularly convenient result of this relationship is that a generalized coupling coefficient with a multiset of intermediate projections $\boldsymbol{M}_{\pm}=(M_1,M_2,M_R)$ may be written in terms of a sum of generalized coupling coefficients with a different multiset of intermediate projections, $\boldsymbol{M}_{\mp}$. Explicitly after distributing in Eq. \eqref{eq:generalized_raise_lower}, one obtains: 
\begin{equation}
\begin{split}
& \begin{pmatrix}
     l_1 & l_2 & l_3 & l_4\\
     m_1 & m_2 & m_3 & m_4
\end{pmatrix}\big(\boldsymbol{L} , (M_1  , M_2, M_R) \big) =\\ 
& \phi(\boldsymbol{l},\boldsymbol{m}) \bigg[ G_1(l_1,L_1,m_1,M_1)G_2(l_3,L_2,m_3,M_2)P_{+}
\begin{pmatrix}
     l_1 & l_2 & l_3 & l_4\\
     m_1 \mp 1 & m_2 & m_3 \pm 1 & m_4 
\end{pmatrix}\big(\boldsymbol{L}, (M_1 \mp 1, M_2 \pm 1, M_R) \big) \\
& + G_1(l_1,L_1,m_1,M_1)H_2(l_4,L_2,m_4,M_2)P_{+} \begin{pmatrix}
     l_1 & l_2 & l_3 & l_4\\
     m_1 \mp 1 & m_2 & m_3 & m_4 \pm 1
\end{pmatrix}\big(\boldsymbol{L}, (M_1 \mp 1 , M_2 \pm 1, M_R \big)  \\
& + H_1(l_2,L_1,m_2,M_1)G_2(l_3,L_2,m_3,M_2)P_{-} \begin{pmatrix}
     l_1 & l_2 & l_3 & l_4\\
     m_1  & m_2  \mp 1 & m_3 \pm 1 & m_4
\end{pmatrix}\big(\boldsymbol{L}, (M_1 \mp 1, M_2 \pm 1, M_R) \big) \\ & +H_1(l_2,L_1,m_2,M_1)H_2(l_4,L_2,m_4,M_2)P_{-}\begin{pmatrix}
     l_1 & l_2 & l_3 & l_4\\
     m_1 & m_2 \mp 1 & m_3 & m_4 \pm 1
\end{pmatrix}\big(\boldsymbol{L},(M_1 \mp 1 , M_2 \pm 1, M_R) \big) \bigg],
\end{split}
\label{eq:generalized_raise_lower2}
\end{equation}
which again holds true for fixed $\boldsymbol{lL}$ in the pairwise coupling scheme. This recursion relationship for generalized Wigner symbols in Eq. \eqref{eq:generalized_raise_lower2} can be simplified in an algebraic form. In general, one may write the expression from Eq. \eqref{eq:generalized_raise_lower2} in terms of a sum over intermediate projection quantum numbers.
\begin{equation}
\begin{split}
&W_{\boldsymbol{l}}^{\boldsymbol{m}_{\pm}}
\big( \boldsymbol{L}, \boldsymbol{M}_{\pm} \big) =\sum_{\boldsymbol{m}_{\mp}} c_{\boldsymbol{m}_{\mp}} W_{\boldsymbol{l}}^{\boldsymbol{m}_{\mp}}
\big(\boldsymbol{L}, \boldsymbol{M}_{\mp} \big)
\end{split}
\label{eq:generalized_raise_lower3}
\end{equation}
Here, $\boldsymbol{M}_{\pm}$ is an incremented multiset of intermediate projections, the same as that in the left hand side of Eq. \eqref{eq:generalized_raise_lower2}, and the $\boldsymbol{m}_{\pm}$ are the corresponding projection quantum numbers that define those intermediates (e.g. $M_{1} = m_1 + m_2)$). The Wigner symbol with $\boldsymbol{m}_{\pm},\boldsymbol{M}_{\pm}$ are related to Wigner symbols with the opposing increments, $\boldsymbol{m}_{\mp},\boldsymbol{M}_{\mp}$, as seen in the right hand side of Eq. \eqref{eq:generalized_raise_lower2}. The $c_{\boldsymbol{m}_{\mp}}$ are the coefficients containing products of $H_k$, $G_k$, and $P_{\mp}$ from Eqs. \eqref{eq:m_wig_recur} and \eqref{eq:m3_cg_recur}, respectively. 

\subsection{Recursion Relationships for Generalized Wigner Symbols with Arbitrary Rank}
These recursion relationships can also be used to show linear relationships between generalized Wigner symbols of higher rank. In the pairwise coupling scheme, the linear relationships between different intermediates may be given for Wigner symbols of arbitrary rank. 
\begin{equation}
\begin{split}
&\begin{pmatrix}
     l_1 & l_2 & l_3 & \cdots & l_N\\
     m_1 & m_2 & m_3 & \cdots & m_N
\end{pmatrix}\big((\mathcal{L}_1 + k_1 , \mathcal{L}_2 + k_2 , \cdots  \mathcal{L}_{(N-2)} +k_{(N-2)}),\boldsymbol{M}\big)=\\ 
& \phi(\boldsymbol{l},\boldsymbol{m},\boldsymbol{k}) f_{12}(l_1,l_2,m_1,m_2,k_1)
\begin{pmatrix}
     l_1 & l_2 & \mathcal{L}_1 \\
     m_1 & m_2 & -(m_1 + m_2)
\end{pmatrix}
f_{34}(l_3,l_4,m_3,m_4,k_2)\begin{pmatrix}
     l_3 & l_4 & \mathcal{L}_2 \\
     m_3 & m_4 & -(m_3 + m_4)
\end{pmatrix} \\
& \cdots f_{(N-1,N)}(l_{(N-1)}, l_N, m_{(N-1)}, m_N)
\begin{pmatrix}
     l_{(N-1)} & l_N & \mathcal{L}_{N/2} \\
     m_{(N-1)} & m_N & -(m_{(N-1)} + m_N)
\end{pmatrix} \cdots \\
&\begin{pmatrix}
     \mathcal{L}_1 + k_1  & \mathcal{L}_2 + k_2 &  \mathcal{L}_{(N/2) +1 } + k_{(N/2) +1 } \\
     (m_1 + m_2) & (m_3 + m_4) &  -(m_1+m_2+m_3+m_4)
\end{pmatrix} \cdots \begin{pmatrix}
     \mathcal{L}_{(N -3)} + k_{(N -3)} & \mathcal{L}_{(N -2)} + k_{(N -2)} & L_R \\
     \sum_i^{m_{div}}m_i & \sum_{i > m_{div}}^{N} m_i & - \sum_i m_i
\end{pmatrix} 
\end{split}
\label{eq:general_rank_b}
\end{equation}
where $m_{div}=p{[floor(N/p)]}$ defines a left/right split in the binary tree given the closest perfect binary tree with $p$ leaves, rounded down. If the recursion relationships for intermediates are applied as in Eq. \eqref{eq:first_rank4e} and generalized we obtain, 
\begin{equation}
\begin{split}
&\begin{pmatrix}
     l_1 & l_2 & l_3 & \cdots & l_N\\
     m_1 & m_2 & m_3 & \cdots & m_N
\end{pmatrix}\big(\boldsymbol{\mathcal{L}} + \boldsymbol{k}, \boldsymbol{M} \big)=\\ 
& \phi(\boldsymbol{l},\boldsymbol{m},\boldsymbol{k}) F(\boldsymbol{l},\boldsymbol{m},\boldsymbol{k}) 
\begin{pmatrix}
     l_1 & l_2 & \mathcal{L}_1 \\
     m_1 & m_2 & -(m_1 + m_2)
\end{pmatrix}
\begin{pmatrix}
     l_3 & l_4 & \mathcal{L}_2 \\
     m_3 & m_4 & -(m_3 + m_4)
\end{pmatrix} \cdots 
\begin{pmatrix}
     l_{(N-1)} & l_N & \mathcal{L}_{N/2} \\
     m_{(N-1)} & m_N & -(m_{(N-1)} + m_N)
\end{pmatrix} \\
&\cdots f_{12}'(l_1,l_2,m_1,m_2,k_1) f_{34}'(l_3,l_4,m_3,m_4,k_2) f'_{(L_1,L_2)} (\mathcal{L}_1,\mathcal{L}_2, M_1,M_2, k_1,k_2) \begin{pmatrix}
     \mathcal{L}_1  & \mathcal{L}_2 &  \mathcal{L}_{(N/2) +1 }  \\
     (m_1 + m_2) & (m_3 + m_4) &  -(M_1+M_2)
\end{pmatrix} \\
&\cdots f'_{L_{(N-5)}L_{(N-4)}}(\mathcal{L}_{(N-5)},\mathcal{L}_{(N-4)},M_{(N-5)},M_{(N-4)},k_{(N-5)},k_{(N-4)}) \\
& f'_{L_{(N-3)},L_{(N-2)}}(\mathcal{L}_{(N-3)},\mathcal{L}_{(N-2)},M_{(N-3)},M_{(N-2)},k_{(N-2)})
\begin{pmatrix}
     \mathcal{L}_{(N -3)}  & \mathcal{L}_{(N -2)}  & L_R \\
     \sum_i^{m_{div}}m_i & \sum_{i > m_{div}}^{N} m_i & - \sum_i m_i
\end{pmatrix}.
\end{split}
\label{eq:general_rank_c}
\end{equation}
The expression in Eq. \eqref{eq:general_rank_c} is obtained after making substitutions for certain products of factors as done in Eq. \eqref{eq:first_rank4_final}, $F(\boldsymbol{l},\boldsymbol{m}, \boldsymbol{k})=f_{12}(l_1,l_2,m_1,m_2,k_1)f_{23}(l_3,l_4,m_3,m_4,k_2) \cdots f_{(N-1,N)}(l_{(N-1)}, l_N, m_{(N-1)}, m_N)$. The $F(\boldsymbol{l},\boldsymbol{m}, \boldsymbol{k})$ is the product of factors associated with raising/lowering the intermediates resulting from coupling the angular function indices, but not other intermediates. In place of incremented traditional Wigner-3j symbols in Eq. \eqref{eq:general_rank_b}, factors and the Wigner-3j symbols for intermediates that have not been incremented have been substituted into Eq. \eqref{eq:general_rank_c}.
The final result is a relationship,
\begin{equation}
\begin{split}
&\begin{pmatrix}
     l_1 & l_2 & l_3 & \cdots & l_N\\
     m_1 & m_2 & m_3 & \cdots & m_N
\end{pmatrix}\big(\boldsymbol{\mathcal{L}} + \boldsymbol{k}, \boldsymbol{M} \big)=\\ 
& F(\boldsymbol{l},\boldsymbol{m}, \boldsymbol{k})F'(\boldsymbol{l},\boldsymbol{m}, \boldsymbol{k})\begin{pmatrix}
     l_1 & l_2 & l_3 & \cdots & l_N\\
     m_1 & m_2 & m_3 & \cdots & m_N
\end{pmatrix}\big(\boldsymbol{\mathcal{L}}, \boldsymbol{M}  \big) \\
& = \mathcal{A}(\boldsymbol{l},\boldsymbol{m},\boldsymbol{k}) \begin{pmatrix}
     l_1 & l_2 & l_3 & \cdots & l_N\\
     m_1 & m_2 & m_3 & \cdots & m_N
\end{pmatrix}\big(\boldsymbol{\mathcal{L}}, \boldsymbol{M} \big) 
\end{split},
\label{eq:general_rank_d}
\end{equation}
where $\mathcal{A}(\boldsymbol{l},\boldsymbol{m},\boldsymbol{k})$ is a factor that may be written entirely in terms of $\boldsymbol{l}=\{l_1,l_2,\cdots l_N \}$, $\boldsymbol{m}=\{m_1,m_2,\cdots m_N \}$, and increments $\boldsymbol{k}=\{k_1,k_2,\cdots k_{(N-2)} \}$. Hence, for generalized Wigner symbols of rank $N$, relationships between symbols with incremented intermediates from the lowest possible values may be defined.

For raising/lowering projection quantum numbers in generalized symbols of arbitrary rank, Eq. \eqref{eq:generalized_raise_lower3} is of the correct form. A Wigner symbol with a multiset of $N-2$ incremented intermediate projections $\boldsymbol{M}_{\pm}= \{ M_1, M_2, \cdots M_{N/2} \}$ may be used in place of the rank 4 case. It is important to note that the $N-2$ intermediate projections are usually incremented up or down such that $M_R$ is held constant. The right hand side will contain more related Wigner symbols with the opposing increments $\boldsymbol{m}_{\mp}$. These and the coefficients $c_{\boldsymbol{m}_{\mp}}$ are obtained by applying Eq. \eqref{eq:full_m_wig_recur} to each constituent Wigner-3j symbol just as in Eqs. \eqref{eq:generalized_raise_lower} and \eqref{eq:generalized_raise_lower2}.

\subsection{Proof: Linearly Dependent RPI Functions}

Using both the permutation properties of the generalized Wigner symbols and their recursion relationships, we may prove linear dependence for many sets of functions when there are duplicate values of $n_i$, $l_i$, and/or $L_k$. We show the process for doing this for some cases of rank 4 and 5, but it may be extended to any other case as well. First, we consider the case of $\boldsymbol{l}=(1111)$, $L_R=0$ for the case where all non-angular indices are duplicates $\boldsymbol{n}=(nnnn)$. 
\begin{equation}
    B_{(nnnn)(1111)(00)} = \sum_{m_1,m_2,m_3,m_4} W_{1,1,1,1}^{m_1,m_2,m_3,m_4} ((0 0 0 ),(0 0 0)) A_{n,1,m_1}A_{n,1,m_2}A_{n,1,m_3} A_{n,1,m_4}
    \label{eq:rank_4_l1_ex}
\end{equation}
For the sake of providing connections to practical cases, we use the products of the atomic base in Eq. \eqref{eq:rank_4_l1_ex}, but we could also use symmetrized cluster functions from Eq. \eqref{eq:pi_product_cluster}. In either case, the $\boldsymbol{m}$ multisets to be summed over are:
\begin{equation}
\begin{split}
&\boldsymbol{m}=(-11-11) \; , \;
\boldsymbol{m}=(-1100)\; , \;
\boldsymbol{m}=(-111-1)\; , \;\\
&\boldsymbol{m}=(00-11)\; , \;
\boldsymbol{m}=(0000)\; , \;
\boldsymbol{m}=(001-1)\; , \;\\
&\boldsymbol{m}=(1-1-11)\; , \;
\boldsymbol{m}=(1-100)\; , \;
\boldsymbol{m}=(1-11-1).
\end{split}
\label{eq:example_m_vectors}
\end{equation}
The product of atomic base functions, $\boldsymbol{A}_{\boldsymbol{nlm}}=A_{n,1,m_1}A_{n,1,m_2}A_{n,1,m_3} A_{n,1,m_4}$ with the corresponding values of $m_i$ are to be multiplied by the appropriate generalized Wigner symbol. It will be convenient to write sums over certain multisets of $\boldsymbol{m}$ , such as those in Eq. \eqref{eq:example_m_vectors}, as sums over $(|M_1|,|M_2|)$. Rewriting Eq.\eqref{eq:rank_4_l1_ex} in these terms gives, 
\begin{equation}
\begin{split}
&B_{(nnnn)(1111)(00)} = \sum_{|M_1|=|M_2|=0}
W_{(1111)}^{\boldsymbol{m}}((000),(000))\boldsymbol{A}_{(nnnn)(1111)\boldsymbol{m}}\; 
\end{split}
    \label{eq:rank_4_l1_ex_a}
\end{equation}
where the values of $M_1=(m_1+m_2)$ and $M_2=(m_3+m_4)$ are implied by the values of $m_1,m_2,m_3,m_4$.
One important note is that each $\boldsymbol{A}_{\boldsymbol{nlm}}$ in the sum is symmetric with respect to permutations of the $A_{n,1,m_i}$. Due to this permutation invariance, Eq. \eqref{eq:rank_4_l1_ex_a} may also be written as:
\begin{equation}
\begin{split}
&B_{(nnnn)(1111)(00)} = \sum_{|M_1|=|M_2|=0\leftarrow \sigma_\circ({\boldsymbol{m}})}
W_{(1111)}^{\boldsymbol{m}}((000),(000)) \boldsymbol{A}_{(nnnn)(1111)\sigma({\boldsymbol{m}})}\; 
\end{split}
    \label{eq:rank_4_l1_ex_b}
\end{equation}
where $\sigma({\boldsymbol{m}})$ is an arbitrary permutation of $\boldsymbol{m}$, which we arbitrarily take to the permutation that preserves the values of $M_1,M_2$ and has strictly ordered children for $M_1$ and $M_2$. For example, if $\boldsymbol{m}=(1-1-11)$, then $\sigma_{\circ}(\boldsymbol{m})=(-11-11)$, which is the same as ordering the $m_i$ in each orbit of the coupling partition, $P_c$. We indicate that these are still the $\boldsymbol{m}$ that yield the $(|M_1|=|M_2|=0)$ using an arrow. For similar reasons, the general evaluation of the ACE descriptors only require that one evaluates $\boldsymbol{A}_{\boldsymbol{nlm}}$ where $\boldsymbol{nlm}$ are lexicographically ordered. We make use of this simplification of this simplified summation in the proof.

Next, we make use of both the permutation symmetries for the generalized Wigner symbols as well as the relationships for incremented projection quantum numbers to relate all $W_{\boldsymbol{l}=1}^{\boldsymbol{m}}$ to $W_{\boldsymbol{l}=1}^{\boldsymbol{0000}}$. The recursion relationships presented in Eq. \eqref{eq:generalized_raise_lower3} allow one to show that Eq. \eqref{eq:rank_4_l1_ex} can be written as,
\begin{equation}
\begin{split}
    &B_{(nnnn)(1111)(00)} = \\
& 4 W_{1111}^{0000}((000),(000)) \, \cdot \, \boldsymbol{A}_{(nnnn)(1111)(-11-11)}\; + \\
& -4 W_{1111}^{0000}((000),(000)) \, \cdot \, \boldsymbol{A}_{(nnnn)(1111)(-1100)}\; +\\
& W_{1111}^{0000}((000),(000)) \, \cdot \, \boldsymbol{A}_{(nnnn)(1111)(0000)}\;.
\end{split}
    \label{eq:rank_4_l1_ex_m0}
\end{equation}
From Eq. \eqref{eq:rank_4_l1_ex_m0}, it may be seen that the sum over 9 terms is replaced by a sum over 3 terms, potentially with different prefactors. The recursion relationships could also be used to relate all of the generalized Wigner symbols in the sum to some other single Wigner symbol; $W^{1111}_{0000}(L_1=L_2=0)$ is chosen for simplicity. It is noted here that expressions such as this in Eq. \eqref{eq:rank_4_l1_ex_m0} may allow for more efficient computation of descriptors since fewer atomic base products need to be evaluated.

To continue the proof, we will repeat this for other values of intermediates. For the current example, the only other multisets of intermediates to consider are $\boldsymbol{L}=(11) \; and \; (22)$. It is straightforward to show that with this degenaracy of non-angular and angular indices, $P_f(\boldsymbol{l})=4$ and $P_f(\boldsymbol{n})=4$ , that the odd-parity intermediates produce zero-valued functions, and one only needs to consider $\boldsymbol{L}=(22)$ for non-trivial relationships. Beginning with an analogue of Eq. \eqref{eq:rank_4_l1_ex_b}, the descriptor corresponding to this multiset of intermediates is
\begin{equation}
\begin{split}
&B_{(nnnn)(1111)(22)} = \\
&\sum_{|M_1|=|M_2|=0\leftarrow \sigma({\boldsymbol{m}})}
W_{1111}^{\boldsymbol{m}}((220),\boldsymbol{M})\boldsymbol{A}_{(nnnn)(1111)\sigma({\boldsymbol{m}})}\; \\
& +\sum_{|M_1|=|M_2|=1\leftarrow \sigma({\boldsymbol{m}})}
W_{1111}^{\boldsymbol{m}}((220),\boldsymbol{M}) \boldsymbol{A}_{(nnnn)(1111)\sigma({\boldsymbol{m}})}\;  \\
&+ \sum_{|M_1|=|M_2|=2\leftarrow \sigma({\boldsymbol{m}})}
W_{1111}^{\boldsymbol{m}}((220,\boldsymbol{M})) \boldsymbol{A}_{(nnnn)(1111)\sigma({\boldsymbol{m}})}.
\end{split}
    \label{eq:rank_4_l1_ex_22}
\end{equation}
Key differences here are that the sum one would see in the typical expression from Eq. \eqref{eq:rank_4_l1_ex}, has now been split into 3 sums over different values of $M_k$. Again, we take advantage of the permutation symmetries for the generalized Wigner symbols as well as the recursion relationships for the $m_i$ to simplify Eq. \eqref{eq:rank_4_l1_ex_22}. If one uses the recursion relationships as well as equivalent permutations of generalized Wigner symbols, all terms in Eq. \eqref{eq:rank_4_l1_ex_22} may be written in terms of the same generalized Wigner symbol, $W^{1111}_{0000}(22)$. Additionally, taking advantage of expressing the indices permutation invariant basis, a symmetrized version of Eq. \eqref{eq:direct_prod} or the atomic base in Eq. \eqref{eq:atomic_base}, with any permutation of $\boldsymbol{nlm}$ that preserves the absolute values: $|M_1|$ and $|M_2|$, we obtain:
\begin{equation}
\begin{split}
&B_{(nnnn)(1111)(22)} = \\
& W_{1111}^{0000}((220),\boldsymbol{M}) \, \cdot \, \boldsymbol{A}_{(nnnn)(1111)(-11-11)}\; + \\
&2 W_{1111}^{0000}((220),\boldsymbol{M}) \, \cdot \, \boldsymbol{A}_{(nnnn)(1111)(-1100)}\; +\\
&W_{1111}^{0000}((220),\boldsymbol{M}) \, \cdot \, \boldsymbol{A}_{(nnnn)(1111)(0000)}\; \\
& -\\
&\frac{2}{3} \big[ ( \frac{9}{4} W_{1111}^{0000}((220),\boldsymbol{M})) \, \cdot \, \boldsymbol{A}_{(nnnn),(1111)(-1100)}\; + \\
& ( \frac{9}{4} W_{1111}^{0000}((220),\boldsymbol{M})) \, \cdot \, \boldsymbol{A}_{(nnnn)(1111)(-1100)}\; + \\
& ( \frac{9}{4} W_{1111}^{0000}((220),\boldsymbol{M})) \, \cdot \, \boldsymbol{A}_{(nnnn)(1111)(-1100)}\; + \\
& ( \frac{9}{4} W_{1111}^{0000}((220),\boldsymbol{M})) \, \cdot \, \boldsymbol{A}_{(nnnn)(1111)(-1100)}\; \big] \\
& + \\
& \frac{1}{3} \big[( \frac{9}{4} W_{1111}^{0000}((220),\boldsymbol{M})) \, \cdot \, \boldsymbol{A}_{(nnnn)(1111)(-11-11)}\; +\\
& ( \frac{9}{4} W_{1111}^{0000}((220),\boldsymbol{M}))\, \cdot \, \boldsymbol{A}_{(nnnn)(1111)(-11-11)}\; + \\
& ( \frac{9}{4} W_{1111}^{0000}((220),\boldsymbol{M})) \, \cdot \, \boldsymbol{A}_{(nnnn)(1111)(-11-11)}\; + \\
& ( \frac{9}{4} W_{1111}^{0000}((220),\boldsymbol{M})) \, \cdot \, \boldsymbol{A}_{(nnnn)(1111)(-11-11)}\;  \big].
\end{split}
    \label{eq:rank_4_l1_ex_M2_e}
\end{equation}
To obtain the expression in Eq. \eqref{eq:rank_4_l1_ex_M2_e},  Note that in Eq. \eqref{eq:rank_4_l1_ex_M2_e}, the bracketed terms correspond to different values of $|M_1|=|M_2|$ from Eq. \eqref{eq:rank_4_l1_ex_22}. This may now be simplified even further.
\begin{equation}
\begin{split}
&B_{(nnnn)(1111)(22)} = \\
& 4 W_{1111}^{0000}((220),\boldsymbol{M}) \, \cdot \, \boldsymbol{A}_{(nnnn)(1111)(-11-11)}\; + \\
& -4 W_{1111}^{0000}((220),\boldsymbol{M}) \, \cdot \, \boldsymbol{A}_{(nnnn)(1111)(-1100)}\; +\\
& W_{1111}^{0000}((220),\boldsymbol{M}) \, \cdot \, \boldsymbol{A}_{(nnnn)(1111)(0000)}\;.
\end{split}
    \label{eq:rank_4_l1_ex_final_aa}
\end{equation}
This simplified form in Eq. \eqref{eq:rank_4_l1_ex_final_aa} follows from many applications of recursion relationships on projection quantum numbers and the permutation symmetries of the generalized Wigner symbols. From Eq. \eqref{eq:rank_4_l1_ex_M2_e} it can be seen that this simplification is heavily reliant on the fact that the atomic base or the cluster basis is permutation invariant, because the $A$ basis indices indices must be reordered. Even though this simplified expression in Eq. \eqref{eq:rank_4_l1_ex_final_aa} is quite similar to the simplified form of the $B_{(nnnn)(1111)(00)}$ case, but not exactly equivalent. Key differences are in the intermediate angular momenta. Even though we have described all generalized Wigner symbols from Eq. \eqref{eq:rank_4_l1_ex_22}, in terms of the generalized Wigner symbol with $\boldsymbol{m}=(0000)$, they are still evaluated for the different intermediate angular momentum quantum numbers $\boldsymbol{L}=(22)$ vs. $\boldsymbol{L}=(00)$. At this stage, one may apply the recursion relationships on the angular momentum quantum numbers as in Eq. \eqref{eq:first_rank4c}. 
\begin{equation}
\begin{split}
&B_{(nnnn)(1111)(22)} = \\
& 4 \frac{2}{\sqrt{5}} W_{1111}^{0000}((000),(000)) \, \cdot \, \boldsymbol{A}_{(nnnn)(1111)(-11-11)}\; + \\
& -4 \frac{2}{\sqrt{5}} W_{1111}^{0000}((000),(000)) \, \cdot \, \boldsymbol{A}_{(nnnn)(1111)(-1100)}\; +\\
& \frac{2}{\sqrt{5}} W_{1111}^{0000}((000),(000)) \, \cdot \, \boldsymbol{A}_{(nnnn)(1111)(0000)}\; .
\end{split}
    \label{eq:rank_4_l1_ex_final_b}
\end{equation}
All terms in Eq. \eqref{eq:rank_4_l1_ex_final_b} are now described in the same ones from Eq. \eqref{eq:rank_4_l1_ex_m0}.

Finally, we may equate both Eq. \eqref{eq:rank_4_l1_ex_final_b} and \eqref{eq:rank_4_l1_ex_m0} yields the proof that their is a linear dependence between these two descriptors with different intermediates, Eq. \eqref{eq:final_rank4_explicit_proof},
\begin{equation}
    B_{(nnnn)(1111)(22)} = \frac{2}{5\sqrt{5}} B_{(nnnn)(1111)(00)},
    \label{eq:final_rank4_explicit_proof}
\end{equation}
which again is for the case of $\boldsymbol{l}=(1111)$ and all $n_i$ equal, $\boldsymbol{n}=(nnnn)$. Though this is a specific case, the application of these relationships may be generalized. Permutation symmetries of the generalized Wigner symbols along with raising/lowering operators may be defined for arbitrary rank and for arbitrary $\boldsymbol{nl}$ degeneracy. This result allows us to prove the first entry in in Table 3 of Ref~[\citenum{dusson_atomic_2022}].

This method of using both the permutation symmetries and the recursion relationships of the generalized Wigner symbols to describe linear dependencies between descriptors is not limited to the case from Eq. \eqref{eq:final_rank4_explicit_proof}. Other examples will be provided, including those for incremented angular indices. The first of which will be $\boldsymbol{l}=(2222)$, all $n_i$ equivalent. Following similar procedures as for the case of $\boldsymbol{l}=(1111)$, one may show that $B_{(nnnn)(2222)(00)}$ may be written as,
\begin{equation}
\begin{split}
&B_{(nnnn)(2222)(00)} = \\
& 4 W_{2222}^{0000}((000),\boldsymbol{M}) \, \cdot \, \boldsymbol{A}_{(nnnn)(2222)(-22-22)}\; + \\
& -8 W_{2222}^{0000}((000),\boldsymbol{M}) \, \cdot \, \boldsymbol{A}_{(nnnn)(2222)(-22-11)}\; + \\
& 4 W_{2222}^{0000}((000),\boldsymbol{M}) \, \cdot \, \boldsymbol{A}_{(nnnn)(2222)(-2200)}\; + \\
& 4 W_{2222}^{0000}((000),\boldsymbol{M}) \, \cdot \, \boldsymbol{A}_{(nnnn)(2222)(-11-11)}\; + \\
& -4 W_{2222}^{0000}((000),\boldsymbol{M}) \, \cdot \, \boldsymbol{A}_{(nnnn)(2222)(-1100)}\; + \\
& W_{2222}^{0000}((000),\boldsymbol{M}) \, \cdot \, \boldsymbol{A}_{(nnnn)(2222)(0000)},
\end{split}
    \label{eq:rank_4_l2_ex}
\end{equation}
where we have already written all $W_{2222}^{\boldsymbol{m}}((000),\boldsymbol{M})$ in terms of $W_{2222}^{0000}((000),\boldsymbol{M})$. Next, we may consider the other descriptors, characterized by different intermediates, $\boldsymbol{L}=(2,2)$ and $\boldsymbol{L}=(4,4)$. What we ultimately obtain for $\boldsymbol{L}=(2,2)$ after applying the treatment above is:
\begin{equation}
    B_{(nnnn)(2222)(22)}  = \frac{2}{7\sqrt{5}}B_{(nnnn)(2222)(00)},
    \label{eq:rank_4_l2_ex_2}
\end{equation}
where certain terms cancel for this to be true. Contributions from $|M_1|=|M_2|=1$ with $\boldsymbol{m}=\{-2101,-10-12\}$ cancel with those from $|M_1|=|M_2|=3$ and $\boldsymbol{m}=\{-2011,-1-102\}$. These multisets of $m_i$ yielding $|M_1|=|M_2|=1$ and $|M_1|=|M_2|=3$ are not found in \textit{any} permutation for any $\boldsymbol{m}$ with intermediate projection quantum numbers equalling zero, $|M_1|=|M_2|=0$, and such multisets of $m_i$ are the only ones contributing to $B_{\boldsymbol{nl}(00)}$, as seen in Eq. \eqref{eq:rank_4_l2_ex}. In a similar way, one may show that terms cancel for the case of $B_{(nnnn)(2222)(44)}$ to yield a relationship with $B_{(nnnn)(2222)(00)}$.
\begin{equation}
    B_{(nnnn)(2222)(44)}  = \frac{2}{21}B_{(nnnn)(2222)(00)},
    \label{eq:rank_4_l2_ex_3}
\end{equation}
As a result of Eq. \eqref{eq:rank_4_l2_ex_3}, we are also able to prove the first entry for the block of $\boldsymbol{l}=(2222)$ in Table 3 of Ref~[\citenum{dusson_atomic_2022}].

Attempting to repeat this for $\boldsymbol{l}=(3333)$ with all $\boldsymbol{n}=n$ degenerate, gives different patterns. The possible intermediates not resulting in trivial functions are $\{ (00), (22), (44), (66)\}$ and when we try to derive a relationship between $B_{\boldsymbol{nl}(00)}$ and the function with the largest intermediates, we obtain:
\begin{equation}
    B_{(nnnn)(3333)(66)}  = \frac{100}{33 (13\sqrt{13})}B_{(nnnn)(3333)(00)} + W_{3333}^{0000}((000),(000)) \sum_{\boldsymbol{m}}^{ordered}c_{\boldsymbol{m}}\boldsymbol{A}_{\boldsymbol{nlm}}
    \label{eq:rank_4_l3_ex_1}
\end{equation}
where $c_{\boldsymbol{m}}$ is a constant derived from regrouping products of the atomic base by their raised/lowered generalized Wigner symbol. It may change per multiset of $m_i$. In this case, we are not left with a constant factor between the $\boldsymbol{L}=(00)$ and the $\boldsymbol{L}=(66)$ descriptors, but a variable one where all terms in the second sum of Eq. \eqref{eq:rank_4_l3_ex_1} do not cancel. These two functions, by themselves, are linearly independent. Using other raising and lowering operations, one may relate $B_{\boldsymbol{nl}(66)}$ to $B_{\boldsymbol{nl}(22)}$ and $B_{\boldsymbol{nl}(44)}$ and many other relationships and ultimately show that only two independent functions are in the block, $\{ B_{(nnnn)(3333)(00)}, B_{(nnnn)(3333)(22)}, B_{(nnnn)(3333)(44)}, B_{(nnnn)(3333)(66)} \}$. It is straightforward generalize this to arbitrary $\boldsymbol{l}$ by incrementing the multiset of $l_i$ while maintaining the multiplicity of $l_i\in\boldsymbol{l}$ and repeating this process. After repeating this process and deriving relationships between regularly incremented multisets of intermediates, one may define a function sequence of independent RPI functions with arbitrary $\boldsymbol{l}=(llll)$. It is important to note that this function series is only defined for a block of RPI functions that have the same multiplicity for the multiset of angular indices as well as the same multiplicity/permutation of non-angular indices. Compact characterization of these conditions and these sets of functions may be given in terms of the frequency partitions of non-angular and angular indices, which are $P_f((nnnn))=(4)$ and $P_{f}((llll)))=(4)$, respectively.

\begin{equation} 
\begin{split}
    &\mathcal{F}_{i}^{OC}(\boldsymbol{l}, L_R) = \{ B_{(nnnn)(llll)(00)} , B_{(nnnn)(llll)(11)}, \cdots B_{(nnnn)(llll)\boldsymbol{L}_i} \cdots  B_{(nnnn)(llll)(2l \, 2l)}\} \\
    & \mathcal{F}_{a}^{PA}(P_f(\boldsymbol{n}), P_f(\boldsymbol{l})) = \{ B_{(nnnn)(llll)(00)},  \cdots B_{(nnnn)(llll)\boldsymbol{L}_{a=7i}}  \}
\end{split}
    \label{eq:ladder_relate2}
\end{equation}

In Eq. \eqref{eq:ladder_relate2}, the over-complete sequence of RPI functions is given in the first line. It is defined in terms of the angular indices, and it is worth noting that the indices $i$ in this equation correspond to one multiset of valid intermediates, $\boldsymbol{L}\in \{ \boldsymbol{L} \}^{\boldsymbol{l}}_{0}$ In general, this also depends on the value of $L_R$, but for brevity it is restricted to the important case of $L_R=0$. In this over-complete set, RPI functions indexed by intermediates with odd parity are included for completeness, even though they are zero-valued in this case. The sequence of independent RPI functions, $\mathcal{F}_{a}^{PA}$, may be obtained by sampling the over-complete function sequence,  $\mathcal{F}_{i}^{OC}$. The correct sampling of $\mathcal{F}_{i}^{OC}$ depends on both the frequency partition of non-angular and angular indices, which are $P_f((nnnn))=(4)$ and $P_{fc}((llll)))=(4)$ in Eq. \eqref{eq:ladder_relate2}. As previously mentioned, obtaining this sampling requires incrementing $\boldsymbol{l}$, deriving all ladder relationships  intermediates through substitutions and recursions. The result is that one may obtain a set of independent RPI functions within a specific block of $\boldsymbol{nl}$. The function sequences generated in this way (e.g., in Eq. \eqref{eq:ladder_relate2}) give function counts consistent with previously tabulated ACE bases using semi-numerical constructions.\cite{dusson_atomic_2022}

A limitation of this approach is that function sequence of independent RPI functions must be defined for each possible type of function block, characterized by $P_f(\boldsymbol{n}),P_f(\boldsymbol{l})$. While our newly derived properties of generalized Wigner symbols allow one to do this in theory for arbitrary $P_f(\boldsymbol{n}),P_f(\boldsymbol{l})$, obtaining the independent function sequence from ladder relationships with increasing $N$ becomes intensive. For this reason, $\mathcal{F}_{a}^{PA}$ have only been derived up to $N=8$. Though this rank is often more than enough for interatomic potentials, it is desirable extend these procedures to arbitrary $P_f(\boldsymbol{n}),P_f(\boldsymbol{l})$ and $N$. 

This procedure and these proofs are not limited to rank four functions. For the case of rank 5 and higher, recursion relationships and permutation automorphisms of the Wigner symbols may be used to show which high rank functions are equivalent, and derive relationships. We will not show the full procedure for rank 5 because it is cumbersome to show in detail, but we do provide the results. This is done for the case where all $n_i$ are equivalent with some arbitrary value, and all $l_i=2$.
\begin{equation}
    B_{(nnnnn)(22222)(022)} = 
    \begin{cases}
        B_{(nnnnn)(22222)(202)} \\[10pt]
        \frac{7}{2} B_{(nnnnn)(22222)(222)} \\[10pt]
        \frac{7}{2} B_{(nnnnn)(22222)(242)} \\[10pt]
        \frac{7}{2} B_{(nnnnn)(22222)(422)} \\[10pt]
        \frac{21}{10} \sqrt{\frac{11}{2}}B_{(nnnnn)(22222)(442)} \\
    \end{cases}
    \label{eq:rank5_ladder}
\end{equation}
This helps show that the method extends to cases where all $L_k$ are not equivalent. Similar series of independent RPI functions, Eq. \eqref{eq:ladder_relate2} can be obtained by incrementing $\boldsymbol{l}$ while maintaining the multiplicity of $l_i\in\boldsymbol{l}$. This approach is not limited to the case where all $n_i$ and $l_i$ are equivalent. One may derive ladder relationships for other amounts of duplicate $n_i$ and $l_i$ such as $\boldsymbol{n}=(11111)$ and $\boldsymbol{l}=(11112)$. It is done in the same general procedure of using permutation and recursion properties of the generalized Wigner symbols to obtain ladder relationships for multisets of $\boldsymbol{L}$ and incrementing $\boldsymbol{l}$ systematically such that multiplicity of $l_i\in\boldsymbol{l}$. For the second example of $\boldsymbol{l}=(11112)$ an incremented value $\boldsymbol{l}=(22224)$. In general, the sampling of the over-complete function series to obtain an independent one is differs between different $\boldsymbol{nl}$ blocks.

The method is not restricted to be used with the generalized Wigner symbols. Generalized Wigner symbols are used for convenient permutation symmetries. For example in Eq. \eqref{eq:rank5_ladder}, equivalent permutations of intermediates yield the same functions up to a phase (e.g. $B_{(nnnnn)(22222)(022)}=\phi B_{(nnnnn)(22222)(202)}$). Observing linear dependence is trivial in these cases. It is less so with the generalized Clebsch-Gordan coefficients. Similar relationships are found with descriptors constructed with generalized Clebsch-Gordan coefficients, however extra factors are often needed with certain permutations of the intermediates (e.g. $B_{(nnnnn)(22222)(022)}^{CG}=\phi c  B_{(nnnnn)(22222)(202)}^{CG}$ where $c$ is some constant resulting from the permutation). These terms are linearly dependent, but one has to consider the extra constant factor. Regardless, the analogous relationships for the important terms may obtained when using the generalized Clebsch-Gordan coefficients.
\begin{equation}
    B_{(nnnnn)(22222)(022)}^{CG} = 
    \begin{cases}
        \frac{7}{2\sqrt{5}} B_{(nnnnn)(22222)(222)} \\[10pt]
        \frac{7}{6} B_{(nnnnn)(22222)(242)} \\[10pt]
        \frac{7}{6}\sqrt{\frac{11}{10}}B_{(nnnnn)(22222)(442)} \\
    \end{cases}
    \label{eq:rank5_ladder_cg}
\end{equation}
Remaining terms for $B_{(nnnnn)(22222)(422)}$ and $B_{(nnnnn)(22222)(202)}$ can be obtained after considering permutation properties of the generalized Clebsch-Gordan coefficients.

\end{document}